\documentclass[11pt,a4paper]{article}

\usepackage[utf8x]{inputenc}

\usepackage{amsmath}
\usepackage{amsfonts}
\usepackage{amssymb}
\usepackage{amsthm}
\usepackage{mathptmx}
\usepackage{stmaryrd}
\usepackage[english]{babel}

\usepackage{subfiles}
\usepackage{hyperref}

\usepackage{authblk}

\usepackage{tikz}
\usetikzlibrary{cd}

\usepackage{mathdots}
\usepackage{yhmath}
\usepackage{cancel}
\usepackage{color}
\usepackage{siunitx}
\usepackage{array}
\usepackage{multirow}
\usepackage{gensymb}
\usepackage{tabularx}
\usepackage{booktabs}
\usetikzlibrary{fadings}

\usepackage{mdframed}

\theoremstyle{definition}
\newtheorem{definition}{Definition}
\newenvironment{defin}{\begin{mdframed}
[hidealllines=true,innerleftmargin=28pt,innerrightmargin=0pt,innertopmargin=-0.5\baselineskip, innerbottommargin=0.1\baselineskip, linewidth=0.7pt] 
\begin{definition}}{\end{definition}\end{mdframed}}

\newtheorem{mydef}{Notion}

\newtheorem{myreq}{Axiom}

\newtheorem{myaxiom}{A}

\newtheorem{theorem}{Theorem}
\newenvironment{theoreme}{\begin{mdframed}
[hidealllines=true,innerleftmargin=28pt,innerrightmargin=0pt,innertopmargin=-0.5\baselineskip, innerbottommargin=0.1\baselineskip, linewidth=0.7pt] 
\begin{theorem}}{\hfill $\blacksquare$ \end{theorem} \end{mdframed}}

\newtheorem{lemma}{\bf Lemma}
\newenvironment{lemme}{\begin{mdframed}
[hidealllines=true,innerleftmargin=28pt,innerrightmargin=0pt,innertopmargin=-0.5\baselineskip, innerbottommargin=0.1\baselineskip, linewidth=0.7pt] 
\begin{lemma}}{\hfill $\blacksquare$\end{lemma}\end{mdframed}} 

\newtheorem{corollary}{Corollary}
\newenvironment{corollaire}{\begin{mdframed}
[hidealllines=true,innerleftmargin=28pt,innerrightmargin=0pt,innertopmargin=-0.5\baselineskip, innerbottommargin=0.1\baselineskip, linewidth=0.7pt] 
\begin{corollary}}{\hfill $\blacksquare$\end{corollary}\end{mdframed}}

\newtheorem{Conject}{Conjecture}

\theoremstyle{remark}
\newtheorem{remark}{Remark}

\newcommand{\sqcoversubset}{\raisebox{1.3ex}{\,\rotatebox{180}{$\sqsupseteq$}}\,}

\renewcommand{\textbf}[1]{\begingroup\bfseries\mathversion{bold}#1\endgroup}

\newcommand{\existunique}{\exists \textit{\bf !}}
\renewcommand{\footnote}[1]{\textsuperscript{\addtocounter{footnote}{1}$\lfloor$\thefootnote$\rfloor$}\footnotetext{#1}}
\makeatletter
\newcommand{\subjclass}[2][2020]{ \let\@oldtitle\@title\gdef\@title{\@oldtitle\footnotetext{#1 \emph{\href{https://zbmath.org/static/msc2020.pdf}{Mathematics subject classification}.} #2}}}
\newcommand{\keywords}[1]{\let\@@oldtitle\@title\gdef\@title{\@@oldtitle\footnotetext{\emph{Keywords:} #1.}}}
\makeatother

\DeclareFontFamily{U}{mathx}{\hyphenchar\font45}
\DeclareFontShape{U}{mathx}{m}{n}{
      <5> <6> <7> <8> <9> <10>
      <10.95> <12> <14.4> <17.28> <20.74> <24.88>
      mathx10
      }{}
\DeclareSymbolFont{mathx}{U}{mathx}{m}{n}
\DeclareFontSubstitution{U}{mathx}{m}{n}
\DeclareMathAccent{\widecheck}{0}{mathx}{"71}
\DeclareMathAccent{\wideparen}{0}{mathx}{"75}

\begin{document}

\title{Generalized possibilistic Theories : \\ the multipartite experiments problem}
\author{Eric Buffenoir\footnote{Email: \href{mailto:eric.buffenoir@cnrs.fr}{eric.buffenoir@cnrs.fr}}}
\affil{\href{http://univ-cotedazur.fr}{Universit\'e de la C\^ote d'Azur}, \href{http://www.cnrs.fr}{CNRS}, \href{https://inphyni.cnrs.fr/fr}{InPhyNi}, FRANCE}

\maketitle

\begin{abstract}
In a recent paper, the author introduced an operational description of physical theories where probabilities are replaced by counterfactual statements belonging to a three-valued (i.e. possibilistic) semantic domain. The complete axiomatic of these Generalized possibilistic Theories is generalized and clarified in the present paper. The problem of bipartite experiments is then addressed as the main skill of this paper.  An axiomatic for the tensor product of our spaces of states is given and different solutions are explicitly constructed.  This description of tensor products of Inf semi-lattices is partly independent from the usual mathematical description of this problem. The nature of the tensor product of orthocomplemented Inf semi-lattices is then also explored. This subject is indeed fundamental for the development of a reconstruction program for quantum theory within our framework.  Our analysis constitutes a first step towards this achievement. 
\end{abstract}

\newpage

\section{Introduction}

{\em General Probabilistic Theory} (GPT) is a framework developed within the foundations of physics (see \cite{Plavala} for a recent review of the abundant literature and an axiomatic construction of GPTs). Promoters of GPT intent to answer the question: what is a physical theory? This study appeared initially in the context of axiomatizations of quantum theory, as many researchers were attempting to derive quantum theory from a set of reasonably motivated axioms. \\
In the current days,  research on GPT is oriented towards operational properties of GPTs, the main skill being to identify what structure is needed to realize certain protocols or constructions known from quantum information theory or classical information theory.  One uses GPTs to get better understanding of what makes different things in quantum information theory work. \\
Despite the indeterministic character of quantum theory, it is an empirical fact that the distinct outcomes of measurements, operated on a large collection of samples of a quantum object,  prepared according to the same experimental procedure, have reproducible relative frequencies. This fundamental fact has led physicists to consider large collections of statistically independent experimental sequences as the basic objects of  physical description, rather than a single experiment on a singular realization of the object under study.  According to GPT, a physical state (corresponding to a class of operationally equivalent preparation procedures) is then defined by a vector of probabilities associated with the outcomes of a maximal and irredundant set of fiducial tests that can be effectuated on collections of samples produced by any of these preparation procedures. In other words, two distinct collections of prepared samples will be considered as operationally equivalent if they lead to the same probabilities for the outcomes of any test on them. The physical description consists, therefore, in a set of prescriptions that allows sophisticated constructs to be defined from elementary ones. In particular, combination rules are defined for the concrete mixtures of states and for the allowed operations/tests. \\ 
It is a basic fact in GPT that this approach is the same as starting with an abstract state space, but instead of using vectors we would describe states in terms of all of the probabilities they can produce. In GPTs, ensembles of objects, conditional probabilities and conditional states can be represented by their respective state spaces and so we can treat them as any other state space and we can use known results, instead of having to prove them {\it ab initio}. Representing all transformations by "channels" allows us to use the constructions from frameworks based on category theory, since one can interpret state spaces as objects and channels as morphisms.\\

Although this probabilistic approach is now accepted as a standard conceptual framework for the reconstruction of quantum theory, the adopted perspective appears puzzling for different reasons.\\
First of all, the observer contributes fundamentally to give an intuitive meaning to the notions of preparation, operation and measurement on physical systems. However, the concrete process of 'acquisition of information' (by the observer / on the system) has no real place in this description. Secondly, the definition of the state has definitively lost its meaning for a singular prepared sample, and the physical state is now intrinsically attached to large collections of similarly prepared samples. The GPT approach adopts the probabilistic description of quantum phenomena without any discussion or attempt to explain why it is necessary. Thirdly, in order to clarify the requirements of the basic set of fiducial tests necessary and sufficient to define the space of states, this approach must proceed along a technical analysis which fundamentally limits this description to 'finite dimensional' systems (finite dimensional Hilbert spaces of states). Lastly, the axioms chosen to characterize quantum theory, among other theories encompassed by the GPT formalism, must exhibit a 'naturality' that - from our point of view - is still missing in the existing proposals.\\

Alternative research programs have tried to overcome some of these conceptual problems. Adopting another perspective, the {\em operational quantum logic} approach tries to avoid the introduction of probabilities and explores the relevant categorical structures underlying the space of states and the set of properties of a quantum system. In this description, probabilities appear only as a derived concept.  Following G. Birkhoff and J. Von Neumann \cite{Birkhoff1936} and G. W. Mackey \cite{MacKey63}, this approach focuses on the structured space of 'testable properties' of a physical system. The mathematical structure associated with the set of quantum propositions defined by the closed subspaces of a Hilbert space is not a Boolean algebra (contrary to the case encountered in classical mechanics). By shifting the attention to the set of closed subspaces instead of the Hilbert space itself, the possibility is open to build an operational approach to quantum mechanics, because the basic elements of this description are yes/no tests.  G.W.  Mackey identified axioms on the set of yes/no tests sufficient to relate this set to the set of closed subspaces of a complex Hilbert space. Later, C. Piron \cite{Piron1972,Piron1976} proposed a set of axioms that (almost) lead back to the general framework of quantum mechanics (see \cite{Coecke2000} for a historical perspective on the abundant literature). Piron’s framework has been developed into a full operational approach and the categories underlying this approach were analyzed. It must be noted that these constructions are established in reference to some general results of projective geometry and are not restricted to a finite-dimensional perspective. \\
Despite some beautiful results (in particular the restriction of the division ring associated to the Hilbert space from Piron's propositional lattices \cite{Holland:447786}) and the attractiveness of a completely categorical approach (see \cite{STUBBE2007477} for an analysis of the main results on propositional systems),  this approach has encountered several problems. Among these problems, we may cite the difficulty of building a consistent description of compound systems due to no-go results related to the existence of a tensor product of Piron's propositional systems \cite{randall1979tensor}\cite{aerts1984construction,Aerts2004}. These problems have cast doubts on the adequacy of Piron's choice of an "orthomodular complete lattice" structure for the set of properties of the system. \\ 

Other categorical formalisms, adapted to the axiomatic study of quantum theory, have been developed more recently \cite{ABRAMSKY2009261} and their relation with the 'operational approach' has been partly explored \cite{Abramsky2012,Abramsky2013, abramsky_heunen_2016}.  In \cite[Theorem 3.15]{Abramsky2012}, S. Abramsky makes explicit the fact that the {\em Projective quantum symmetry groupoid} $PSymmH$\footnote{The objects of this category are the natural space of states in quantum mechanics, i.e., the Hilbert spaces of dimension greater than two, and the morphisms are the orbits on semi-unitary maps (i.e. unitary or anti-unitary) under the $U(1)$ group action, which are the relevant symmetries of Hilbert spaces from the point of view of quantum mechanics.} is fully and faithfully represented by the category $bmChu_{[0,1]}$, i.e., by the sub-category of the category of bi-extensional Chu spaces associated {{}}{with} the evaluation set $[0,1]$ obtained by restricting {{}}{it} to Chu morphisms $(f_\ast,f^\ast)$ for which $f_\ast$ is injective. 
This result suggests that Chu categories could have a central role in the construction of axiomatic quantum mechanics as they provide a natural characterization of the automorphisms of the theory.  More surprisingly, and interestingly for us,  S. Abramsky shows that the aforementioned representation of $PSymmH$ is 'already' full and faithfull if we replace the evaluation space of the Chu category by a three-element set, where the three values represent "definitely yes", "definitely no" and "maybe" \cite[Theorem 4.4]{Abramsky2012}. S. Abramsky did not affirm that a three valued semantic is sufficient to found a complete axiomatic quantum theory, close to Piron's program or alternative to it, and allowing a complete reconstruction of the usual Hilbert formalism, although its result was clearly leading to this prospect.  It was the purpose of our last paper \cite{Buffenoir2021} to explore this question for the first time. This paper was devoted to present the basic elements of this 'possibilistic'\footnote{In the rest of this paper we refer to this construction, based on a three-valued Chu space, as a 'possibilistic' approach to distinguish it from the 'probabilistic' one.} semantic formalism.  \\ 

In the present paper, we continue to develop an analog of GPT based on this three-valued Chu space operational description of physical systems (as a sort of word game, we will designate this attempt as Generalized possibilistic Theory (GpT)). To allow for the same degree of generality as GPT, we present in Section 2 a set of axioms for the spaces of states and the spaces of effects of single systems which appears more general than in \cite{Buffenoir2021}.  In Section 3, we intentionally focus our study on the problem of bipartite experiments (this question had been led untouched in \cite{Buffenoir2021}). To complete this description, we exhibit a construction of the tensor product of complete semi-lattices which necessarily differs from the traditional construction of this tensor product, present in the mathematical literature.  This can be considered as a significant byproduct of the present paper, which deserves further investigations.  We finally address the problem of the orthocomplementation of the tensor product of orthocomplemented Inf semi-lattices. This problem appears fundamental for a reconstruction program of quantum theory,  as mentioned in \cite{Buffenoir2021}. 

\section{Generalized possibilistic Theories (GpT)}

Adopting the operational perspective on quantum experiments, we will introduce the following definitions.\\
A {\em preparation process} is an objectively defined, and thus 'repeatable', experimental sequence that allows singular samples of a certain physical system to be produced, in such a way that we are able to submit them to tests. 
We will denote by ${ \mathfrak{P}}$ the set of preparation processes (each element of ${ \mathfrak{P}}$ can be equivalently considered as the collection of samples produced through this preparation procedure). \footnote{The information corresponding to macroscopic events/operations describing the procedure depend on an observer $O$. 
If this dependence has to be made explicit, we will adopt the notation ${ \mathfrak{P}}^{{}^{(O)}}$ to denote the set of preparation processes defined by the observer $O$. This mention of the observer will be also attached to the different quotients associated to the space of preparations.}\\
For each  {\em property}, that the observer aims to test macroscopically on {\em any particular sample} of the considered micro-system, it will be assumed that the observer is able to define (i) some detailed 'procedure', in reference to the modes of use of some experimental apparatuses chosen to perform the operation/test, and (ii) a 'rule' allowing the answer 'yes' to be extracted if the macroscopic outcome of the experiment conforms with the expectation of the observer, when the test is performed on any input sample (as soon as this experimental procedure can be opportunely applied to this particular sample). 
These operations/tests, designed to determine the occurrence of a given property for a given sample, will be called {\em yes/no tests} {\em associated with this property}.  The set of 'yes/no tests' at the disposal of the observer will be denoted by ${ \mathfrak{T}}$. \footnote{If the dependence with respect to the observer $O$ has to be made explicit, we will adopt the notation ${ \mathfrak{T}}^{{}^{(O)}}$ to denote the set of tests defined by the observer $O$.  This mention of the observer will be also attached to the different quotients associated to the space of yes/no tests.}\\
A yes/no test ${ \mathfrak{t}}\in { \mathfrak{T}}$ will be said to be {\em positive with certainty} (resp. {\em negative with certainty}) relatively to a preparation process ${ \mathfrak{p}}\in { \mathfrak{P}}$ iff the observer is led to affirm that the result of this test, realized on any of the particular samples that could be prepared according to this preparation process, would be 'positive with certainty' (resp. would be 'negative with certainty'), 'should' this test be effectuated. If the yes/no test can not be stated as 'certain', this yes/no test will be said to be {\em indeterminate}. Concretely, the observer can establish the 'certainty' of the result of a given yes/no test on any given sample issued from a given preparation procedure, by running the same test on a sufficiently large (but finite) collection of samples issued from this same preparation process: if the outcome is always the same, the observer will be led to claim that similarly prepared 'new' samples would also produce the same result, if the experiment was effectuated. To summarize, for any preparation process ${ \mathfrak{p}}$ and any yes/no test ${ \mathfrak{t}}$, the element ${ \mathfrak{e}}({ \mathfrak{p}},{ \mathfrak{t}})\in { \mathfrak{B} } := \{{ \bot}, { \rm \bf Y}, { \rm \bf N}\}$ will be defined to be ${ \bot}$ (alternatively, ${ \rm \bf Y}$ or ${ \rm \bf N}$) if the outcome of the yes/no test ${ \mathfrak{t}}$ on any sample prepared according to the preparation procedure ${ \mathfrak{p}}$ is judged as 'indeterminate' ('positive with certainty' or 'negative with certainty', respectively) by the observer. 
\begin{eqnarray}
\begin{array}{rcrcl}
{ \mathfrak{e}} & : &{ \mathfrak{P}} \times { \mathfrak{T}} & \longrightarrow & { \mathfrak{B} } :=\{{ \bot}, { \rm \bf Y}, { \rm \bf N}\} \\
& &({ \mathfrak{p}},{ \mathfrak{t}}) & \mapsto &{ \mathfrak{e}}({ \mathfrak{p}},{ \mathfrak{t}}).
\end{array}
\end{eqnarray}


When the determinacy of a yes/no test is established for an observer, we can consider that this observer possesses some elementary 
'information' about the state of the system, whereas, in the 'indeterminate case', the observer has none (relatively to the occurrence of the considered property). \\ 
The set ${ \mathfrak{B} }$ will then be equipped with the following poset structure, characterizing the 'information' gathered by the observer: 
\begin{eqnarray}
\forall u,v\in { \mathfrak{B} },&& (u\leq v)\; :\Leftrightarrow\; (u={ \bot}\;\;\textit{\rm or}\;\; u=v).
\end{eqnarray} 
$({ \mathfrak{B}},\leq)$ is also an Inf semi-lattice which infima will be denoted $\bigwedge$. We have
\begin{eqnarray}
\forall x,y\in { \mathfrak{B} },&&  x \wedge y = \left\{\begin{array}{ll} x & \textit{\rm if}\;\;\; x=y\\
\bot & \textit{\rm if}\;\;\; x\not= y\end{array}\right.
\end{eqnarray}
We will also introduce a commutative monoid law denoted $\bullet$ and defined by
\begin{eqnarray}
\forall x\in { \mathfrak{B}},&& x \bullet \textit{\bf Y}=x, \;\;\;\;\; x \bullet \textit{\bf N}=\textit{\bf N}, \;\;\;\;\; \bot \bullet \bot = \bot. \label{expressionbullet}
\end{eqnarray}
$x \bullet y$ will be called {\em the product of the determinations $x$ and $y$}.\\
This law verifies the following properties
\begin{eqnarray}
\forall x\in { \mathfrak{B}},\forall B\subseteq { \mathfrak{B}} && x \bullet \bigwedge B=\bigwedge{}_{b\in B}(x\bullet b),\label{distributivitybullet}\\
\forall x\in { \mathfrak{B}},\forall C\subseteq_{Chain} { \mathfrak{B}} && x \bullet \bigvee B=\bigvee{}_{b\in B}(x\bullet b).\label{distributivitybullet2}
\end{eqnarray}

$({ \mathfrak{B} },\leq)$ will be also equipped with the following involution map :
\begin{eqnarray}
\overline{{ \bot}}:={ \bot} \;\;\;\;\;\;\;\; \overline{ \rm \bf Y}:={ \rm \bf N}\;\;\;\;\;\;\;\; \overline{ \rm \bf N}:={ \rm \bf Y}.
\end{eqnarray}

\subsection{The space of states}\label{subsectionstates}

A pre-order relation can be defined on the set ${ \mathfrak{P}}$ of preparation processes.  A preparation process ${ \mathfrak{p}}_2\in { \mathfrak{P}}$ is said to be {\em sharper} than another preparation process ${ \mathfrak{p}}_1\in { \mathfrak{P}}$ (this fact will be denoted ${ \mathfrak{p}}_1\sqsubseteq_{{}_{ \mathfrak{P}}} { \mathfrak{p}}_2$) iff any yes/no test ${ \mathfrak{t}}\in { \mathfrak{T}}$ that is 'determinate' for the samples prepared through ${ \mathfrak{p}}_1$ is also necessarily 'determinate' with the same determination for the samples prepared through ${ \mathfrak{p}}_2$, i.e.,
\begin{eqnarray}
 && \forall { \mathfrak{p}}_1, { \mathfrak{p}}_2\in { \mathfrak{P}},\;\;\;\;\;\;\;\; (\; { \mathfrak{p}}_1\sqsubseteq_{{}_{ \mathfrak{P}}} { \mathfrak{p}}_2\;) \;\; :\Leftrightarrow \;\; 
(\; \forall { \mathfrak{t}}\in { \mathfrak{T}},\; { \mathfrak{e}}({ \mathfrak{p}}_1,{ \mathfrak{t}})\leq { \mathfrak{e}}({ \mathfrak{p}}_2,{ \mathfrak{t}}) \;),\label{orderP}
\end{eqnarray}
If ${ \mathfrak{p}}_1\sqsubseteq_{{}_{ \mathfrak{P}}} { \mathfrak{p}}_2$ (i.e., ${ \mathfrak{p}}_2$ is 'sharper' than ${ \mathfrak{p}}_1$), ${ \mathfrak{p}}_1$ is said to be 'coarser' than ${ \mathfrak{p}}_2$.\\

An equivalence relation, denoted $\sim_{{}_{ \mathfrak{P}}}$, is defined on the set of preparations ${ \mathfrak{P}}$ from this pre-order relation.  Two preparation processes are identified iff the statements established by the observer about the corresponding prepared samples are identical.  A {\em state} of the physical system is an equivalence class of preparation processes corresponding to the same informational content.
The set of equivalence classes, modulo $\sim_{{}_{ \mathfrak{P}}}$, will be called {\em space of states} and denoted ${ { \mathfrak{S}}}$. In other words,
\begin{eqnarray}
\forall { \mathfrak{p}}_1,{ \mathfrak{p}}_2\in { \mathfrak{P}}, \;\; ({ \mathfrak{p}}_1\sim_{{}_{ \mathfrak{P}}} { \mathfrak{p}}_2)\;  & :\Leftrightarrow & 
(\; \forall { \mathfrak{t}}\in { \mathfrak{T}},\; { \mathfrak{e}}({ \mathfrak{p}}_1,{ \mathfrak{t}})= { \mathfrak{e}}({ \mathfrak{p}}_2,{ \mathfrak{t}}) \;) \Leftrightarrow  (\,{ \mathfrak{p}}_1\sqsubseteq_{{}_{ \mathfrak{P}}} { \mathfrak{p}}_2 \;\;\textit{\rm and}\;\; { \mathfrak{p}}_1\sqsupseteq_{{}_{ \mathfrak{P}}} { \mathfrak{p}}_2\;),\label{equivrelpreparations}\\
\lceil {{ \mathfrak{p}}} \rceil & := & \{\, { \mathfrak{p}}'\in { \mathfrak{P}}\;\vert\; { \mathfrak{p}}'\sim_{{}_{ \mathfrak{P}}} { \mathfrak{p}}\,\},\\
{ { \mathfrak{S}}} & := & \{\, \lceil {{ \mathfrak{p}}} \rceil \;\vert\; { \mathfrak{p}}\in { \mathfrak{P}}\,\}.
\end{eqnarray}

The space of states ${ { \mathfrak{S}}}$ is partially ordered.  Explicitly
\begin{eqnarray}
\forall \sigma_1,\sigma_2\in { { \mathfrak{S}}}, (\, \sigma_1 \sqsubseteq_{{}_{ { \mathfrak{S}}}} \sigma_2\,) \;\;:\Leftrightarrow\;\; (\, \forall { \mathfrak{p}}_1,{ \mathfrak{p}}_2 \in  { \mathfrak{P}}, \;\; 
(\, \sigma_1=\lceil {{ \mathfrak{p}}_1}\rceil , \sigma_2=\lceil {{ \mathfrak{p}}_2}\rceil \,)\;\;\Rightarrow\;\; (\,{ \mathfrak{p}}_1\sqsubseteq_{{}_{ \mathfrak{P}}} { \mathfrak{p}}_2\,) \,).
\end{eqnarray}

We will derive a map ${\epsilon}$ according to the following definition :
\begin{eqnarray}
\begin{array}{rcrcl}
{\epsilon} & : &{ \mathfrak{T}} & \rightarrow & { \mathfrak{B} }{}^{ \mathfrak{S}} \\
& &{ \mathfrak{t}} & \mapsto & {\epsilon}_{ { \mathfrak{t}}} \;\;\;\;\vert\;\;\;\;\; {\epsilon}_{ { \mathfrak{t}}}(\lceil { \mathfrak{p}} \rceil):={ \mathfrak{e}}({ \mathfrak{p}},{ \mathfrak{t}}),\; \forall { \mathfrak{p}}\in { \mathfrak{P}}.\label{defetilde}
\end{array}
\end{eqnarray}
For any ${ \mathfrak{t}}\in { \mathfrak{T}}$, ${\epsilon}_{ { \mathfrak{t}}}$ is an order-preserving  map on ${ \mathfrak{S}}$
\begin{eqnarray}
 && \forall \sigma_1, \sigma_2\in { \mathfrak{S}},\;\;\;\;\;\;\;\; (\; \sigma_1\sqsubseteq_{{}_{ \mathfrak{S}}} \sigma_2\;) \;\; :\Leftrightarrow \;\; 
(\; \forall { \mathfrak{t}}\in { \mathfrak{T}},\; \epsilon_{ \mathfrak{t}}(\sigma_1)\leq \epsilon_{ \mathfrak{t}}(\sigma_2) \;),\label{orderS}
\end{eqnarray}

If we consider a collection of preparation processes $P\subseteq { \mathfrak{P}}$, we can define a new preparation procedure, called {\em mixture} and denoted $\bigsqcap{}^{{}^{ \mathfrak{P}}} P$, as follows. The samples produced from the preparation procedure $\bigsqcap{}^{{}^{ \mathfrak{P}}} P$ are obtained by a random mixing of the samples issued from the preparation processes of the collection $P$ indiscriminately.  As a consequence, the statements that the observer can establish after a sequence of tests ${ \mathfrak{t}}\in { \mathfrak{T}}$ on these samples produced through the procedure $\bigsqcap{}^{{}^{ \mathfrak{P}}} P$ is given as the infimum of the statements that the observer can establish for the elements of $P$  separately. In other words, 
\begin{eqnarray}
&& \forall P\subseteq { \mathfrak{P}},\;\; \exists\; \bigsqcap{}^{{}^{ \mathfrak{P}}} P\in { \mathfrak{P}} \;\;\;\vert\;\;\;(\,\forall { \mathfrak{t}}\in { \mathfrak{T}},\; { \mathfrak{e}}(\bigsqcap{}^{{}^{ \mathfrak{P}}} P,{ \mathfrak{t}})=\bigwedge{}_{{}_{{ \mathfrak{p}}\in P}}{ \mathfrak{e}}({ \mathfrak{p}},{ \mathfrak{t}})\,).\;\;\;\;\;\;\;\;\;\label{propconjunctiveprop}
\end{eqnarray} 

The space of states inherits a notion of {\em mixed states} by defining
\begin{eqnarray}
\forall P\subseteq { \mathfrak{P}},&&\bigsqcap{}^{{}^{ \mathfrak{S}}}_{{}_{{ \mathfrak{p}}\in P}} \lceil  { \mathfrak{p}}\rceil :=\lceil \bigsqcap{}^{{}^{ \mathfrak{P}}}P \rceil.
\end{eqnarray}
As a result, the space of states inherits a structure of {\em down-complete Inf semi-lattice}.
In other words,
\begin{eqnarray}  
\textit{\bf (A1)}&&\forall S\subseteq {\mathfrak{S}},\;\; (\bigsqcap{}^{{}^{ \mathfrak{S}}} S) \;\; \textit{\rm exists in}\; {\mathfrak{S}}, \;\;\textit{\rm and}\;\; \forall { \mathfrak{t}}\in { \mathfrak{T}},\;{\epsilon}_{ { \mathfrak{t}}}(\bigsqcap{}^{{}^{ \mathfrak{S}}} S)=\bigwedge{}_{{}_{\sigma\in S}}\; {\epsilon}_{ { \mathfrak{t}}}(\sigma).
 \label{axiomsigmainfsemilattice}
\end{eqnarray}
As a direct consequence, the space of states is then also {\em bounded-complete}, i.e. 
\begin{eqnarray}
\forall S\subseteq { \mathfrak{S}}\;\vert\; \widehat{\;S\;}{}^{{}^{ \mathfrak{S}}},&& (\bigsqcup{}^{{}^{ \mathfrak{S}}} S) \; \textit{\rm exists in ${ \mathfrak{S}}$.}\label{Sboundedcomplete}
\end{eqnarray}
where 
\begin{eqnarray}
\forall{ \mathfrak{S}'}\subseteq { \mathfrak{S}}, \forall S\subseteq { \mathfrak{S}'}, && \widehat{\; S\;}{}^{{}^{{ \mathfrak{S}'}}}  :\Leftrightarrow  \exists \sigma'\in { \mathfrak{S}}'\;\vert\; \sigma \sqsubseteq_{{}_{ \mathfrak{S}}} \sigma',\forall \sigma\in S.\label{defwidehat}
\end{eqnarray}
We will adopt the shortened notation $\forall \sigma,\sigma'\in { \mathfrak{S}}, \widehat{\sigma\sigma'}{}^{{}^{ \mathfrak{S}}}:=\widehat{\{\sigma,\sigma'\}}{}^{{}^{ \mathfrak{S}}}$.\\

We will also assume that there exists a preparation process, unique from the point of view of the statements that can be produced about it, that can be interpreted as a 'randomly-selected' collection of 'un-prepared samples'. This element leads to complete indeterminacy for any yes/no test realized on it. 
\begin{eqnarray} 
&& \exists \;{ \mathfrak{p}}_{{ \bot}}\in { \mathfrak{P}}\;\;\vert \;\; (\,\forall { \mathfrak{t}}\in { \mathfrak{T}},\;\; { \mathfrak{e}}({ \mathfrak{p}}_{{ \bot}},{ \mathfrak{t}})={ \bot}\,).
\end{eqnarray} 
Hence, the partial order $({ { \mathfrak{S}}},\sqsubseteq_{{}_{ { \mathfrak{S}}}})$ admits {\em a bottom element}, denoted ${ \bot}_{{}_{ { \mathfrak{S}}}}:= \lceil  {{ \mathfrak{p}}_{ \bot}}\rceil$. In other words,
\begin{eqnarray}
\textit{\bf (A2)}&& \exists\; {\bot}_{{}_{ \mathfrak{S}}}\in {\mathfrak{S}}\;\vert \;\forall \sigma\in {\mathfrak{S}},\;{\bot}_{{}_{ \mathfrak{S}}}\sqsubseteq_{{}_{ \mathfrak{S}}} \sigma, \label{axiomsigmapointed}
\end{eqnarray}

\subsection{The space of effects}\label{subsectioneffects}

We can introduce a pre-order relation on the space of yes/no tests ${ \mathfrak{T}}$ as well :
\begin{eqnarray}
 && \forall { \mathfrak{t}}_1, { \mathfrak{t}}_2\in { \mathfrak{T}},\;\;\;\;\;\;\;\; (\; { \mathfrak{t}}_1\sqsubseteq_{{}_{ \mathfrak{T}}} { \mathfrak{t}}_2\;) \;\; :\Leftrightarrow \;\; 
(\; \forall \sigma\in { \mathfrak{S}},\; \epsilon_{{ \mathfrak{t}}_1}(\sigma)\leq \epsilon_{{ \mathfrak{t}}_2}(\sigma) \;),\label{orderT}
\end{eqnarray}
and an equivalence relation, denoted $\sim_{{}_{ \mathfrak{T}}}$, can be derived from this pre-order on the set of yes/no tests ${ \mathfrak{T}}$, i.e. ${ \mathfrak{t}}_1 \sim_{{}_{ \mathfrak{T}}} { \mathfrak{t}}_2$ is equivalent to $({ \mathfrak{t}}_1\sqsubseteq_{{}_{ \mathfrak{T}}} { \mathfrak{t}}_2 \;\textit{\rm and}\; { \mathfrak{t}}_1\sqsupseteq_{{}_{ \mathfrak{T}}} { \mathfrak{t}}_2 )$.  An {\em effect} of the physical system is an equivalence class of yes/no tests, i.e., a class of yes/no tests that are not distinguished from the point of view of the statements that the observer can produce by using these yes/no tests on finite collections of samples. The set of equivalence classes of yes/no tests, modulo the relation $\sim_{{}_{ \mathfrak{T}}}$, will be denoted ${ \mathfrak{E}}$.  In other words,
\begin{eqnarray}
\forall { \mathfrak{t}}_1, { \mathfrak{t}}_2\in { \mathfrak{T}},\;\; (\; { \mathfrak{t}}_1 \sim_{{}_{ \mathfrak{T}}} { \mathfrak{t}}_2\;) & :\Leftrightarrow &
(\; \forall \sigma\in { \mathfrak{S}},\; \epsilon_{{ \mathfrak{t}}_1}(\sigma)= \epsilon_{{ \mathfrak{t}}_2}(\sigma) \;),\label{simTchu}\\
\lfloor { \mathfrak{t}} \rfloor & := & \{\, { \mathfrak{t}}'\in { \mathfrak{T}}\;\vert\; { \mathfrak{t}}'\sim_{{}_{ \mathfrak{T}}} { \mathfrak{t}}\,\},\\
{ \mathfrak{E}} & := & \{\,\lfloor { \mathfrak{t}} \rfloor \;\vert\; { \mathfrak{t}}\in { \mathfrak{T}}\,\}.
\end{eqnarray}
The set of effects ${ \mathfrak{E}}$ is then equipped naturally with a partial order denoted $\sqsubseteq_{{}_{ \mathfrak{E}}} $.\\
We will adopt the following abuse of notation ${\epsilon}_{ \lfloor{ \mathfrak{t}}\rfloor }:={\epsilon}_{ { \mathfrak{t}}},\;\forall { \mathfrak{t}}\in { \mathfrak{T}}$. \\
We have by construction 
\begin{eqnarray}
\textit{\bf (Extensionality)}&& \forall { \mathfrak{l}},{ \mathfrak{l}}'\in { \mathfrak{E}},\;\;\;\;\;\;\;\;\;\;\;\;(\, \forall \sigma\in { \mathfrak{S}},\; {\epsilon}_{ { \mathfrak{l}}}(\sigma)={\epsilon}_{ { \mathfrak{l}}'}(\sigma) \,) \Leftrightarrow  (\, { \mathfrak{l}}= { \mathfrak{l}}' \,),\label{Chuextensional}\\
\textit{\bf (Separation)} && \forall \sigma,\sigma'\in { \mathfrak{S}},\;\;\;\;\;\;\;\;\;\;\;\;(\, \forall { \mathfrak{l}}\in { \mathfrak{E}},\; {\epsilon}_{ { \mathfrak{l}}}(\sigma)={\epsilon}_{ { \mathfrak{l}}}(\sigma') \,) \Leftrightarrow  (\, \sigma= \sigma' \,).\label{Chuseparated}
\end{eqnarray}
We note that $({ \mathfrak{S}},{ \mathfrak{E}},\epsilon)$ forms a bi-extensional Chu space \cite{Pratt1999}.  \\

If we consider a collection of tests $T\subseteq { \mathfrak{T}}$, we can define a new test, called {\em mixture} and denoted $\bigsqcap{}^{{}^{ \mathfrak{T}}} T$, as follows. The result obtained for the test $\bigsqcap{}^{{}^{ \mathfrak{T}}} T$ is obtained by a random mixing of the results issued from the tests of the collection $T$ indiscriminately.  As a consequence, the statements that the observer can establish after a sequence of tests is given as the infimum of the statements that the observer can establish for each test separately. In other words, 
\begin{eqnarray}
&& \forall T\subseteq { \mathfrak{T}},\;\; \exists\; \bigsqcap{}^{{}^{ \mathfrak{T}}} T\in { \mathfrak{T}} \;\;\;\vert\;\;\;(\,\forall \sigma\in { \mathfrak{S}},\; \epsilon_{\bigsqcap{}^{{}^{ \mathfrak{T}}} T}(\sigma)=\bigwedge{}_{{}_{{ \mathfrak{t}}\in T}}\epsilon_{ \mathfrak{t}}(\sigma)\,).\;\;\;\;\;\;\;\;\;\label{epsiloninft}
\end{eqnarray} 

The space of effects inherits a notion of {\em mixed effects} by defining
\begin{eqnarray}
\forall T\subseteq { \mathfrak{T}},&&\bigsqcap{}^{{}^{ \mathfrak{E}}}_{{}_{{ \mathfrak{t}}\in T}} \lceil  { \mathfrak{t}}\rceil :=\lceil \bigsqcap{}^{{}^{ \mathfrak{T}}}T \rceil.
\end{eqnarray}
As a result, the space of effects inherits a structure of {\em down-complete Inf semi-lattice}.
In other words,
\begin{eqnarray}  
\textit{\bf (A3)}&&\forall E\subseteq {\mathfrak{E}},\;\; (\bigsqcap{}^{{}^{ \mathfrak{E}}} E) \;\; \textit{\rm exists in}\; {\mathfrak{E}}, \;\;\textit{\rm and}\;\; \forall \sigma\in { \mathfrak{S}},\;{\epsilon}_{\bigsqcap{}^{{}^{ \mathfrak{E}}} E}(\sigma)=\bigwedge{}_{{}_{{ \mathfrak{l}}\in E}}\; {\epsilon}_{ { \mathfrak{l}}}(\sigma).
 \label{axiomEinfsemilattice}
\end{eqnarray}

The conjugate of a yes/no test ${{ \mathfrak{t}}}\in { \mathfrak{T}}$ is the yes/no test denoted $\overline{{ \mathfrak{t}}}$ and defined from ${{ \mathfrak{t}}}$ by exchanging the roles of {\bf Y} and {\bf N} in every {{}}{result} obtained by applying ${{ \mathfrak{t}}}$ to any given input sample. In other words,
\begin{eqnarray} \forall  { \mathfrak{t}}\in { \mathfrak{T}}, \forall  \sigma\in { \mathfrak{S}},&& \epsilon_{\overline{{ \mathfrak{t}}}}(\sigma):=\overline{ \epsilon_{{{ \mathfrak{t}}}}(\sigma) }.\label{etbar} \end{eqnarray} 
We note the following definition of the conjugate of an effect
\begin{eqnarray}\forall { \mathfrak{l}}\in { \mathfrak{E}},&& \overline{{ \mathfrak{l}}}=\{\, \overline{{ \mathfrak{t}}}\;\vert\; { \mathfrak{l}}=\lfloor{ \mathfrak{t}}\rfloor\,\}.\end{eqnarray} 

We will sometimes use a particular effect called "partial trace", denoted  ${ \mathfrak{Y}}_{ \mathfrak{E}}$ and defined by 
\begin{eqnarray}
\forall \sigma\in { \mathfrak{S}},&& \epsilon_{\;{ \mathfrak{Y}}_{ \mathfrak{E}}}(\sigma) := \textit{\bf Y}.
\end{eqnarray} 

An effect ${ \mathfrak{l}} \in { \mathfrak{E}}$ will be said to be {\em testable} iff it can be revealed as 'certain' at least for some collections of prepared samples.  In other words,  '${ \mathfrak{l}}$ is testable' means ${\epsilon}_{{ \mathfrak{l}}}^{\;-1}(\textit{\bf Y})\not= \varnothing$. 
\begin{lemme}
For any testable effect ${ \mathfrak{l}}$,  {{}}{there} exists an element ${\Sigma}_{{ \mathfrak{l}}}:=\bigsqcap{}^{{}^{ \mathfrak{S}}}{\epsilon}_{{ \mathfrak{l}}}^{\;-1}(\textit{\bf Y}) \;\in { \mathfrak{S}}$ 
, called {\em effect-state},  such that the filter ${\epsilon}_{{ \mathfrak{l}}}{}^{-1}(\textit{\bf Y})$ is the principal filter $(\uparrow^{{}^{ \mathfrak{S}}}\!\!{\Sigma}_{{ \mathfrak{l}}})$. 
\end{lemme}

We will allow for a generalized definition of effects.  Let us consider $\Sigma,\Sigma' \in { \mathfrak{S}}$ such that $\neg \widehat{\Sigma\Sigma'}{}^{{}^{ \mathfrak{S}}}$.  We denote ${ \mathfrak{l}}_{{}_{(\Sigma,\Sigma')}}$ the effect defined by 
\begin{eqnarray}
{\epsilon}_{\,{ \mathfrak{l}}_{{}_{(\Sigma,\Sigma')}}}^{\;-1}(\textit{\bf Y}) :=  \uparrow^{{}^{ \mathfrak{S}}}\!\!\Sigma & \textit{\rm and} & {\epsilon}_{\,{ \mathfrak{l}}_{{}_{(\Sigma,\Sigma')}}}^{\;-1}(\textit{\bf N}) :=  \uparrow^{{}^{ \mathfrak{S}}}\!\!\Sigma'. \label{expressioneffects}
\end{eqnarray}
By extension, we denote ${ \mathfrak{l}}_{{}_{(\Sigma,\cdot)}}$ the effect defined by 
\begin{eqnarray}
{\epsilon}_{\,{ \mathfrak{l}}_{{}_{(\Sigma,\cdot)}}}^{\;-1}(\textit{\bf Y}) :=  \uparrow^{{}^{ \mathfrak{S}}}\!\!\Sigma & \textit{\rm and} & {\epsilon}_{\,{ \mathfrak{l}}_{{}_{(\Sigma,\cdot)}}}^{\;-1}(\textit{\bf N}) :=  \varnothing
\end{eqnarray}
and by 
${ \mathfrak{l}}_{{}_{(\cdot,\Sigma')}}$ the effect defined by 
\begin{eqnarray}
&&{ \mathfrak{l}}_{{}_{(\cdot,\Sigma')}} := \overline{{ \mathfrak{l}}_{{}_{(\Sigma',\cdot)}}}
\end{eqnarray}

We note that the order on these effects is "inversed" with respect to the order on states. More precisely,
\begin{eqnarray}
\left\{
\begin{array}{l}
({ \mathfrak{l}}_{{}_{(\Sigma_1,\Sigma_1')}} \sqsubseteq_{{}_{ \mathfrak{E}}}
{ \mathfrak{l}}_{{}_{(\Sigma_2,\Sigma_2')}}) \;\Leftrightarrow\; (\Sigma_1 \sqsupseteq_{{}_{ \mathfrak{S}}}\Sigma_2 \;\textit{\rm and}\; \Sigma_1' \sqsupseteq_{{}_{ \mathfrak{S}}}\Sigma_2')\\
({ \mathfrak{l}}_{{}_{(\Sigma_1,\cdot)}} \sqsubseteq_{{}_{ \mathfrak{E}}}
{ \mathfrak{l}}_{{}_{(\Sigma_2,\Sigma_2')}}) \;\Leftrightarrow\; (\Sigma_1 \sqsupseteq_{{}_{ \mathfrak{S}}}\Sigma_2 )\\
({ \mathfrak{l}}_{{}_{(\cdot,\Sigma_1')}} \sqsubseteq_{{}_{ \mathfrak{E}}}
{ \mathfrak{l}}_{{}_{(\Sigma_2,\Sigma_2')}}) \;\Leftrightarrow\; ( \Sigma_1' \sqsupseteq_{{}_{ \mathfrak{S}}}\Sigma_2')
\end{array}
\right.
\label{expressionordereffects}
\end{eqnarray}

\begin{theoreme}  \label{aepsilonsigma}
Let us consider a map $(A : { \mathfrak{S}}\longrightarrow { \mathfrak{B}},  \sigma \mapsto { \mathfrak{a}}_{\sigma})$ satisfying
\begin{eqnarray}
\forall \sigma,\sigma'\in { \mathfrak{S}},&& (\sigma \sqsubseteq_{{}_{{ \mathfrak{S}}}} \sigma')\Rightarrow ({ \mathfrak{a}}_{\sigma} \leq { \mathfrak{a}}_{\sigma'}),\label{theorema1}\\
\forall \{\sigma_i\;\vert\; i\in I\}\subseteq { \mathfrak{S}},&&  { \mathfrak{a}}_{{\bigsqcap{}^{{}^{ \mathfrak{S}}}_{{}_{i\in i}}\sigma_i}} = \bigwedge{}_{{i\in I}} \;{ \mathfrak{a}}_{{\sigma_i}},\label{theorema2}
\end{eqnarray}
Then, we have
\begin{eqnarray}
\existunique \;{ \mathfrak{l}}\in { \mathfrak{E}} & \vert & \forall \sigma\in { \mathfrak{S}}, \; \epsilon_{{ \mathfrak{l}}}(\sigma)={ \mathfrak{a}}_{\sigma}.
\end{eqnarray}
\end{theoreme}
\begin{proof}
Straightforward. If $\{\, \sigma\;\vert\; { \mathfrak{a}}_{\sigma}=\textit{\bf Y}\,\}$ and $\{\, \sigma\;\vert\; { \mathfrak{a}}_{\sigma}=\textit{\bf N}\,\}$ are not empty, it suffices to define $\Sigma_A:=\bigsqcap^{{}^{{ \mathfrak{S}}}}\{\, \sigma\;\vert\; { \mathfrak{a}}_{\sigma}=\textit{\bf Y}\,\}$,  $\Sigma_A':=\bigsqcap^{{}^{{ \mathfrak{S}}}}\{\, \sigma\;\vert\; { \mathfrak{a}}_{\sigma}=\textit{\bf N}\,\}$ and ${ \mathfrak{l}}:={ \mathfrak{l}}_{(\Sigma_A,\Sigma_A')} $ (the case where some or all of these subsets are empty is treated immediately).
\end{proof}

\begin{corollaire}\label{corollaryfsigma}
For any $f:{ \mathfrak{S}} \longrightarrow { \mathfrak{S}}'$ satisfying $\forall \{\sigma_i\;\vert\; i\in I\}\subseteq { \mathfrak{S}}, \; f({\bigsqcap{}^{{}^{ \mathfrak{S}}}_{{}_{i\in i}}\sigma_i}) = \bigsqcap{}^{{}^{{ \mathfrak{S}}'}}_{{i\in I}} \;f(\sigma_i)$, there exists a map $f^\ast: { \mathfrak{E}}' \longrightarrow { \mathfrak{E}}$ such that $\forall { \mathfrak{l}} \in { \mathfrak{E}},\; \epsilon^{{ \mathfrak{S}}}_{f^\ast({ \mathfrak{l}})}(\sigma)=\epsilon^{{ \mathfrak{S}}'}_{ \mathfrak{l}}(f(\sigma))$.
\end{corollaire}

\begin{theoreme}  \label{blepsilonsigma}
Let us consider a map $(B : { \mathfrak{E}} \longrightarrow { \mathfrak{B}},{ \mathfrak{l}}\mapsto { \mathfrak{b}}_{{ \mathfrak{l}}})$ satisfying
\begin{eqnarray}
\forall { \mathfrak{l}}, { \mathfrak{l}}'\in { \mathfrak{E}},&& ({ \mathfrak{l}} \sqsubseteq_{{}_{{ \mathfrak{E}}}} { \mathfrak{l}}')\Rightarrow ({ \mathfrak{b}}_{{ \mathfrak{l}}} \leq { \mathfrak{b}}_{{ \mathfrak{l}}'}),\label{theorembl1}\\
\forall \{{ \mathfrak{l}}_i\;\vert\; i\in I\}\subseteq { \mathfrak{E}},&&  { \mathfrak{b}}_{{\bigsqcap{}^{{}^{ \mathfrak{E}}}_{{}_{i\in i}}{ \mathfrak{l}}_i}} = \bigwedge{}_{{i\in I}} \;{ \mathfrak{b}}_{{{ \mathfrak{l}}_i}},\label{theorembl2}\\
\forall { \mathfrak{l}}\in { \mathfrak{E}},&& { \mathfrak{b}}_{\overline{ \mathfrak{l}}}=\overline{{ \mathfrak{b}}_{{ \mathfrak{l}}}},\label{theorembl3}\\
&& { \mathfrak{b}}_{{ \mathfrak{Y}}_{ \mathfrak{E}}}=\textit{\bf Y}\label{theorembl4}
\end{eqnarray}
Then, we have
\begin{eqnarray}
\existunique \; \sigma\in { \mathfrak{S}} & \vert & \forall { \mathfrak{l}}\in { \mathfrak{E}}, \; \epsilon_{{ \mathfrak{l}}}(\sigma)={ \mathfrak{b}}_{{ \mathfrak{l}}}.
\end{eqnarray}
\end{theoreme}
\begin{proof}
Straightforward. It suffices to define ${ \mathfrak{l}}_B:=\bigsqcap^{{}^{{ \mathfrak{E}}}}\{\, { \mathfrak{l}}\in { \mathfrak{E}} \;\vert\; { \mathfrak{b}}_{{ \mathfrak{l}}}=\textit{\bf Y}\,\}$ and $\sigma:=\Sigma_{{}_{{ \mathfrak{l}}_B}}=\bigsqcap{}^{{}^{ \mathfrak{S}}}{\epsilon}_{{ \mathfrak{l}}_B}^{\;-1}(\textit{\bf Y})$.
\end{proof}

\begin{corollaire}\label{corollarychaincontinuous}
\begin{eqnarray}
\forall \{\sigma_i\;\vert\; i\in I\} \subseteq_{Chain} { \mathfrak{S}},\;\;\exists \sigma \in { \mathfrak{S}} & \vert & \forall { \mathfrak{l}}\in { \mathfrak{E}}, \; \epsilon_{{ \mathfrak{l}}}(\sigma)=\bigvee{}_{\!\!\! i\in I}\; \epsilon_{{ \mathfrak{l}}}(\sigma_i),\label{demochaincontinuous1}\\
\sigma & = & \bigsqcup{}^{{}^{{ \mathfrak{S}}}}_{{i\in I}}\;\sigma_i.
\end{eqnarray} 
\end{corollaire}
\begin{proof}
First of all, we note that $\{\sigma_i\;\vert\; i\in I\} \subseteq_{Chain} { \mathfrak{S}}$ and property (\ref{orderT}) implies that $\{\epsilon_{{ \mathfrak{l}}}(\sigma_i)\;\vert\; i\in I\} \subseteq_{Chain} { \mathfrak{B}}$ for any ${ \mathfrak{l}}\in { \mathfrak{E}}$ and then $\bigvee{}_{\!\!\! i\in I}\; \epsilon_{{ \mathfrak{l}}}(\sigma_i)$ exists for any ${ \mathfrak{l}}\in { \mathfrak{E}}$ due to the chain-completeness of ${ \mathfrak{B}}$.\\
Using the properties (\ref{orderT})(\ref{epsiloninft})(\ref{etbar}) of the map $\epsilon$ and the complete-distributivity properties satisfied by ${ \mathfrak{B}}$, we can check easily that the map ${ \mathfrak{l}}\mapsto \bigvee{}_{\!\!\! i\in I}\; \epsilon_{{ \mathfrak{l}}}(\sigma_i)$ satisfies properties (\ref{theorembl1}) (\ref{theorembl2}) (\ref{theorembl3}) (\ref{theorembl4}). As a consequence, the property (\ref{demochaincontinuous1}) is a direct consequence of Theorem \ref{blepsilonsigma}. \\
By definition of the poset structure (\ref{orderS}), we deduce, from the property $(\, \forall { \mathfrak{l}}\in { \mathfrak{E}}, \;\epsilon_{{ \mathfrak{l}}}(\sigma)=\bigvee{}_{\!\!\! i\in I}\; \epsilon_{{ \mathfrak{l}}}(\sigma_i) \,)$, that $\sigma \sqsupseteq_{{}_{{ \mathfrak{S}}}} \sigma_i,\; \forall i\in I$ and $(\sigma'\sqsupseteq_{{}_{{ \mathfrak{S}}}} \sigma_i,\; \forall i\in I)\Rightarrow (\sigma \sqsupseteq_{{}_{{ \mathfrak{S}}}} \sigma')$. In other words, $\sigma  =  \bigsqcup{}^{{}^{{ \mathfrak{S}}}}_{{}_{i\in I}}\sigma_i$.
\end{proof}

\subsection{Pure states}\label{subsectionpurestates}

A state is said to be {\em a pure state} iff it cannot be built as a mixture of other states (the set of pure states will be denoted ${ \mathfrak{S}}^{{}^{pure}}$). More explicitly,
\begin{eqnarray}
\sigma \in { \mathfrak{S}}^{{}^{pure}} & :\Leftrightarrow & (\,\forall S \subseteq^{\not= \varnothing} { \mathfrak{S}},\;\;  (\,\sigma = \bigsqcap{}^{{}^{ { \mathfrak{S}}}}S\,) \;\; \Rightarrow \;\; (\, \sigma \in S\,) \,).\label{purestates1} 
\end{eqnarray}
In other words, pure states are associated with completely meet-irreducible elements in ${ \mathfrak{S}}$. \footnote{We note that complete meet-irreducibility implies meet-irreducibility. In other words,
\begin{eqnarray}
\sigma \in { \mathfrak{S}}_{{}_{pure}} & \Rightarrow & (\, \forall \sigma_1, \sigma_2\in { \mathfrak{S}}, \;\;(\, \sigma =
\sigma_1 \sqcap_{{}_{ { \mathfrak{S}}}} \sigma_2\,) \;\; \Rightarrow \;\; (\, \sigma =  \sigma_1  \;\;\textit{\rm or}\;\; \sigma =  \sigma_2  \,) \,).
\end{eqnarray}
}

We will moreover assume that every state can be written as a mixture of pure states. In other words, 
\begin{eqnarray}
\textit{\bf (A4)}&&\forall \sigma \in { \mathfrak{S}}, \;\; \sigma= \bigsqcap{}^{{}^{ { \mathfrak{S}}}}  \underline{\sigma}_{{}_{ { \mathfrak{S}}}} ,\;\;\textit{\rm where}\;\;
\underline{\sigma}_{{}_{ { \mathfrak{S}}}}=
({ \mathfrak{S}}^{{}^{pure}} \cap (\uparrow^{{}^{ { \mathfrak{S}}}}\!\!\!\! \sigma) ).\label{completemeetirreducibleordergenerating}
\end{eqnarray}
\begin{remark}
If ${ \mathfrak{S}}$ is a bounded-complete algebraic domain (here, ${ \mathfrak{S}}$ is already assumed to be a  bounded-complete and chain-complete Inf semi-lattice), previous property is a direct consequence of \cite[Theorem I-4.26 p.126]{gierz_hofmann_keimel_lawson_mislove_scott_2003}. 
\end{remark}
\begin{remark}
We note that ${ \mathfrak{S}}^{{}^{pure}}=\sqcap-Irr({ \mathfrak{S}})$ is the unique smallest subset of ${ \mathfrak{S}}$ satisfying property (\ref{completemeetirreducibleordergenerating}). This point is mentioned in \cite[Remark I-4.22 p.125]{gierz_hofmann_keimel_lawson_mislove_scott_2003}. 
\end{remark}

A simple characterization of completely meet-irreducible elements within posets is given in \cite[Definition I-4.21]{gierz_hofmann_keimel_lawson_mislove_scott_2003} : 
\begin{eqnarray}
\sigma \in { \mathfrak{S}}^{{}^{pure}} & \Leftrightarrow & 
\left\{ \begin{array}{ll}
\sigma\in Max({ \mathfrak{S}}) & \textit{\rm (Type 1)}\\
(\uparrow^{{}^{ { \mathfrak{S}}}}\!\!\!\! \sigma)\!\smallsetminus\! \{\sigma\} \;\;\textit{\rm admits a minimum element} & \textit{\rm (Type 2)}
\end{array}\right.
\end{eqnarray}
This characterization is equivalent to the basic definition (\ref{purestates1}) for {a} bounded{-}complete Inf semi-lattice like ${ \mathfrak{S}}$.\\

From Corollary \ref{corollarychaincontinuous}, using Zorn's Lemma, we deduce that 
\begin{eqnarray}
\forall \sigma\in { \mathfrak{S}}, && \exists \sigma'\in Max({ \mathfrak{S}})\;\vert\; \sigma \sqsubseteq_{{}_{{ \mathfrak{S}}}}\sigma'.\label{existsmaxabove}
\end{eqnarray}
From that remark, we can decide to eliminate Type 2 pure states. Indeed, it is clear that 'Type 2' pure states have no physical meaning. Indeed, for any 'Type 2' pure states, it exists some 'Type 1' pure states sharper than it (and, then, containing more information than it). The existence of 'Type 2' pure states in the space of states leads then to a redundant description of the system.  We will then require that ${ \mathfrak{S}}^{{}^{pure}}$, i.e. the set of completely meet-irreducible elements $\sqcap-Irr({ \mathfrak{S}})$,  be constituted exclusively of maximal elements of ${ \mathfrak{S}}$. In other words, we require the space of states to be such that
\begin{eqnarray} 
\textit{\bf (A5)}&&\sqcap-Irr({ \mathfrak{S}})=Max({ \mathfrak{S}}).\label{axiomsigmameetirreducible}
\end{eqnarray}

From now on, Chu spaces $({ \mathfrak{S}},{ \mathfrak{E}},\epsilon)$ which elements satisfy the axioms {\bf (A1) $-$ (A5)} will be called {\em States/Effects Chu spaces}.

\subsection{The space of effects as a space of states}

From axiom {\bf (A3)}, analog to the axiom {\bf (A1)}, we know that the space of effects looks like a space of states.  ${ \mathfrak{E}}$ is indeed a down-complete Inf semi-lattice (and then a bounded-complete Inf semi-lattice).\\
We need to complete this information to be able to consider spaces of effects plainly as spaces of states. \\ 

First of all, we have the following result analog to the axiom {\bf (A2)}.
\begin{lemme}
The space of effects has a bottom element denoted $\bot_{{}_{ \mathfrak{E}}}$ and defined by
\begin{eqnarray}
\forall \sigma\in { \mathfrak{S}}, && \epsilon_{\bot_{{}_{ \mathfrak{E}}}}(\sigma) = \bot. 
\end{eqnarray}
\end{lemme}
\begin{proof}
Obvious.
\end{proof}

Secondly, ${ \mathfrak{E}}$ appears to satisfy the following chain-completeness property
\begin{lemme}
\begin{eqnarray}
\forall \{{ \mathfrak{l}}_i\;\vert\; i\in I\} \subseteq_{Chain} { \mathfrak{E}},\;\;\exists { \mathfrak{l}} \in { \mathfrak{E}} & \vert & \forall \sigma\in { \mathfrak{S}}, \; \epsilon_{{ \mathfrak{l}}}(\sigma)=\bigvee{}_{\!\!\! i\in I}\; \epsilon_{{ \mathfrak{l}}_i}(\sigma),\label{demochaincontinuous1}\\
{ \mathfrak{l}} & = & \bigsqcup{}^{{}^{{ \mathfrak{E}}}}_{{i\in I}}\;{ \mathfrak{l}}_i.
\end{eqnarray} 
\end{lemme}
\begin{proof}
Immediate consequence of axiom {\bf (A1)} using the general expression of effects introduced in (\ref{expressioneffects}) and subsequents, because of the properties (\ref{expressionordereffects}). 

\end{proof}
Moreover, the domain ${ \mathfrak{E}}$ appears to be algebraic.
\begin{lemme}
The poset ${ \mathfrak{E}}$ is atomic.
\begin{eqnarray}
\exists { \mathcal{A}}_{{ \mathfrak{E}}}\subseteq { \mathfrak{E}} & \vert &
\left\{
\begin{array}{l}\forall { \mathfrak{l}}\in { \mathcal{A}}_{{ \mathfrak{E}}},\;\; \bot_{{}_{ \mathfrak{E}}} \sqcoversubset_{{}_{{ \mathfrak{E}}}} { \mathfrak{l}} \\
\forall { \mathfrak{l}} \in { \mathfrak{E}}\smallsetminus \{\bot_{{}_{ \mathfrak{E}}}\}, \;\; \exists { \mathfrak{l}}'\in { \mathcal{A}}_{{ \mathfrak{E}}}\;\vert\; { \mathfrak{l}}' \sqsubseteq_{{}_{{ \mathfrak{E}}}} { \mathfrak{l}}
\end{array}\right.\label{defatomeffects}\\
{ \mathcal{A}}_{{ \mathfrak{E}}} &=& \{{ \mathfrak{l}}_{{}_{(\Sigma,\cdot)}}\vert \Sigma\in { \mathfrak{S}}^{{}^{pure}}\}\cup \{{ \mathfrak{l}}_{{}_{(\cdot,\Sigma)}}\vert \Sigma\in { \mathfrak{S}}^{{}^{pure}}\}\label{expressionatomeffects}
\end{eqnarray}
The poset ${ \mathfrak{E}}$ is atomistic, i.e. 
\begin{eqnarray}
\forall { \mathfrak{l}}\in { \mathfrak{E}}, && { \mathfrak{l}}= \bigsqcup{}^{{}^{ \mathfrak{E}}} \{\, { \mathfrak{l}}'\in { \mathcal{A}}_{{ \mathfrak{E}}} \;\vert\; { \mathfrak{l}}' \sqsubseteq_{{}_{ \mathfrak{E}}}{ \mathfrak{l}}\}.\label{atomisticeffects}
\end{eqnarray}
and then algebraic (i.e. compactly generated).
\end{lemme}
\begin{proof}
The properties (\ref{existsmaxabove}) implies directly the second condition of (\ref{defatomeffects}). The first condition of (\ref{defatomeffects}) is easy to check using the expression of the order (\ref{expressionordereffects}).\\
The property (\ref{atomisticeffects}) is a direct consequence of axiom {\bf (A4)}.\\
The atoms being trivially compact, the algebraicity follows.
\end{proof}

Thirdly, we have the following result analog to the axiom {\bf (A4)}.  As before, we define the "pure effects" as follows:
 \begin{eqnarray}
{ \mathfrak{l}} \in { \mathfrak{E}}^{{}^{pure}} & :\Leftrightarrow & (\,\forall E \subseteq^{\not= \varnothing} { \mathfrak{E}},\;\;  (\,{ \mathfrak{l}} = \bigsqcap{}^{{}^{ { \mathfrak{E}}}}E\,) \;\; \Rightarrow \;\; (\, { \mathfrak{l}} \in E\,) \,).\label{pureeffects1} 
\end{eqnarray}
In other words, pure states are associated with completely meet-irreducible elements in ${ \mathfrak{E}}$. \\
In order to complete our comparison, we have to check that every effect can be written as a mixture of pure effects. 
\begin{theoreme}
The space of effects satisfies
\begin{eqnarray}
&&\forall { \mathfrak{l}} \in { \mathfrak{E}}, \;\; { \mathfrak{l}}= \bigsqcap{}^{{}^{ { \mathfrak{E}}}}  \underline{\;{ \mathfrak{l}}\;}_{{}_{ { \mathfrak{E}}}},\;\;\textit{\rm where}\;\;
\underline{\;{ \mathfrak{l}}\;}_{{}_{ { \mathfrak{E}}}}=
({ \mathfrak{E}}^{{}^{pure}} \cap (\uparrow^{{}^{ { \mathfrak{E}}}}\!\!\!\! { \mathfrak{l}}) ).\label{completemeetirreducibleordergeneratingeffects}
\end{eqnarray}
\end{theoreme}
\begin{proof}
From previous Lemmas  ${ \mathfrak{E}}$ is a bounded-complete algebraic domain, The property (\ref{completemeetirreducibleordergeneratingeffects}) is then a direct consequence of \cite[Theorem I-4.26 p.126]{gierz_hofmann_keimel_lawson_mislove_scott_2003}. 
\end{proof}
As before, we have the following characterization of elements of ${ \mathfrak{E}}^{{}^{pure}}$ \cite[Definition I-4.21]{gierz_hofmann_keimel_lawson_mislove_scott_2003} : 
\begin{eqnarray}
\sigma \in { \mathfrak{E}}^{{}^{pure}} & \Leftrightarrow & 
\left\{ \begin{array}{ll}
\sigma\in Max({ \mathfrak{E}}) & \textit{\rm (Type 1)}\\
(\uparrow^{{}^{ { \mathfrak{E}}}}\!\! { \mathfrak{l}})\!\smallsetminus\! \{{ \mathfrak{l}}\} \;\;\textit{\rm admits a minimum element} & \textit{\rm (Type 2)}
\end{array}\right.
\end{eqnarray}
This characterization is equivalent to the basic definition (\ref{pureeffects1}) for a bounded-complete Inf semi-lattice like ${ \mathfrak{E}}$.\\
We observe it is natural to eliminate Type 2 pure effects.Indeed, for any 'Type 2' pure effects, it exists some 'Type 1' pure effects extracting more information than it. The existence of 'Type 2' pure states in the space of states leads then to a redundant description of the system.  Happily, this condition is already a simple consequence of Axiom {\bf (A5)}.
\begin{lemme}
\begin{eqnarray} 
&&\sqcap-Irr({ \mathfrak{E}})=Max({ \mathfrak{E}}).\label{axiomeffectsmeetirreducible}
\end{eqnarray}
\end{lemme}
\begin{proof}
This result is obtained directly from the previous characterization of pure effects. Indeed, a Type 2 pure effect is of the form ${ \mathfrak{l}}_{{}_{(\Sigma,\Sigma')}}$ with either $\Sigma$ or $\Sigma'$ being a Type 2 pure state. The absence of Type 2 pure state closes the proof.
\end{proof}

As a final conclusion of previous results and axioms, we obtain the complete characterization of spaces of effects.
\begin{theoreme}
A space of effects is nothing else than a space of states.
\end{theoreme}

We can go further by characterizing explicitly the elements of $Max({ \mathfrak{E}})$. According to \cite[Section 3.5]{Buffenoir2021}, we introduce the following binary relation, denoted $\widecheck{\bowtie}_{{}_{ \mathfrak{S}}}$  and defined on ${ \mathfrak{S}}$ by
\begin{eqnarray}
\forall (\sigma,\sigma')\in { \mathfrak{S}}^{\times 2},\;\;\;\;\;\sigma\, \widecheck{\bowtie}_{{}_{ \mathfrak{S}}}\sigma'  & :\Leftrightarrow & (\,\forall \sigma'' \sqsubset_{{}_{ \mathfrak{S}}}\sigma', \widehat{\sigma\sigma''}{}^{{}^{ \mathfrak{S}}}\;\;\textit{\rm and}\;\; \forall \sigma'' \sqsubset_{{}_{ \mathfrak{S}}}\sigma, \widehat{\sigma'\sigma''}{}^{{}^{ \mathfrak{S}}}\;\textit{\rm and not}\; \widehat{\sigma\sigma'}{}^{{}^{ \mathfrak{S}}}\,).\;\;\;\;\;\;\;\;\;\;\label{quasiantipodal}
\end{eqnarray}
(here we have used the notation introduced in (\ref{defwidehat})).
\begin{lemme}
\begin{eqnarray}
{ \mathfrak{E}}^{{}^{pure}}=Max({ \mathfrak{E}}) &=& \{\, { \mathfrak{l}}_{{}_{(\Sigma,\Sigma')}}\;\vert\; \Sigma\, \widecheck{\bowtie}_{{}_{ \mathfrak{S}}}\Sigma'\,\} \cup \{{ \mathfrak{Y}}_{{}_{ \mathfrak{E}}} \}\cup \{\overline{{ \mathfrak{Y}}_{{}_{ \mathfrak{E}}}} \}
\end{eqnarray}
\end{lemme}
\begin{proof}
Obvious.
\end{proof}

\subsection{Particular types of spaces of states}

According to \cite[definition p.117 and Section 11 Lemma 1 p.118]{Gratzer1971}, we introduce the following notion.
\begin{defin} \label{defdistrib}  
An Inf semi-lattice ${ \mathfrak{S}}$ is said to be {\em distributive} iff 
\begin{eqnarray}
\hspace{-1.8cm}\forall \sigma,\sigma_1,\sigma_2\in { \mathfrak{S}}\;\vert\; \sigma\not= \sigma_1,\sigma_2 ,\;\;\;\;\;\; (\sigma_1 \sqcap_{{}_{ \mathfrak{S}}} \sigma_2) \sqsubseteq_{{}_{ \mathfrak{S}}}  \sigma  &\Rightarrow & \exists \sigma'_1,\sigma'_2\in { \mathfrak{S}}\;\vert\; (\,\sigma_1 \sqsubseteq_{{}_{ \mathfrak{S}}} \sigma'_1,\;\;\; \sigma_2 \sqsubseteq_{{}_{ \mathfrak{S}}} \sigma'_2\;\;\;\textit{\rm and}\;\;\; \sigma = \sigma'_1 \sqcap_{{}_{ \mathfrak{S}}} \sigma'_2\,).\;\;\;\;\;\;\;\;\;\;\;\;
\end{eqnarray}
When ${ \mathfrak{S}}$ is distributive, we have the following standard properties satisfied, as soon as the implied suprema are well defined
\begin{eqnarray}
\sigma_1 \sqcap_{{}_{ \mathfrak{S}}} (\sigma_2 \sqcup_{{}_{ \mathfrak{S}}} \sigma_3) &=& (\sigma_1 \sqcap_{{}_{ \mathfrak{S}}} \sigma_2) \sqcup_{{}_{ \mathfrak{S}}} (\sigma_1 \sqcap_{{}_{ \mathfrak{S}}} \sigma_3)\\
\sigma_1 \sqcup_{{}_{ \mathfrak{S}}} (\sigma_2 \sqcap_{{}_{ \mathfrak{S}}} \sigma_3) &=& (\sigma_1 \sqcup_{{}_{ \mathfrak{S}}} \sigma_2) \sqcap_{{}_{ \mathfrak{S}}} (\sigma_1 \sqcup_{{}_{ \mathfrak{S}}} \sigma_3).
\end{eqnarray}
\end{defin}

\begin{defin}
The space of states ${ \mathfrak{S}}$ is said to be {\em equipped with a star} iff there exists a map $\star : { \mathfrak{S}}\smallsetminus \{\bot_{{}_{ \mathfrak{S}}}\} \rightarrow { \mathfrak{S}}\smallsetminus \{\bot_{{}_{ \mathfrak{S}}}\}$ such that
\begin{eqnarray}
&&\forall \sigma\in { \mathfrak{S}}, \;\; \sigma^{\star\star}=\sigma \label{involutive}\\
&& \forall \sigma_1,\sigma_2\in { \mathfrak{S}}, \;\; \sigma_1\sqsubseteq_{{}_{{ \mathfrak{S}}}}\sigma_2\;\;\Rightarrow\;\; \sigma_2^\star \sqsubseteq_{{}_{{ \mathfrak{S}}}} \sigma_1^\star \label{orderreversing}\\
&&\forall \sigma\in { \mathfrak{S}}, \;\; \neg\; \widehat{\sigma \sigma^\star}{}^{{}^{{ \mathfrak{S}}}} \label{inconsistent}\\
&&\forall \sigma\in { \mathfrak{S}}^{{}^{pure}}, \;\; \sigma^\star \in At({ \mathfrak{S}}).\label{staratom}
\end{eqnarray}
If we have moreover 
\begin{eqnarray}
\forall \sigma\in { \mathfrak{S}},&&\sigma\, \widecheck{\bowtie}_{{}_{ \mathfrak{S}}}\sigma^{\star},\label{starcomplement}
\end{eqnarray}
then the space of states ${ \mathfrak{S}}$ is said to be {\em orthocomplemented}.\\
We say that an Inf semi-lattice ${ \mathfrak{S}}$ is {\em orthogonal} iff ${ \mathfrak{S}}$ is equipped with a star map such that 
\begin{eqnarray}
\forall \sigma\in { \mathfrak{S}},&& { \mathfrak{S}}^{{}^{pure}}= \underline{\sigma}_{{}_{ { \mathfrak{S}}}}\cup \underline{\sigma^\star}_{{}_{ { \mathfrak{S}}}}.
\end{eqnarray}
If ${ \mathfrak{S}}$ is orthogonal, it is tautologically orthocomplemented.
\end{defin}


\begin{defin}
As long as the space ${ \mathfrak{S}}$ is equipped with a star map, we can build a map $\langle\cdot,\cdot\rangle$ from ${ \mathfrak{S}}\times { \mathfrak{S}}$ to ${ \mathfrak{B}}$ defined as follows
\begin{eqnarray}
\forall \sigma,\sigma'\in { \mathfrak{S}},\;\;\;\; \langle\sigma,\sigma'\rangle & := & \left\{\begin{array}{lcl}
\textit{\bf Y} & \textit{\rm iff} & \sigma\in { \mathfrak{S}}^{{}^{pure}} \;\textit{\rm and}\; \sigma=\sigma'\\
\textit{\bf N} & \textit{\rm iff} & \sigma\not=\bot_{{}_{{ \mathfrak{S}}}} \;\textit{\rm and}\; \sigma' \sqsupseteq_{{}_{{ \mathfrak{S}}}}\sigma^\star\\
\bot & \textit{\rm otherwise}
\end{array}\right.\label{explicitomega1}
\end{eqnarray}
Note first of all that to have a consistent definition we use explicitly the property (\ref{inconsistent}).\\
We note secondly the following fundamental property :
\begin{eqnarray}
\langle \sigma,\sigma'\rangle = \langle \sigma', \sigma\rangle \label{symmetrylangle}
\end{eqnarray} 
To check this symmetry, we have to use the properties (\ref{involutive}) and (\ref{orderreversing}) of the star operation on ${ \mathfrak{S}}$.\\
We note endly that the map $\langle\cdot,\cdot\rangle$ from ${ \mathfrak{S}}\times { \mathfrak{S}}$ to ${ \mathfrak{B}}$ is bimorphic, i.e.
\begin{eqnarray}
\forall \{\,\sigma_i\;\vert\; i\in I\,\}\subseteq { \mathfrak{S}},&&  \langle\sigma',\bigsqcap{}^{{}^{{ \mathfrak{S}}}}_{i\in I}\,\sigma_i\rangle=\bigwedge{}_{i\in I}\, \langle\sigma',\sigma_i\rangle. 
\end{eqnarray}
\end{defin}
\begin{defin}
We can define as usual an orthogonality relation on ${ \mathfrak{S}}$
\begin{eqnarray}
\forall \sigma,\sigma'\in { \mathfrak{S}},\;\; \sigma {\underline{\perp}} \sigma' & :\Leftrightarrow & \sigma' \sqsupseteq_{{}_{{ \mathfrak{S}}}}  \sigma^\star \\
&\Leftrightarrow &  \langle \sigma,\sigma'\rangle =\textit{\bf N}.
\end{eqnarray}
The binary relation ${\underline{\perp}}$ is symmetric and irreflexive.\\
We note that the star map can be recovered from the relation $\perp$ or from the bracket by
\begin{eqnarray}
\forall \sigma\in { \mathfrak{S}},\;\; \sigma^\star &=& \bigsqcap{}^{{}^{{ \mathfrak{S}}}}\{\sigma'\;\vert\; \sigma'{\underline{\perp}} \sigma\}= \bigsqcap{}^{{}^{{ \mathfrak{S}}}}\{\sigma'\;\vert\; \langle\sigma',\sigma\rangle=\textit{\bf N}\}.
\end{eqnarray}
\end{defin}

\begin{lemme}\label{equivalentorthomodular}
The orthocomplementation condition (\ref{starcomplement}) is equivalent to the following condition
\begin{eqnarray}
\forall \sigma\in { \mathfrak{S}},\forall \sigma'\in { \mathfrak{S}},&& \sigma^{\underline{\perp}}\; \cap \; \uparrow^{{}^{{ \mathfrak{S}}}}\!\!\!\!\! (\sigma \sqcap_{{}_{{ \mathfrak{S}}}} \sigma') \not= \varnothing.
\end{eqnarray}
\end{lemme}

\begin{defin}
We say that a family $\{\,\alpha_i\;\vert\; i\in I\,\}\subseteq { \mathfrak{S}}^{{}^{pure}}$ is an orthonormal basis of ${ \mathfrak{S}}$ iff 
\begin{eqnarray}
&&\forall i\in I,\;\; \langle\alpha_i,\alpha_i\rangle = \textit{\bf Y}\\
&& \forall i,j\in I,\;\; i\not=j \;\;\Rightarrow \;\;\langle\alpha_i,\alpha_j\rangle = \textit{\bf N}\\
&&\forall \sigma\in { \mathfrak{S}},\;\; \exists \{\,\alpha_j\;\vert\; j\in J\,\}\subseteq \{\,\alpha_i\;\vert\; i\in I\,\} \;\; \vert \;\; \sigma \sqsupseteq_{{}_{{ \mathfrak{S}}}} \bigsqcap{}^{{}^{{ \mathfrak{S}}}}_{j\in J}\alpha_j.
\end{eqnarray}
\end{defin}

\begin{lemme}
If ${\mathcal{A}}:=\{\,\alpha_i\;\vert\; i\in I\,\}\subseteq { \mathfrak{S}}^{{}^{pure}}$ is an orthonormal basis of ${ \mathfrak{S}}$, then the sub Inf semi-lattice generated by the family $\{\,\alpha_i\;\vert\; i\in I\,\}$ in ${ \mathfrak{S}}$, denoted ${ \mathfrak{S}}_{{\mathcal{A}}}$, is orthogonal. More precisely, the star map $\star$ defined by
\begin{eqnarray}
\forall \sigma:=\bigsqcap{}^{{}^{ \mathfrak{S}}}_{j\in J}\alpha_j \in { \mathfrak{S}}_{{\mathcal{A}}},&& \sigma^\star:=\bigsqcap{}^{{}^{ \mathfrak{S}}}_{k\in I\smallsetminus J}\alpha_k \in { \mathfrak{S}}_{{\mathcal{A}}}
\end{eqnarray}
satisfies
\begin{eqnarray}
\forall \sigma\in { \mathfrak{S}}_{{\mathcal{A}}}, && \{\alpha\in {\mathcal{A}}\;\vert\; \alpha\sqsupseteq_{{}_{{ \mathfrak{S}}}} \sigma\} \cup \{\alpha\in {\mathcal{A}}\;\vert\; \alpha\sqsupseteq_{{}_{{ \mathfrak{S}}}} \sigma^\star\} = {\mathcal{A}}.
\end{eqnarray}
Moreover,  ${ \mathfrak{S}}_{{\mathcal{A}}}$ is distributive.
\end{lemme}

\begin{lemme}
If ${ \mathfrak{S}}$ is orthocomplemented and finite, an orthonormal basis of ${ \mathfrak{S}}$ necessarily exists.
\end{lemme}
\begin{proof}
We begin with a random pure state $\alpha_{0}$. We have $\langle\alpha_0,\alpha_0\rangle = \textit{\bf Y}$ by definition.  Let us consider any $\sigma\in { \mathfrak{S}}^{{}^{pure}}$ with $\alpha_{0}\not=\sigma$. We know from Lemma \ref{equivalentorthomodular} that $\alpha_{0}\sqcap_{{}_{{ \mathfrak{S}}}} \sigma$ and $\alpha_{0}^\star$ admit a common upper bound. In other words, $\exists \alpha_{1}\in { \mathfrak{S}}^{{}^{pure}}$ such that $(\alpha_{0}\sqcap_{{}_{{ \mathfrak{S}}}} \sigma) \sqsubseteq_{{}_{{ \mathfrak{S}}}} \alpha_{1}$ and $\alpha_{0}^\star \sqsubseteq_{{}_{{ \mathfrak{S}}}} \alpha_{1}$ (i.e.  $\langle\alpha_0,\alpha_1\rangle = \textit{\bf N}$ and then obviously $\alpha_{0}\not=\alpha_1$). We have $\langle\alpha_1,\alpha_1\rangle = \textit{\bf Y}$ by definition. \\
We proceed now to the second step. Let us suppose that $(\alpha_{0}\sqcap_{{}_{{ \mathfrak{S}}}} \alpha_1)\not= \bot_{{}_{{ \mathfrak{S}}}}$ or, in other words, that there exists $\sigma\in { \mathfrak{S}}^{{}^{pure}}$ such that $(\alpha_{0}\sqcap_{{}_{{ \mathfrak{S}}}} \alpha_1) \not\sqsubseteq_{{}_{{ \mathfrak{S}}}} \sigma$. We know from Lemma \ref{equivalentorthomodular} that $(\alpha_{0}\sqcap_{{}_{{ \mathfrak{S}}}} \alpha_1)\sqcap_{{}_{{ \mathfrak{S}}}} \sigma$ and $(\alpha_{0}\sqcap_{{}_{{ \mathfrak{S}}}} \alpha_1)^\star$ admit a common upper bound. 
In other words, $\exists \alpha_{2}\in { \mathfrak{S}}^{{}^{pure}}$ such that $((\alpha_{0}\sqcap_{{}_{{ \mathfrak{S}}}} \alpha_1)\sqcap_{{}_{{ \mathfrak{S}}}} \sigma) \sqsubseteq_{{}_{{ \mathfrak{S}}}} \alpha_{2}$ and $(\alpha_{0}\sqcap_{{}_{{ \mathfrak{S}}}} \alpha_1)^\star \sqsubseteq_{{}_{{ \mathfrak{S}}}} \alpha_{2}$ (i.e.  $\langle(\alpha_{0}\sqcap_{{}_{{ \mathfrak{S}}}} \alpha_1),\alpha_2\rangle = \textit{\bf N}$ and then obviously $\alpha_{2}\not\sqsupseteq_{{}_{{ \mathfrak{S}}}} (\alpha_{0}\sqcap_{{}_{{ \mathfrak{S}}}} \alpha_1)$). We have $\langle\alpha_2,\alpha_2\rangle = \textit{\bf Y}$ by definition. \\
This algorithm stops at a moment because ${ \mathfrak{S}}^{{}^{pure}}$ is finite.\\
The three conditions are then satisfied by construction.
\end{proof}

\subsection{GpT and Quantum theory}\label{subsectionquantum}

Let ${ \mathcal{H}}$ be a finite-dimensional complex Hilbert space. We will denote ${ \mathcal{H}}^\ast:={ \mathcal{H}}\smallsetminus \{\overrightarrow{0}\}$.  We will adopt the bra–ket notation to denote the vectors as $\mid\!\!\! \psi\rangle$ and the inner product of $\mid \!\!\!\psi \rangle, \mid \!\!\!\phi\rangle\in { \mathcal{H}}$ as $\langle \phi \!\!\!\mid\!\!\! \psi\rangle$.  $\mid\!\!\! \psi\rangle^{{}^{ray}}$ will denote the ray associated to $\mid\!\!\! \psi\rangle\in { \mathcal{H}}^\ast$, i.e. the set $\{\,\lambda.\mid\!\!\! \psi\rangle\;\vert\; \lambda\in \mathbb{C}^\ast\,\}$.\\
$B({ \mathcal{H}})$ will denote the real vector space of self-adjoint operators.  And for any $U \in B({ \mathcal{H}})$, $\textit{\rm Tr}(U)$ will denote the trace of $U$.  $B^+({ \mathcal{H}})$ will be the set of positive semi-definite operators, i.e. the set of operators $U$ satisfying $0\leq \langle \psi \!\!\!\mid\!\! U \!\!\mid\!\!\! \psi\rangle$ for any $\mid \!\!\!\psi \rangle \in { \mathcal{H}}^\ast$. \\
For any $T\in B({ \mathcal{H}})$, we will adopt the following notation 
\begin{eqnarray}
 \textit{\rm spec}(T, { \mathcal{H}}; \lambda)& := & \{\, \mid \!\!\!\psi \rangle \in { \mathcal{H}}\;\vert\;\; T\!\!\mid\!\!\! \psi\rangle =\lambda. \!\!\mid\!\!\! \psi\rangle\,\}.
 \end{eqnarray}
 
\begin{defin}
According to Von Neumann's formalism of Quantum mechanics, we will adopt the following choice for the space of preparations ${ \mathfrak{P}}$ :
\begin{eqnarray}
{ \mathfrak{P}} & := & {\mathcal{D}}({ \mathcal{H}})=\{\, P\in B^+({ \mathcal{H}}) \;\vert\; \textit{\rm Tr}(P)=1\,\}.
\end{eqnarray} 
${\mathcal{D}}({ \mathcal{H}})$ is the set of density operators on ${ \mathcal{H}}$.
\end{defin}
\begin{lemme}
${ \mathfrak{P}}$ is a convex set. The set of extremal elements (i.e. the pure states  of quantum mechanics), denoted ${ \mathfrak{P}}^{{}^{pure}}$, is given by 
\begin{eqnarray}
{ \mathfrak{P}}^{{}^{pure}} & = & \{\, L_{{}_{\mid \psi\rangle}} \;\vert\; \mid\!\!\! \psi\rangle \in { \mathcal{H}}^\ast\,\}\cong \{\, \mid\!\!\! \psi\rangle^{{}^{ray}}\;\vert\; \mid\!\!\! \psi\rangle\in { \mathcal{H}}^\ast \,\},
\end{eqnarray} 
where $L_{{}_{\mid \psi\rangle}}$ denotes the rank-one projector $\frac{\mid \psi\rangle \langle \psi \mid}{\langle \psi \mid \psi\rangle}$.
\end{lemme}
\begin{defin}
According to Von Neumann's formalism of Quantum mechanics,  the space of tests ${ \mathfrak{T}}$ will be chosen as follows
\begin{eqnarray}
{ \mathfrak{T}} & := & \{\, T\in B({ \mathcal{H}}) \;\vert\; \forall P\in { \mathfrak{P}},\; 0\leq \textit{\rm Tr}(PT) \leq 1\,\}\\
&=&\{\, T\in B({ \mathcal{H}}) \;\vert\; \forall \mid \!\!\!\psi \rangle \in { \mathcal{H}}^\ast,\; 0\leq \textit{\rm Tr}(L_{{}_{\mid \psi\rangle}}T) \leq 1\,\}
\end{eqnarray} 
\end{defin}
\begin{lemme}
${ \mathfrak{T}}$ is a convex set. The set of extremal elements, denoted ${ \mathfrak{T}}^{{}^{pure}}$, is given by 
\begin{eqnarray}
{ \mathfrak{T}}^{{}^{pure}} & = & \{\, L_{{}_{ \mathcal{G}}}\;\vert\; { \mathcal{G}}\;\textit{\rm closed subspace of ${ \mathcal{H}}$}\,\}, 
\end{eqnarray}
where, for any closed subspace ${ \mathcal{G}}$ of the Hilbert space ${ \mathcal{H}}$, we denote by $L_{{}_{ \mathcal{G}}}$ the self-adjoint projector whose spectral decomposition is ${ \mathcal{G}} \oplus { \mathcal{G}}^\perp \cong { \mathcal{H}}$ (more precisely,
$\textit{\rm spec}(L_{{}_{ \mathcal{G}}}, { \mathcal{H}}; 1)={ \mathcal{G}}$ and $\textit{\rm spec}(L_{{}_{ \mathcal{G}}}, { \mathcal{H}}; 0)={ \mathcal{G}}^{\perp}$).
\end{lemme}
\begin{defin}
The evaluation map ${ \mathfrak{e}}$ will be defined as follows
\begin{eqnarray}
\forall P\in { \mathfrak{P}},\forall T\in { \mathfrak{T}},\;\;{ \mathfrak{e}}(P,T) & := & \left\{ \begin{array}{lcl}
\textit{\bf Y} & \textit{\rm iff} & \textit{\rm Tr}(PT)=1\\
\textit{\bf N} & \textit{\rm iff} & \textit{\rm Tr}(PT)=0\\
\bot & \textit{\rm otherwise} & 
\end{array}\right.
\end{eqnarray} 
\end{defin}

\begin{lemme}
We have already
\begin{eqnarray}
\forall L,L'\in { \mathfrak{T}}^{{}^{pure}},&& (\,\forall U \in { \mathfrak{P}}^{{}^{pure}},  { \mathfrak{e}}(U,L)={ \mathfrak{e}}(U,L'))\,) \;\; \Leftrightarrow \;\; (\, L=L' \,)\\
\forall U,U'\in { \mathfrak{P}}^{{}^{pure}},&& (\,\forall L \in { \mathfrak{T}}^{{}^{pure}},  { \mathfrak{e}}(U,L)={ \mathfrak{e}}(U',L))\,) \;\; \Leftrightarrow \;\; (\, U=U' \,).
\end{eqnarray}
\end{lemme}

\begin{theoreme}
Following our construction, the set of states is given by
\begin{eqnarray}
{ \mathfrak{S}}  & := & \{\,\;\textit{\rm closed subspaces of ${ \mathcal{H}}$}\,\}
\end{eqnarray}
${ \mathfrak{S}}$ is a down-complete Inf semi-lattice with
\begin{eqnarray}
\forall { \mathcal{G}}_1,{ \mathcal{G}}_2\in { \mathfrak{S}},  & & { \mathcal{G}}_1 \sqcap_{{}_{{ \mathfrak{S}}}} { \mathcal{G}}_2 := { \mathcal{G}}_1 \oplus { \mathcal{G}}_2.
\end{eqnarray}
The evaluation map $\epsilon$ is then defined as a map from ${ \mathfrak{T}}$ to ${ \mathfrak{B}}^{ \mathfrak{S}}$ by
\begin{eqnarray}
\forall T\in { \mathfrak{T}},\;\forall { \mathcal{G}}\in { \mathfrak{S}},&&\epsilon^{ \mathfrak{S}} _{T}({ \mathcal{G}}):=\left\{\begin{array}{lcl} 
\textit{\bf Y} & \textit{\rm iff} & { \mathcal{G}} \subseteq \textit{\rm spec}(T,{ \mathcal{H}};1)\\
\textit{\bf N} & \textit{\rm iff} & { \mathcal{G}} \subseteq \textit{\rm spec}(T,{ \mathcal{H}};0)\\
\bot & \textit{\rm otherwise} &
\end{array}\right.
\end{eqnarray}
\begin{eqnarray}
{ \mathfrak{S}}^{{}^{pure}}  & = & \{\,\textit{\rm one-dimensional closed subspaces of ${ \mathcal{H}}$}\,\}\\
\bot_{{}_{{ \mathfrak{S}}}} & = & {\mathcal{H}}.
\end{eqnarray}
\end{theoreme}
\begin{remark}
${ \mathfrak{S}}$ is ortho-complemented with ${ \mathcal{G}}^\star:={ \mathcal{G}}^{\underline{\perp}}$ (here, ${\underline{\perp}}$ denotes the orthogonality of vector spaces) for any ${ \mathcal{G}}$ in ${ \mathfrak{S}}$.\\
The involutive and order-reversing properties of $\star$ are trivial to check. The property (\ref{starcomplement}) is in fact a direct consequence of the property 
\begin{eqnarray}
\forall { \mathcal{G}} \in { \mathfrak{S}},\forall x\in { \mathcal{H}}\smallsetminus { \mathcal{G}},&& (\{x\}\oplus { \mathcal{G}})\cap { \mathcal{G}}^{\underline{\perp}} \not= \varnothing
\end{eqnarray}
satisfied by any Hilbert space ${ \mathcal{H}}$.
\end{remark}

\begin{theoreme}
The set of effects is then defined by
\begin{eqnarray}
{ \mathfrak{E}} & := & \{\, ({ \mathcal{G}}_1,{ \mathcal{G}}_2)\in { \mathfrak{S}}^{\times 2}\;\vert\; { \mathcal{G}}_1  {\underline{\perp}}  { \mathcal{G}}_2\,\}
\end{eqnarray}
with the following down-complete Inf semi-lattice structure
\begin{eqnarray}
\forall  ({ \mathcal{G}}_1,{ \mathcal{G}}_2), ({ \mathcal{G}}'_1,{ \mathcal{G}}'_2)\in { \mathfrak{E}}, && ({ \mathcal{G}}_1,{ \mathcal{G}}_2) \sqcap_{{}_{{ \mathfrak{L}}}} ({ \mathcal{G}}'_1,{ \mathcal{G}}'_2) := ({ \mathcal{G}}_1\cap { \mathcal{G}}_1',{ \mathcal{G}}_2\cap { \mathcal{G}}'_2).
\end{eqnarray}
And the evaluation map is given by
\begin{eqnarray}
\forall ({ \mathcal{G}}_1,{ \mathcal{G}}_2) \in { \mathfrak{E}},\;\forall { \mathcal{G}}\in { \mathfrak{S}},&&\epsilon^{ \mathfrak{S}}_{({ \mathcal{G}}_1,{ \mathcal{G}}_2)}({ \mathcal{G}}):=\left\{\begin{array}{lcl} 
\textit{\bf Y} & \textit{\rm iff} & { \mathcal{G}} \subseteq { \mathcal{G}}_1\\
\textit{\bf N} & \textit{\rm iff} & { \mathcal{G}} \subseteq { \mathcal{G}}_2\\
\bot & \textit{\rm otherwise} &
\end{array}\right.
\end{eqnarray}
\begin{eqnarray}
{ \mathfrak{E}}^{{}^{pure}}  & := & \{\,({ \mathcal{G}},{ \mathcal{G}}^{\underline{\perp}})\;\vert\; { \mathcal{G}}\oplus { \mathcal{G}}^{\underline{\perp}} = { \mathcal{H}}\,\}.
\end{eqnarray}
\end{theoreme}

\subsection{Symmetries}\label{subsectionsymmetries}

Observer $O_1$ has prepared a state $\sigma_1\in { \mathfrak{S}}^{{}^{(O_1)}}$ and intends to describe it to observer $O_2$.  Observer $O_2$ is able to interpret the macroscopic data defining $\sigma_1$ in terms of the elements of ${ \mathfrak{S}}^{{}^{(O_2)}}$ using a map $f_{{}_{(12)}} : { \mathfrak{S}}^{{}^{(O_1)}} \rightarrow { \mathfrak{S}}^{{}^{(O_2)}}$ (i.e., $O_2$ knows how to identify a state $f_{{}_{(12)}}(\sigma_1)$ corresponding to any $\sigma_1$). 
Observer $O_2$ has selected an effect ${ \mathfrak{l}}_2\in { \mathfrak{E}}^{{}^{(O_2)}}$ and intends to address the corresponding question to $O_1$.  Observer $O_1$ is able to interpret the macroscopic data defining ${ \mathfrak{l}}_2$ in terms of the elements of ${ \mathfrak{E}}^{{}^{(O_1)}}${{}}{ } using a map $f^{{}^{(21)}} : { \mathfrak{E}}^{{}^{(O_2)}} \rightarrow { \mathfrak{E}}^{{}^{(O_1)}}$ (i.e., $O_1$ {{}}{ }knows how to fix an effect $f^{{}^{(21)}}({ \mathfrak{l}}_2)$ corresponding to any ${ \mathfrak{l}}_2$). The pair of maps $(f_{{}_{(12)}},f^{{}^{(21)}})$ where  $f_{{}_{(12)}} : { \mathfrak{S}}^{{}^{(O_1)}} \rightarrow { \mathfrak{S}}^{{}^{(O_2)}}$ and $f^{{}^{(21)}} : { \mathfrak{E}}^{{}^{(O_2)}}\rightarrow { \mathfrak{E}}^{{}^{(O_1)}}$ defines {\em a dictionary} formalizing the transaction from $O_1$ to $O_2$.  The main task these observers want to accomplish is to confront their knowledge, i.e., to compare their 'statements' about the system. As soon as the transaction is formalized using a dictionary,  the two observers can formulate their statements and each confront them with the statements of the other.  First, observer $O_1$ can interpret the macroscopic data defining ${ \mathfrak{l}}_2$ using the map $f^{{}^{(21)}}$. Then, he produces the statement $\epsilon^{{}^{(O_1)}}_{f^{{}^{(21)}}({ \mathfrak{l}}_2)}(\sigma_1)$ concerning the results associated to this effect on the chosen state.  Secondly, observer $O_2$ can interpret the macroscopic data defining $\sigma_1$ using the map $f_{{}_{(12)}}$. Then, observer $O_2$ pronounces her statement $\epsilon^{{}^{(O_2)}}_{{ \mathfrak{l}}_2}(f_{{}_{(12)}}(\sigma_1))$ concerning the results associated to the effect ${ \mathfrak{l}}_2$ on the correspondingly prepared state.  The two observers, $O_1$ and $O_2$, are said to {\em agree about all their statements} iff 
\begin{eqnarray} \forall \sigma_1\in { \mathfrak{S}}^{{}^{(O_1)}}, \forall { \mathfrak{l}}_2 \in { \mathfrak{E}}^{{}^{(O_2)}},&&\epsilon^{{}^{(O_2)}}_{{ \mathfrak{l}}_2}(f_{{}_{(12)}}(\sigma_1))=\epsilon^{{}^{(O_1)}}_{f^{{}^{(21)}}({ \mathfrak{l}}_2)}(\sigma_1).\label{defchumorphism}\end{eqnarray}

To summarize, we will define symmetries of the system as follows.

\begin{defin}
The symmetries of the system are defined as Chu morphisms \cite{Pratt1999} from a States/Effects Chu space $({ \mathfrak{S}}^{{}^{(O_1)}},{ \mathfrak{E}}^{{}^{(O_1)}},\epsilon^{{}^{(O_1)}})$ defining the space of states and effects associated to the observer $O_1$, to another States/Effects Chu space $({ \mathfrak{S}}^{{}^{(O_2)}},{ \mathfrak{E}}^{{}^{(O_2)}},\epsilon^{{}^{(O_2)}})$ associated to the observer $O_2$, i.e. as pairs of bijective maps $f_{{}_{(12)}} : { \mathfrak{S}}^{{}^{(O_1)}} \rightarrow { \mathfrak{S}}^{{}^{(O_2)}}$ and $f^{{}^{(21)}} : { \mathfrak{E}}^{{}^{(O_2)}}\rightarrow { \mathfrak{E}}^{{}^{(O_1)}}$ satisfying property (\ref{defchumorphism}).
\end{defin}

\begin{defin}
The composition of a symmetry $(f_{{}_{(12)}} ,f^{{}^{(21)}})$ from $({ \mathfrak{S}}^{{}^{(O_1)}},{ \mathfrak{E}}^{{}^{(O_1)}},\epsilon^{{}^{(O_1)}})$ to $({ \mathfrak{S}}^{{}^{(O_2)}},{ \mathfrak{E}}^{{}^{(O_2)}},\epsilon^{{}^{(O_2)}})$ by another symmetry $(g_{{}_{(23)}} ,g^{{}^{(32)}})$ defined from $({ \mathfrak{S}}^{{}^{(O_2)}},{ \mathfrak{E}}^{{}^{(O_2)}},\epsilon^{{}^{(O_2)}})$ to $({ \mathfrak{S}}^{{}^{(O_3)}},{ \mathfrak{E}}^{{}^{(O_3)}},\epsilon^{{}^{(O_3)}})$ is given by the pair of bijective maps 
$( g_{{}_{(23)}}\circ f_{{}_{(12)}} ,f^{{}^{(21)}} \circ g^{{}^{(32)}})$ defining a valid symmetry from $({ \mathfrak{S}}^{{}^{(O_1)}},{ \mathfrak{E}}^{{}^{(O_1)}},\epsilon^{{}^{(O_1)}})$ to $({ \mathfrak{S}}^{{}^{(O_3)}},{ \mathfrak{E}}^{{}^{(O_3)}},\epsilon^{{}^{(O_3)}})$.
\end{defin}

As noted in \cite{Buffenoir2021}, the duality property (\ref{defchumorphism}) suffices to deduce the following properties.
\begin{theoreme} $f_{{}_{(12)}}$ and $f^{{}^{(21)}}$ are maps satisfying 
\begin{eqnarray}
\forall S_1\subseteq { \mathfrak{S}}^{(O_1)},&& f_{{}_{(12)}}(\bigsqcap{}^{{}^{{ \mathfrak{S}}^{(O_1)}}} \!\! S_1) = \bigsqcap{}^{{}^{{ \mathfrak{S}}^{(O_2)}}}_{{}_{\sigma_1\in S_1}} \; f_{{}_{(12)}}(\sigma_1)\label{f12cap}\\
\forall C_1\subseteq_{Chain} { \mathfrak{S}}^{(O_1)},&& f_{{}_{(12)}}(\bigsqcup{}^{{}^{{ \mathfrak{S}}^{(O_1)}}} \!\! C_1) = \bigsqcup{}^{{}^{{ \mathfrak{S}}^{(O_2)}}}_{{}_{\sigma_1\in C_1}} \; f_{{}_{(12)}}(\sigma_1)\label{f12cupchain}
\end{eqnarray}
As a consequence of (\ref{f12cupchain}), it is in particular order-preserving. \\
We have moreover
\begin{eqnarray}
\forall E_2\subseteq { \mathfrak{E}}^{(O_2)},&& f^{{}^{(21)}}(\bigsqcap{}^{{}^{{ \mathfrak{E}}^{(O_2)}}} \!\! E_2) = \bigsqcap{}^{{}^{{ \mathfrak{E}}^{(O_1)}}}_{{}_{{ \mathfrak{l}}_2\in E_2}} \; f^{{}^{(21)}}({ \mathfrak{l}}_2)\label{f21cap}\\
\forall C_2\subseteq_{Chain} { \mathfrak{E}}^{(O_2)},&& f^{{}^{(21)}}(\bigsqcup{}^{{}^{{ \mathfrak{E}}^{(O_2)}}} \!\! C_2) = \bigsqcup{}^{{}^{{ \mathfrak{E}}^{(O_1)}}}_{{}_{{ \mathfrak{l}}_2\in C_2}} \; f^{{}^{(21)}}({ \mathfrak{l}}_2)\label{f21cupchain}\\
\forall { \mathfrak{l}}_2 \in { \mathfrak{E}}^{(O_2)},&& f^{{}^{(21)}}(\; \overline{{ \mathfrak{l}}_2} \;) = \overline{f^{{}^{(21)}}({ \mathfrak{l}}_2)}\label{f21bar}\\
&& f^{{}^{(21)}}(\;{ \mathfrak{Y}}^{(O_2)}_{ \mathfrak{E}}\;) = { \mathfrak{Y}}^{(O_1)}_{ \mathfrak{E}}.\label{f21Y}
\end{eqnarray}
Note that, due to properties (\ref{f12cap}) (\ref{f21cap}), as long as ${ \mathfrak{S}}^{(O_1)}$ satisfies axioms {\bf (A1) (A2) (A3)}, ${ \mathfrak{S}}^{(O_2)}$ satisfies axioms {\bf (A1) (A2) (A3)} as well.
\end{theoreme}
\begin{proof}
All proofs follow the same trick. For example, for any $S_1\subseteq { \mathfrak{S}}^{(O_1)}$ and any ${ \mathfrak{l}}_2 \in { \mathfrak{E}}^{{}^{(O_2)}}$, we have, using (\ref{defchumorphism}) and (\ref{axiomEinfsemilattice}) :
\begin{eqnarray}
\epsilon^{{}^{(O_2)}}_{{ \mathfrak{l}}_2}(f_{{}_{(12)}}(\bigsqcap{}^{{}^{{ \mathfrak{S}}^{(O_1)}}} \!\! S_1)) &=& \epsilon^{{}^{(O_1)}}_{f^{{}^{(21)}}({ \mathfrak{l}}_2)}(\bigsqcap{}^{{}^{{ \mathfrak{S}}^{(O_1)}}} \!\! S_1)\nonumber\\
&=& \bigwedge{}_{\sigma_1\in S_1} \; \epsilon^{{}^{(O_1)}}_{f^{{}^{(21)}}({ \mathfrak{l}}_2)}(\sigma_1)\nonumber\\
&=&  \bigwedge{}_{\sigma_1\in S_1} \; \epsilon^{{}^{(O_2)}}_{{ \mathfrak{l}}_2}(f_{{}_{(12)}}(\sigma_1))\nonumber\\
&=&  \epsilon^{{}^{(O_2)}}_{{ \mathfrak{l}}_2}(\bigsqcap{}^{{}^{{ \mathfrak{S}}^{(O_2)}}}_{\sigma_1\in S_1} f_{{}_{(12)}}(\sigma_1))
\end{eqnarray}
We now use the property (\ref{Chuseparated}) to conclude on (\ref{f12cap}).
\end{proof}

\begin{theoreme}
Due to the Corollary \ref{corollaryfsigma}, as soon as a map $f_{{}_{(12)}}$ which is monotonic and satisfies (\ref{f12cap}) is given, we can define unambiguously the map $f^{{}^{(21)}}$ satisfying with $f_{{}_{(12)}}$ the duality relation (\ref{defchumorphism}).
\end{theoreme}

The couple of maps $(f_{{}_{(12)}},f^{{}^{(21)}})$ defining a channel from $({ \mathfrak{S}}^{{}^{(O_1)}},{ \mathfrak{E}}^{{}^{(O_1)}},\epsilon^{{}^{(O_1)}})$ to $({ \mathfrak{S}}^{{}^{(O_2)}},{ \mathfrak{E}}^{{}^{(O_2)}},\epsilon^{{}^{(O_2)}})$ can then be reduced to the single data $f_{{}_{(12)}}$. We will then speak shortly of "the symmetry $f_{{}_{(12)}}$ from the space of states ${ \mathfrak{S}}^{{}^{(O_1)}}$ to the space of states ${ \mathfrak{S}}^{{}^{(O_2)}}$".

\begin{defin}
The space of channels from the space of states ${ \mathfrak{S}}^{{}^{(O_1)}}$ to the space of states ${ \mathfrak{S}}^{{}^{(O_2)}}$ will be denoted ${ \mathfrak{C}}({{ \mathfrak{S}}^{{}^{(O_1)}}},{{ \mathfrak{S}}^{{}^{(O_2)}}})$. It is the space of maps from ${ \mathfrak{S}}^{{}^{(O_1)}}$ to ${ \mathfrak{S}}^{{}^{(O_2)}}$ that is order-preserving and satisfies (\ref{f12cap}).
\end{defin}

\begin{defin}
We define the infimum of two maps $f_{{}_{(12)}}$ and $g_{{}_{(12)}}$ satisfying (\ref{f12cap}) (resp. two maps $f^{{}^{(12)}}$ and $g^{{}^{(12)}}$  satisfying (\ref{f21cap})) by $\forall \sigma \in { \mathfrak{S}}^{{}^{(O_1)}}, (f_{{}_{(12)}}\sqcap g_{{}_{(12)}})(\sigma):= f_{{}_{(12)}}(\sigma) \sqcap_{{}_{{ \mathfrak{S}}^{{}^{(O_2)}}}} g_{{}_{(12)}}(\sigma)$ (resp. $\forall { \mathfrak{l}} \in { \mathfrak{E}}^{{}^{(O_2)}}, (f^{{}^{(12)}}\sqcap g^{{}^{(12)}})({ \mathfrak{l}}):= f^{{}^{(12)}}({ \mathfrak{l}}) \sqcap_{{}_{{ \mathfrak{E}}^{{}^{(O_1)}}}} g^{{}^{(12)}}({ \mathfrak{l}})$).
\end{defin}

\begin{theoreme}
The infimum of a channel $(f_{{}_{(12)}},f^{{}^{(21)}})$ from $({ \mathfrak{S}}^{{}^{(O_1)}},{ \mathfrak{E}}^{{}^{(O_1)}},\epsilon^{{}^{(O_1)}})$ to $({ \mathfrak{S}}^{{}^{(O_2)}},{ \mathfrak{E}}^{{}^{(O_2)}},\epsilon^{{}^{(O_2)}})$ with another channel $(g_{{}_{(12)}},g^{{}^{(21)}})$ defined from $({ \mathfrak{S}}^{{}^{(O_1)}},{ \mathfrak{E}}^{{}^{(O_1)}},\epsilon^{{}^{(O_1)}})$ to $({ \mathfrak{S}}^{{}^{(O_2)}},{ \mathfrak{E}}^{{}^{(O_2)}},\epsilon^{{}^{(O_2)}})$ is given by the pair of maps 
$( f_{{}_{(12)}}\sqcap g_{{}_{(12)}} ,f^{{}^{(21)}} \sqcap g^{{}^{(21)}})$ defining a valid channel (i.e. a Chu morphism) from $({ \mathfrak{S}}^{{}^{(O_1)}},{ \mathfrak{E}}^{{}^{(O_1)}},\epsilon^{{}^{(O_1)}})$ to $({ \mathfrak{S}}^{{}^{(O_2)}},{ \mathfrak{E}}^{{}^{(O_2)}},\epsilon^{{}^{(O_2)}})$.
\end{theoreme}
\begin{proof}Using two times the duality property and the homomorphic property of $\epsilon$, we obtain
\begin{eqnarray} 
\epsilon^{{}^{(O_2)}}_{{ \mathfrak{l}}_2}((f\sqcap g) (\sigma_1))&=&
\epsilon^{{}^{(O_2)}}_{{ \mathfrak{l}}_2}(f (\sigma_1))\wedge \epsilon^{{}^{(O_2)}}_{{ \mathfrak{l}}_2}(g (\sigma_1))=
\epsilon^{{}^{(O_1)}}_{f^{{}^{(21)}}({ \mathfrak{l}}_2)}(\sigma_1) \wedge \epsilon^{{}^{(O_1)}}_{g^{{}^{(21)}}({ \mathfrak{l}}_2)}(\sigma_1)=
\epsilon^{{}^{(O_1)}}_{(f^{{}^{(21)}}\sqcap g^{{}^{(21)}})({ \mathfrak{l}}_2)}(\sigma_1).
\end{eqnarray}
\end{proof}

\begin{theoreme}
Pure states in ${ \mathfrak{S}}^{(O_2)}$ are exactly the direct images by the symmetry $f_{{}_{(12)}}$ of pure states in ${ \mathfrak{S}}^{(O_1)}$.  Moreover, as long as ${ \mathfrak{S}}^{(O_1)}$ satisfies axiom {\bf (A4)}, ${ \mathfrak{S}}^{(O_2)}$ satisfies axiom {\bf (A4)} as well.
\end{theoreme}
\begin{proof}
Let us consider a state $\sigma_2$ in ${ \mathfrak{S}}^{(O_2)}$ such that $f_{{}_{(12)}}^{-1}(\sigma_2)$ is a pure state in ${ \mathfrak{S}}^{(O_1)}$.  For any $S_2\subseteq { \mathfrak{S}}^{(O_2)}$ satisfying $\sigma_2=\bigsqcap{}^{{}^{{ \mathfrak{S}}^{(O_2)}}} S_2$, we have $f_{{}_{(12)}}^{-1}(\sigma_2)=f_{{}_{(12)}}^{-1}(\bigsqcap{}^{{}^{{ \mathfrak{S}}^{(O_2)}}} S_2)=\bigsqcap{}^{{}^{{ \mathfrak{S}}^{(O_1)}}}_{\sigma'_2\in S_2} f_{{}_{(12)}}^{-1}(\sigma'_2)$ using (\ref{f12cap}), and then $f_{{}_{(12)}}^{-1}(\sigma_2)\in f_{{}_{(12)}}^{-1}(S_2)$ (due to complete irreducibility of $f_{{}_{(12)}}^{-1}(\sigma_2)$), and then $\sigma_2\in S_2$.  As a conclusion, $\sigma_2$ is completely meet-irreducible in ${ \mathfrak{S}}^{(O_2)}$, i.e.  it is a pure state of ${ \mathfrak{S}}^{(O_2)}$. \\
Conversely, let us consider $\sigma_2$ a pure state in ${ \mathfrak{S}}^{(O_2)}$ and let us consider $S_1\subseteq { \mathfrak{S}}^{(O_1)}$ such that $f_{{}_{(12)}}^{-1}(\sigma_2)=\bigsqcap{}^{{}^{{ \mathfrak{S}}^{(O_1)}}} S_1$, we have $\sigma_2 =f_{{}_{(12)}}(\bigsqcap{}^{{}^{{ \mathfrak{S}}^{(O_1)}}} S_1)=\bigsqcap{}^{{}^{{ \mathfrak{S}}^{(O_1)}}}_{\sigma'_1\in S_1} f_{{}_{(12)}}(\sigma'_1)$ using (\ref{f12cap}). Now, using complete irreducibility of $\sigma_2$, we deduce that there exists $\sigma_1\in S_1$ such that $\sigma_2=f_{{}_{(12)}}(\sigma_1)$, i.e. $f_{{}_{(12)}}^{-1}(\sigma_2)\in S_1$. Hence, $f_{{}_{(12)}}^{-1}(\sigma_2)$ is a pure state in ${ \mathfrak{S}}^{(O_1)}$. \\ 
Secondly, let us consider that ${ \mathfrak{S}}^{(O_1)}$ satisfies axiom {\bf (A4)}. We note that, due to the property $f_{{}_{(12)}}({ \mathfrak{S}}^{(O_1)}_{{}_{pure}})={ \mathfrak{S}}^{(O_2)}_{{}_{pure}}$ and the monotonic character of the map $f_{{}_{(12)}}$, we have $f_{{}_{(12)}}(\underline{\sigma})=\underline{f_{{}_{(12)}}(\sigma)}$. Using this result, axiom {\bf (A4)} and property (\ref{f12cap}), we obtain for any  $\sigma_1 \in { \mathfrak{S}}^{(O_1)}$,  $f_{{}_{(12)}}(\sigma_1)=f_{{}_{(12)}}(\bigsqcap{}^{{}^{{ \mathfrak{S}}^{(O_1)}}}  \underline{\sigma_1})=\bigsqcap{}^{{}^{{ \mathfrak{S}}^{(O_1)}}}  f_{{}_{(12)}}(\underline{\sigma_1})=\bigsqcap{}^{{}^{{ \mathfrak{S}}^{(O_1)}}}  \underline{f_{{}_{(12)}}(\sigma_1)}$. In other words, ${ \mathfrak{S}}^{(O_2)}=f_{{}_{(12)}}({ \mathfrak{S}}^{(O_1)})$ satisfies axiom {\bf (A4)}.
\end{proof}

\begin{theoreme}
As long as ${ \mathfrak{S}}^{(O_1)}$ satisfies axiom {\bf (A5)},  
${ \mathfrak{S}}^{(O_2)}$ satisfies axiom {\bf (A5)} as well and $f_{{}_{(12)}}(Max({ \mathfrak{S}}^{(O_1)}))= Max({ \mathfrak{S}}^{(O_2)})$.
\end{theoreme}
\begin{proof}
For any $\sigma_2$ completely meet-irreducible element in ${ \mathfrak{S}}^{(O_2)}$,  $f_{{}_{(12)}}^{-1}(\sigma_2)$ is a completely meet-irreducible element in ${ \mathfrak{S}}^{(O_1)}$ and then $f_{{}_{(12)}}^{-1}(\sigma_2)\in Max({ \mathfrak{S}}^{(O_1)})$ because ${ \mathfrak{S}}^{(O_1)}$ satisfies axiom {\bf (A5)}. Let us imagine that there exists $\sigma'_2 \sqsupseteq_{{}_{ { \mathfrak{S}}^{(O_2)}}} \sigma_2$, we have necessarily $f_{{}_{(12)}}^{-1}(\sigma'_2)\sqsupseteq_{{}_{ { \mathfrak{S}}^{(O_2)}}} f_{{}_{(12)}}^{-1}(\sigma_2)$ because $f_{{}_{(12)}}$ is bijective and order-preserving, and then $f_{{}_{(12)}}^{-1}(\sigma_2)= f_{{}_{(12)}}^{-1}(\sigma'_2)$ because $f_{{}_{(12)}}^{-1}(\sigma_2)\in Max({ \mathfrak{S}}^{(O_1)})$.  As a result, $\sigma_2\in Max({ \mathfrak{S}}^{(O_2)})$. We conclude that ${ \mathfrak{S}}^{(O_2)}$ satisfies axiom {\bf (A5)}. \\
Let us consider $\sigma_1\in Max({ \mathfrak{S}}^{(O_1)})$ and let us consider that there exists $\sigma_2 \sqsupseteq_{{}_{ { \mathfrak{S}}^{(O_2)}}}  f_{{}_{(12)}}(\sigma_1)$. We have necessarily $f_{{}_{(12)}}^{-1}(\sigma_2)\sqsupseteq_{{}_{ { \mathfrak{S}}^{(O_2)}}} \sigma_1$ because $f_{{}_{(12)}}$ is bijective and order-preserving, and then $f_{{}_{(12)}}^{-1}(\sigma_2)= \sigma_1$ because $\sigma_1\in Max({ \mathfrak{S}}^{(O_1)})$. As a result, $f_{{}_{(12)}}(\sigma_1)\in Max({ \mathfrak{S}}^{(O_2)})$. We conclude that $f_{{}_{(12)}}(Max({ \mathfrak{S}}^{(O_1)}))\subseteq Max({ \mathfrak{S}}^{(O_2)})$.  
On another part, for any $\sigma_2\in Max({ \mathfrak{S}}^{(O_2)})$, $f_{{}_{(12)}}^{-1}(\sigma_2)$ is a completely meet-irreducible element of ${ \mathfrak{S}}^{(O_1)}$, i.e. an element of $Max({ \mathfrak{S}}^{(O_1)})$, and then $\sigma_2=f_{{}_{(12)}}(f_{{}_{(12)}}^{-1}(\sigma_2))\in f_{{}_{(12)}}(Max({ \mathfrak{S}}^{(O_1)}))$. As a final conclusion,  $f_{{}_{(12)}}(Max({ \mathfrak{S}}^{(O_1)}))= Max({ \mathfrak{S}}^{(O_2)})$. 
\end{proof}

As a conclusion of all results of this subsection, the defined symmetries relate fully and faithfully the States/Effects Chu spaces. 

\begin{remark}\label{idandYchannel}
We note that the identity map $(id_{{}_{{ \mathfrak{S}}}},id_{{}_{{ \mathfrak{E}}}})$ is a symmetry from the States/Effects Chu space $({ \mathfrak{S}},{ \mathfrak{E}},\epsilon)$ to itself.
\end{remark}

From now on, we will consider this category of States/Effects Chu spaces equipped with Chu morphisms and denote it $Chu_{ \mathfrak{B}}^{S/E}$. \\
We will also have to consider the category ${\mathcal{S}}$ which objects are spaces of states (Axioms {\bf A1-A5}) and which morphisms are bijective order-preserving maps satisfying 
(\ref{f12cap})(\ref{f12cupchain}).

\section{Multipartite experiments}

\subsection{An axiomatic proposal}\label{subsectionaxiomatic}

We now intent to describe an experiment on compound systems, implying two parties : Alice and Bob.  The bipartite state space will be formed from two given spaces of states ${ \mathfrak{S}}_{A}$ and ${ \mathfrak{S}}_{B}$. It will be clear later on that this notion of bipartite space of states is ambiguous and different constructions can be proposed. \\

We now begin with a basic axiomatic proposal for the description of bipartite experiments (see \cite[Section 5]{Plavala} for an analogue proposal in GPT's perspective). We will denote by ${ \mathfrak{S}}_{AB}={ \mathfrak{S}}_{A}\boxtimes { \mathfrak{S}}_{B}$ the corresponding space of states.\footnote{Throughout this short axiomatic introduction, we adopt the inadequate notation $\boxtimes$ for the tensor product in order to allow for different candidates for this tensor product. These different candidates will be denoted $\otimes$, $\widetilde{\otimes}$,...} We will also denote by ${ \mathfrak{E}}_{AB}={ \mathfrak{E}}_{A}\boxtimes { \mathfrak{E}}_{B}$ the bipartite effect space formed from two given effect spaces ${ \mathfrak{E}}_{A}$ and ${ \mathfrak{E}}_{B}$. We will denote ${\epsilon}\,{}^{AB}$ the corresponding bipartite evaluation map from ${ \mathfrak{E}}_{AB}$ to ${ \mathfrak{B}}^{{ \mathfrak{S}}_{AB}}$. We will assume the following requirements about these elements.\\

First of all, we have to build $({ \mathfrak{S}}_{AB},{\mathfrak{E}}_{AB},{\epsilon}\,{}^{AB})$ as a valid Spaces/Effects Chu space. \\

In particular, we will assume that ${ \mathfrak{S}}_{AB}$ admits mixed bipartite states. In other words,  
\begin{eqnarray}
\textit{\bf (B1)} &&
\begin{array}{l}
\forall \{\,\sigma_{i,AB}\;\vert\; i\in I\,\}\subseteq {{ \mathfrak{S}}_{AB}}, \;\;\;\;\;\; \bigsqcap{}^{{}^{{ \mathfrak{S}}_{AB}}}_{i\in I} \sigma_{i,AB}\;\;\textit{\rm exists in}\;\; { \mathfrak{S}}_{AB},\\
 \forall \{\,\sigma_{i,AB}\;\vert\; i\in I\,\}\subseteq {{ \mathfrak{S}}_{AB}},\forall {\mathfrak{l}}_{AB}\in { \mathfrak{E}}_{AB},\;\;
{\epsilon}\,{}^{AB}_{{\mathfrak{l}}_{AB}} ( \bigsqcap{}^{{}^{{ \mathfrak{S}}_{AB}}}_{i\in I} \sigma_{i,AB}) =
\bigwedge{}_{i\in I}\;
{\epsilon}\,{}^{AB}_{{\mathfrak{l}}_{AB}} (\sigma_{i,AB}).
\end{array}\;\;\;\;\;\;\label{tensorinfimum}
\end{eqnarray}

\vspace{0.3cm}
In the same logic, we will assume that ${ \mathfrak{E}}_{AB}$ admits mixed bipartite effects. In other words,  
\begin{eqnarray}
\textit{\bf (B2)} && 
\begin{array}{l}
\forall \{\,{ \mathfrak{l}}_{i,AB}\;\vert\; i\in I\,\}\subseteq {{ \mathfrak{E}}_{AB}}, \;\;\;\;\;\; \bigsqcap{}^{{}^{{ \mathfrak{E}}_{AB}}}_{i\in I} { \mathfrak{l}}_{i,AB}\;\;\textit{\rm exists in}\;\; { \mathfrak{E}}_{AB},\\
 \forall \{\,{ \mathfrak{l}}_{i,AB}\;\vert\; i\in I\,\}\subseteq {{ \mathfrak{E}}_{AB}},\forall \sigma_{AB}\in { \mathfrak{S}}_{AB},\;\;
{\epsilon}\,{}^{AB}_{\bigsqcap{}^{{}^{{ \mathfrak{E}}_{AB}}}_{i\in I} {\mathfrak{l}}_{i,AB}} ( \sigma_{AB}) =
\bigwedge{}_{i\in I}\;
{\epsilon}\,{}^{AB}_{{\mathfrak{l}}_{i,AB}} (\sigma_{AB}).
\end{array} \;\;\;\;\;\;\label{tensorinfimumL}
\end{eqnarray}

\vspace{0.3cm}
Secondly, for every effects ${ \mathfrak{l}}_A$ and ${ \mathfrak{l}}_B$ realized independently by Alice and Bob respectively, we will assume that there must exist a unique associated bipartite effect in ${ \mathfrak{E}}_{A B}$.  As a consequence, we will assume that there are maps $\iota^{ \mathfrak{E}}_{AB}:{ \mathfrak{E}}_{A}\times { \mathfrak{E}}_{B} \longrightarrow { \mathfrak{E}}_{A B}$  which describe the inclusion of 'pure tensors' in ${ \mathfrak{E}}_{A B}$ (for readability, we shall write ${ \mathfrak{l}}_A \boxtimes { \mathfrak{l}}_B$ rather than $\iota^{ \mathfrak{E}}_{AB}({ \mathfrak{l}}_A,{ \mathfrak{l}}_B)$).  This axiom will be denoted {\bf (B3)}. Moreover, if Alice (or Bob) chooses a mixture of effects, then this results in a mixture of the respective bipartite effects.  
\begin{eqnarray}
\textit{\bf (B3')} && (\bigsqcap{}^{{}^{{ \mathfrak{E}}_{A}}}_{i\in I}\,{ \mathfrak{l}}_{i,A})\boxtimes { \mathfrak{l}}_B =  \bigsqcap{}^{{}^{{ \mathfrak{E}}_{AB}}}_{i\in I} ({ \mathfrak{l}}_{i,A}\boxtimes { \mathfrak{l}}_B),\label{etensor=tensore1}\\
\textit{\bf (B3'')} && { \mathfrak{l}}_A \boxtimes  (\bigsqcap{}^{{}^{{ \mathfrak{E}}_{B}}}_{i\in I}\,{ \mathfrak{l}}_{i,B}) =  \bigsqcap{}^{{}^{{ \mathfrak{E}}_{AB}}}_{i\in I} ({ \mathfrak{l}}_A \boxtimes { \mathfrak{l}}_{i,B}).\label{etensor=tensore2}
\end{eqnarray}
In the same logic,  for every states $\sigma_A\in { \mathfrak{S}}_{A}$ and $\sigma_B\in { \mathfrak{S}}_{B}$, prepared independently by Alice and Bob, we will assume that there must exist a unique associated bipartite state in ${ \mathfrak{S}}_{A B}$.  As a consequence,  we will assume that there are maps
$\iota^{ \mathfrak{S}}_{AB}:{ \mathfrak{S}}_{A}\times { \mathfrak{S}}_{B} \longrightarrow { \mathfrak{S}}_{A B}$ which describe the inclusion of 'pure tensors' in ${ \mathfrak{S}}_{A B}$ (for readability, we shall write $\sigma_A \boxtimes \sigma_B$ rather than $\iota^{ \mathfrak{S}}_{AB}(\sigma_A,\sigma_B)$).  This axiom will be denoted {\bf (B4)}. Moreover, if Alice (or Bob) prepares a mixture of states, then this results in a mixture of the respective bipartite states.  
\begin{eqnarray}
\textit{\bf (B4')} && (\bigsqcap{}^{{}^{{ \mathfrak{S}}_{A}}}_{i\in I}\,\sigma_{i,A})\boxtimes \sigma_B =  \bigsqcap{}^{{}^{{ \mathfrak{S}}_{AB}}}_{i\in I} (\sigma_{i,A}\boxtimes \sigma_B),\label{pitensor=tensorpi1}\\
\textit{\bf (B4'')} && \sigma_A \boxtimes  (\bigsqcap{}^{{}^{{ \mathfrak{S}}_{B}}}_{i\in I}\,\sigma_{i,B}) =  \bigsqcap{}^{{}^{{ \mathfrak{S}}_{AB}}}_{i\in I} (\sigma_A \boxtimes \sigma_{i,B}).\label{pitensor=tensorpi2}
\end{eqnarray}

Endly,  for every $\sigma_{AB},\sigma'_{AB} \in {{ \mathfrak{S}}_{AB}}$ such that $\sigma_{AB}\not= \sigma'_{AB}$, we will assume that there must exist effects ${\mathfrak{l}}_A\in { \mathfrak{E}}_{A}$ and ${\mathfrak{l}}_B\in { \mathfrak{E}}_{B}$ such that when Alice and Bob prepare $\sigma_{AB}$ and apply ${\mathfrak{l}}_A$ and ${\mathfrak{l}}_B$ respectively,  the resulting determination is different from the experiment where Alice and Bob prepare $\sigma'_{AB}$ and apply ${\mathfrak{l}}_A$ and ${\mathfrak{l}}_B$ respectively. As a summary, applying effects locally is sufficient to distinguish all of the states in ${{ \mathfrak{S}}_{AB}}$ (this principle is called "tomographic locality"), i.e.
\begin{eqnarray}
\textit{\bf (B5)} &&\forall \sigma_{AB},\sigma'_{AB} \in {{ \mathfrak{S}}_{AB}},\;\;\;\; (\,\forall {\mathfrak{l}}_A\in { \mathfrak{E}}_{A},{\mathfrak{l}}_B\in { \mathfrak{E}}_{B},\;\;
{\epsilon}\,{}^{AB}_{{\mathfrak{l}}_A\boxtimes {\mathfrak{l}}_B} (\sigma_{AB})={\epsilon}\,{}^{AB}_{{\mathfrak{l}}_A\boxtimes {\mathfrak{l}}_B} (\sigma'_{AB})\,) \;\; \Leftrightarrow \;\; (\,\sigma_{AB} = \sigma'_{AB}\,).\;\;\;\;\;\;\;\;\;\;\;\;\label{tensorseparated}
\end{eqnarray}

\vspace{0.3cm}
We will eventually adopt a complementary axiom.  Let us consider that Alice and Bob realize their experiments on a pure tensor state.  In the simplest scenario, Alice applies ${\mathfrak{l}}_A\in { \mathfrak{E}}_{A}$ and Bob applies ${\mathfrak{l}}_B\in { \mathfrak{E}}_{B}$ independently.  Since these two experiments are independent, the resulting determination has to be the 'product' of the respective determinations, i.e.
\begin{eqnarray}
\textit{\bf (C)}  && \forall \sigma_A\in { \mathfrak{S}}_{A},\forall \sigma_B\in { \mathfrak{S}}_{B}, \forall {\mathfrak{l}}_A\in { \mathfrak{E}}_{A},\forall {\mathfrak{l}}_B\in { \mathfrak{E}}_{B}, \;\;\;\;{\epsilon}\,{}^{AB}_{{\mathfrak{l}}_A\boxtimes {\mathfrak{l}}_B} (\sigma_A\boxtimes \sigma_B)  =  {\epsilon}\,{}^{A}_{{\mathfrak{l}}_A} (\sigma_A) \bullet {\epsilon}\,{}^{B}_{{\mathfrak{l}}_B} (\sigma_B). \label{evaltensorbullet}
\end{eqnarray}
This axiom appears to be partly redundant with previous axioms. Indeed, using properties (\ref{evaltensorbullet}) (\ref{axiomsigmainfsemilattice}) (\ref{distributivitybullet}) (\ref{tensorinfimum}),  we deduce that, for any ${\mathfrak{l}}_A\in { \mathfrak{E}}_{A}, {\mathfrak{l}}_B\in { \mathfrak{E}}_{B}$, $\{\,\sigma_{i,A}\;\vert\; i\in I\,\}\subseteq { \mathfrak{S}}_{A}$ and $\sigma_B\in { \mathfrak{S}}_{B}$
\begin{eqnarray}
{\epsilon}\,{}^{AB}_{{\mathfrak{l}}_A\boxtimes {\mathfrak{l}}_B} ((\bigsqcap{}^{{}^{{ \mathfrak{S}}_{A}}}_{i\in I}\sigma_{i,A})\boxtimes \sigma_B)&=&
{\epsilon}\,{}^{A}_{{\mathfrak{l}}_A} (\bigsqcap{}^{{}^{{ \mathfrak{S}}_{A}}}_{i\in I}\sigma_{i,A}) \bullet {\epsilon}\,{}^{B}_{{\mathfrak{l}}_B} (\sigma_B) \nonumber\\
&=&
(\bigwedge{}_{i\in I}\;{\epsilon}\,{}^{A}_{{\mathfrak{l}}_A} (\sigma_{i,A})) \bullet {\epsilon}\,{}^{B}_{{\mathfrak{l}}_B} (\sigma_B) \nonumber\\
&=&
\bigwedge{}_{i\in I}\;
(\,{\epsilon}\,{}^{A}_{{\mathfrak{l}}_A} (\sigma_{i,A})\bullet {\epsilon}\,{}^{B}_{{\mathfrak{l}}_B} (\sigma_B)) \nonumber\\
&=& \bigwedge{}_{i\in I}\;
 {\epsilon}\,{}^{AB}_{{\mathfrak{l}}_A\boxtimes {\mathfrak{l}}_B} (\sigma_{i,A}\boxtimes \sigma_B)
\nonumber\\
&=& {\epsilon}\,{}^{AB}_{{\mathfrak{l}}_A\boxtimes {\mathfrak{l}}_B} (\bigsqcap{}^{{}^{{ \mathfrak{S}}_{AB}}}_{i\in I} (\sigma_{i,A}\boxtimes \sigma_B)),\label{demobifilterproperty}
\end{eqnarray}
and then, using property (\ref{tensorseparated}), we obtain the property (\ref{pitensor=tensorpi1}). We obtain the property (\ref{pitensor=tensorpi2}) along the same lines of proof.  \\

In the following, we intent to identify potential candidates for this bipartite space of states ${ \mathfrak{S}}_{AB}$ and space of effects ${ \mathfrak{E}}_{AB}$ and posit it with respect to the standard construction of tensor products of Inf semi-lattices.  But before that, we complete the previous axiomatic by a discussion of the symmetries of the multipartite experiments.

\subsection{Symmetries of the bipartite experiments}\label{subsectionsymmetriesbipartite}

\begin{defin}
Let us consider a symmetry $(f_{{}_{(12)}}, f^{{}^{(21)}})$ from a States/Effects Chu space $({ \mathfrak{S}}_{A_1},{ \mathfrak{E}}_{A_1},\epsilon^{A_1})$ associated a first observer, to another States/Effects Chu space $({ \mathfrak{S}}_{A_2},{ \mathfrak{E}}_{A_2},\epsilon^{A_2})$ associated to another observer. Let us also consider a symmetry $(g_{{}_{(12)}}, g^{{}^{(21)}})$ from the Chu space $({ \mathfrak{S}}_{B_1},{ \mathfrak{E}}_{B_1},\epsilon^{B_1})$ to the Chu space $({ \mathfrak{S}}_{B_2},{ \mathfrak{E}}_{B_2},\epsilon^{B_2})$.  We define the pair of maps $((f\boxtimes g)_{{}_{(12)}}, (f\boxtimes g)^{{}^{(21)}})$ from the Chu space $({ \mathfrak{S}}_{A_1B_1},{ \mathfrak{E}}_{A_1B_1},\epsilon^{A_1B_1})$ to the Chu space $({ \mathfrak{S}}_{A_2B_2},{ \mathfrak{E}}_{A_2B_2},\epsilon^{A_2B_2})$ by
\begin{eqnarray}
(f\boxtimes g)_{{}_{(12)}}(\bigsqcap{}^{{}^{{ \mathfrak{S}}_{A_1B_1}}}_{i\in I} \sigma_{i,A_1}\boxtimes \sigma_{i,B_1}) & := & \bigsqcap{}^{{}^{{ \mathfrak{S}}_{A_2B_2}}}_{i\in I} f_{{}_{(12)}}(\sigma_{i,A_1})\boxtimes g_{{}_{(12)}}(\sigma_{i,B_1})\\
(f\boxtimes g)^{{}^{(21)}}(\bigsqcap{}^{{}^{{ \mathfrak{E}}_{A_2B_2}}}_{j\in J} { \mathfrak{l}}_{j,A_2}\boxtimes { \mathfrak{l}}_{j,B_2}) & := & \bigsqcap{}^{{}^{{ \mathfrak{E}}_{A_1B_1}}}_{j\in J} f^{{}^{(21)}}({ \mathfrak{l}}_{j,A_1})\boxtimes g^{{}^{(21)}}({ \mathfrak{l}}_{j,B_1})
\end{eqnarray}
\end{defin}

\begin{theoreme}
The pair of maps $((f\boxtimes g)_{{}_{(12)}}, (f\boxtimes g)^{{}^{(21)}})$ is a well defined symmetry, i.e.  a Chu morphism from the Chu space $({ \mathfrak{S}}_{A_1B_1},{ \mathfrak{E}}_{A_1B_1},\epsilon^{A_1B_1})$ to the Chu space $({ \mathfrak{S}}_{A_2B_2},{ \mathfrak{E}}_{A_2B_2},\epsilon^{A_2B_2})$.
\end{theoreme}
\begin{proof}
\begin{eqnarray}
\epsilon^{A_2B_2}_{{}_{\bigsqcap{}^{{}^{{ \mathfrak{E}}_{A_2B_2}}}_{j\in J} { \mathfrak{l}}_{j,A_2}\boxtimes { \mathfrak{l}}_{j,B_2}}}\left( (f\boxtimes g)_{{}_{(12)}}(\bigsqcap{}^{{}^{{ \mathfrak{S}}_{A_1B_1}}}_{i\in I} \sigma_{i,A_1}\boxtimes \sigma_{i,B_1})\right) &=& 
\epsilon^{A_2B_2}_{{}_{\bigsqcap{}^{{}^{{ \mathfrak{E}}_{A_2B_2}}}_{j\in J} { \mathfrak{l}}_{j,A_2}\boxtimes { \mathfrak{l}}_{j,B_2}}}\left( 
\bigsqcap{}^{{}^{{ \mathfrak{S}}_{A_2B_2}}}_{i\in I} f_{{}_{(12)}}(\sigma_{i,A_1})\boxtimes g_{{}_{(12)}}(\sigma_{i,B_1})
\right) \nonumber\\
&=& \bigwedge{}_{j\in J}
\bigwedge{}_{i\in I}\;
\epsilon^{A_2}_{{}_{ { \mathfrak{l}}_{j,A_2}}}(f_{{}_{(12)}}(\sigma_{i,A_1}))
\bullet
\epsilon^{B_2}_{{}_{{ \mathfrak{l}}_{j,B_2}}}(g_{{}_{(12)}}(\sigma_{i,B_1})) \nonumber\\
&=& \bigwedge{}_{j\in J}
\bigwedge{}_{i\in I}\;
\epsilon^{A_1}_{{}_{ f^{{}^{(21)}}({ \mathfrak{l}}_{j,A_2})}}(\sigma_{i,A_1})
\bullet
\epsilon^{B_1}_{{}_{ g^{{}^{(21)}}({ \mathfrak{l}}_{j,B_2})}}(\sigma_{i,B_1})\nonumber\\
&=& \epsilon^{A_1B_1}_{{}_{\bigsqcap{}^{{}^{{ \mathfrak{E}}_{A_1B_1}}}_{j\in J} f^{{}^{(21)}}({ \mathfrak{l}}_{j,A_2})\boxtimes g^{{}^{(21)}}({ \mathfrak{l}}_{j,B_2})}}\left( 
\bigsqcap{}^{{}^{{ \mathfrak{S}}_{A_2B_2}}}_{i\in I} \sigma_{i,A_1}\boxtimes \sigma_{i,B_1})
\right) \nonumber\\
&=& \epsilon^{A_1B_1}_{{}_{(f\boxtimes g)^{{}^{(21)}}\left(\bigsqcap{}^{{}^{{ \mathfrak{E}}_{A_2B_2}}}_{j\in J} { \mathfrak{l}}_{j,A_2}\boxtimes { \mathfrak{l}}_{j,B_2}\right)}}\left( 
\bigsqcap{}^{{}^{{ \mathfrak{S}}_{A_2B_2}}}_{i\in I} \sigma_{i,A_1}\boxtimes \sigma_{i,B_1})
\right).\;\;\;\;\;\;\;\;\;\;\;\;
\end{eqnarray}
\end{proof}

From this result, we deduce that the categories $Chu_{ \mathfrak{B}}^{S/E}$ and ${\mathcal{S}}$ are equipped with a tensor product.


\subsection{The canonical tensor product construction}\label{subsectionbasic}

We begin to introduce the classical construction of G.A. Fraser for the tensor product of semi-lattices \cite{Fraser1976,Fraser1978}. As it will be clarified in the next subsection new proposals for the tensor product of semi-lattices have to be made in order to complete our work.  

\begin{defin} Let $A, B$ and $C$ be semilattices. 
A function $f:A \times B \longrightarrow C$ is a bi-homomorphism if the functions $g_a:B \longrightarrow C$ defined by $g_a(b) = f(a, b)$ and $h_b:A\longrightarrow C$ defined by $h_b(a)=f(a, b)$ are homomorphisms for all $a\in A$ and $b\in B$.
\end{defin}

\begin{theoreme}{\bf \cite[Definition 2.2 and Theorem 2.3]{Fraser1976}}\label{theoremtensorbasic}\\
The tensor product ${S}_{AB}:={ \mathfrak{S}}_{A} \otimes { \mathfrak{S}}_{B}$ of the two Inf semi-lattices ${ \mathfrak{S}}_{A}$ and ${ \mathfrak{S}}_{B}$  is obtained as a solution of the following universal problem : there exists a bi-homomorphism, denoted $\iota$ from ${ \mathfrak{S}}_{A} \times { \mathfrak{S}}_{B}$ to ${S}_{AB}$, such that, for any Inf semi-lattice ${ \mathfrak{S}}$ and any bi-homomorphism $f$ from ${ \mathfrak{S}}_{A} \times { \mathfrak{S}}_{B}$ to ${ \mathfrak{S}}$,  there is a unique homomorphism $g$ from ${S}_{AB}$ to ${ \mathfrak{S}}$ with $f = g \circ \iota$. We denote $\iota(\sigma,\sigma')=\sigma \otimes \sigma'$ for any $\sigma\in { \mathfrak{S}}_{A}$ and $\sigma'\in { \mathfrak{S}}_{B}$.  \\
The tensor product ${S}_{AB}$ exists and is unique up to isomorphism, it is built as the homomorphic image of the free $\sqcap$ semi-lattice generated by the set ${ \mathfrak{S}}_{A} \times { \mathfrak{S}}_{B}$ under the congruence relation determined by identifying $(\sigma_1 \sqcap_{{}_{{ \mathfrak{S}}_{A}}}\sigma_2, \sigma')$ with $(\sigma_1,  \sigma')\sqcap (\sigma_2, \sigma')$ for all $\sigma_1,\sigma_2\in { \mathfrak{S}}_{A}, \sigma'\in { \mathfrak{S}}_{B}$ and identifying  $(\sigma, \sigma'_1 \sqcap_{{}_{{ \mathfrak{S}}_{B}}}\sigma'_2)$ with $(\sigma,  \sigma'_1)\sqcap (\sigma, \sigma'_2)$ for all $\sigma\in { \mathfrak{S}}_{A},\sigma'_1,\sigma'_2\in { \mathfrak{S}}_{B}$. \\
In other words, ${S}_{AB}$ is the Inf semi-lattice (the infimum of $S\subseteq {S}_{AB}$ will be denoted $\bigsqcap{}^{{}^{{S}_{AB}}} S$) generated by the elements $\sigma_A \otimes \sigma_B$ with $\sigma_A \in { \mathfrak{S}}_{A}, \sigma_B \in { \mathfrak{S}}_{B}$ and subject to the conditions 
\begin{eqnarray}
(\sigma_A\sqcap_{{}_{{ \mathfrak{S}}_{A}}}\sigma'_A)\otimes \sigma_B = (\sigma_A\otimes \sigma_B)\sqcap_{{}_{{S}_{AB}}}(\sigma'_A\otimes \sigma_B),&& \sigma_A \otimes (\sigma_B\sqcap_{{}_{{ \mathfrak{S}}_{B}}}\sigma'_B) = (\sigma_A\otimes \sigma_B)\sqcap_{{}_{{S}_{AB}}}(\sigma_A\otimes \sigma'_B). \;\;\;\;\;\;\;\;\;\;\;\;\;\;
\end{eqnarray}
The elements of ${S}_{AB}$ can be written $(\bigsqcap{}^{{}^{{S}_{AB}}}_{i\in I} \sigma_{i,A}\otimes \sigma_{i,B})$ with $I$ finite and $\sigma_{i,A}\in { \mathfrak{S}}_{A},  \sigma_{i,B}\in { \mathfrak{S}}_{B},$ for any $i\in I$.
\end{theoreme}

\begin{defin}
The space ${S}_{AB}={ \mathfrak{S}}_{A} \otimes { \mathfrak{S}}_{B}$ is turned into a partially ordered set with the following binary relation
\begin{eqnarray}
\forall \sigma_{AB},\sigma'_{AB} \in {S}_{AB},\;\;\;\; 
(\,\sigma_{AB} \sqsubseteq_{{}_{{S}_{AB}}} \sigma'_{AB}\,) & :\Leftrightarrow &
(\,\sigma_{AB} \sqcap_{{}_{{S}_{AB}}} \sigma'_{AB} = \sigma_{AB}\,).\;\;\;\;\;\;\;\;\;\;\;\;
\end{eqnarray}
\end{defin}

\begin{defin}
A non-empty subset ${ \mathfrak{R}}$ of ${ \mathfrak{S}}_{A} \times { \mathfrak{S}}_{B}$ is called a bi-filter of ${ \mathfrak{S}}_{A} \times { \mathfrak{S}}_{B}$ iff 
\begin{eqnarray}
&&\forall \sigma_A,\sigma_{1,A},\sigma_{2,A}\in { \mathfrak{S}}_{A},\forall \sigma_B,\sigma_{1,B},\sigma_{2,B}\in { \mathfrak{S}}_{B},\nonumber\\
&&(\,  (\sigma_{1,A},\sigma_{1,B})\leq (\sigma_{2,A},\sigma_{2,B}) \;\;\textit{\rm and}\;\; (\sigma_{1,A},\sigma_{1,B})\in { \mathfrak{R}} \,)\;\;\Rightarrow\;\; (\sigma_{2,A},\sigma_{2,B})\in { \mathfrak{R}},\label{defbifilter1}\\
&&(\sigma_{1,A},\sigma_{B}),(\sigma_{2,A},\sigma_{B})\in { \mathfrak{R}}\;\;\Rightarrow\;\; (\sigma_{1,A}\sqcap_{{}_{{ \mathfrak{S}}_A}}\sigma_{2,A},\sigma_{B})\in { \mathfrak{R}},\label{defbifilter2}\\
&&(\sigma_{A},\sigma_{1,B}),(\sigma_{A},\sigma_{2,B})\in { \mathfrak{R}}\;\;\Rightarrow\;\; (\sigma_{A},\sigma_{1,B}\sqcap_{{}_{{ \mathfrak{S}}_B}}\sigma_{2,B})\in { \mathfrak{R}}.\label{defbifilter3}
\end{eqnarray}
\end{defin}
\begin{defin}
If $\{(\sigma_{1,A},\sigma_{1,B}),\cdots,(\sigma_{n,A},\sigma_{n,B})\}$ is a non-empty finite subset of ${ \mathfrak{S}}_{A} \times { \mathfrak{S}}_{B}$, then the intersection of the collection of all bi-filters of ${ \mathfrak{S}}_{A} \times { \mathfrak{S}}_{B}$ which contain $(\sigma_{1,A},\sigma_{1,B}),\cdots,(\sigma_{n,A},\sigma_{n,B})$ is a bi-filter, which we denote by ${ \mathfrak{F}}\{(\sigma_{1,A},\sigma_{1,B}),\cdots,(\sigma_{n,A},\sigma_{n,B})\}$.
\end{defin}

\begin{lemme}\label{FalphaF}
If $F$ is a filter of ${S}_{AB}={ \mathfrak{S}}_{A} \otimes { \mathfrak{S}}_{B}$ then the set $\alpha(F):=\{\,(\sigma_{A},\sigma_{B})\in { \mathfrak{S}}_{A} \times { \mathfrak{S}}_{B}\;\vert\; \sigma_{A}\otimes \sigma_{B} \in F\;\}$ is a bi-filter of ${ \mathfrak{S}}_{A} \times { \mathfrak{S}}_{B}$.
\end{lemme}

\begin{lemme}{\bf \cite[Lemma 1]{Fraser1978}}\label{sigmainFinequality}
Let us choose $\sigma_A,\sigma_{1,A},\cdots,\sigma_{n,A}\in { \mathfrak{S}}_{A}$ and $\sigma_B,\sigma_{1,B},\cdots,\sigma_{n,B}\in { \mathfrak{S}}_{B}$. Then, 
\begin{eqnarray}
(\sigma_{A},\sigma_{B})\in { \mathfrak{F}}\{(\sigma_{1,A},\sigma_{1,B}),\cdots,(\sigma_{n,A},\sigma_{n,B})\} & \Leftrightarrow & \left( \bigsqcap{}^{{}^{{S}_{AB}}}_{1\leq i\leq n}\; \sigma_{i,A}\otimes \sigma_{i,B}\right) \sqsubseteq_{{}_{{S}_{AB}}} \sigma_{A}\otimes \sigma_{B}.
\end{eqnarray}
\end{lemme}
\begin{proof} Let us suppose that $(\sigma_{A},\sigma_{B})\in { \mathfrak{F}}\{(\sigma_{1,A},\sigma_{1,B}),\cdots,(\sigma_{n,A},\sigma_{n,B})\} $. Let $F$ be the principal filter in ${ \mathfrak{S}}_{A} \otimes { \mathfrak{S}}_{B}$ generated by $\left( \bigsqcap{}^{{}^{{S}_{AB}}}_{1\leq i\leq n}\; \sigma_{i,A}\otimes \sigma_{i,B}\right)$. Then $\sigma_{i,A}\otimes \sigma_{i,B} \in F$ for any $1\leq i\leq n$, and then $(\sigma_{i,A}, \sigma_{i,B})\in \alpha(F)$ for any $1\leq i\leq n$. Hence, ${ \mathfrak{F}}\{(\sigma_{1,A},\sigma_{1,B}),\cdots,(\sigma_{n,A},\sigma_{n,B})\}\subseteq \alpha(F)$, and then $(\sigma_{A},\sigma_{B})\in \alpha(F)$. As a result, $\sigma_{A}\otimes \sigma_{B} \in F$ and then $\left( \bigsqcap{}^{{}^{{S}_{AB}}}_{1\leq i\leq n}\; \sigma_{i,A}\otimes \sigma_{i,B}\right) \sqsubseteq_{{}_{{S}_{AB}}} \sigma_{A}\otimes \sigma_{B}$.\\
Let us now suppose that $\left( \bigsqcap{}^{{}^{{S}_{AB}}}_{1\leq i\leq n}\; \sigma_{i,A}\otimes \sigma_{i,B}\right) \sqsubseteq_{{}_{{S}_{AB}}} \sigma_{A}\otimes \sigma_{B}$. Let $u: { \mathfrak{S}}_{A} \times { \mathfrak{S}}_{B} \rightarrow \{0,1\}$ be such that  \begin{eqnarray} u(\sigma_{A}, \sigma_{B})=1\;\Leftrightarrow\; (\sigma_{A}, \sigma_{B})\in { \mathfrak{F}}\{(\sigma_{1,A},\sigma_{1,B}),\cdots,(\sigma_{n,A},\sigma_{n,B})\} \end{eqnarray} $u$ is a bi-homomorhism. Then,  there exists a homomorphism $v:{ \mathfrak{S}}_{A} \otimes { \mathfrak{S}}_{B} \rightarrow \{0,1\}$ such that $u(\sigma_{A}, \sigma_{B})=v(\sigma_{A}\otimes \sigma_{B})$ for any $\sigma_A\in { \mathfrak{S}}_{A}$ and $\sigma_B\in { \mathfrak{S}}_{B}$. We have then $u(\sigma_{A}, \sigma_{B})=v(\sigma_{A}\otimes \sigma_{B}) \geq v(\bigsqcap{}^{{}^{{S}_{AB}}}_{1\leq i\leq n}\; \sigma_{i,A}\otimes \sigma_{i,B})= \bigwedge{}_{1\leq i\leq n}\;v( \sigma_{i,A}\otimes \sigma_{i,B})=\bigwedge{}_{1\leq i\leq n}\;u( \sigma_{i,A},\sigma_{i,B})$. Since  $u( \sigma_{i,A},\sigma_{i,B})=1$ for any $1\leq i\leq n$, we deduce that $u(\sigma_{A}, \sigma_{B})=1$ and then $(\sigma_{A}, \sigma_{B})\in { \mathfrak{F}}\{(\sigma_{1,A},\sigma_{1,B}),\cdots,(\sigma_{n,A},\sigma_{n,B})\}$.\end{proof}

\begin{lemme}{\bf \cite[Theorem 1]{Fraser1978}}\label{orderimpliespolynomial}\\
Let us choose $\sigma_A,\sigma_{1,A},\cdots,\sigma_{n,A}\in { \mathfrak{S}}_{A}$ and $\sigma_B,\sigma_{1,B},\cdots,\sigma_{n,B}\in { \mathfrak{S}}_{B}$. Then, 
\begin{eqnarray}
\hspace{-1cm} \left( \bigsqcap{}^{{}^{{S}_{AB}}}_{1\leq i\leq n}\; \sigma_{i,A}\otimes \sigma_{i,B}\right) \sqsubseteq_{{}_{{S}_{AB}}} \sigma_{A}\otimes \sigma_{B} & \Leftrightarrow &  \;\textit{\rm there exists a n$-$ary lattice polynomial $p$}\;\vert\; \sigma_{A}\sqsupseteq_{{}_{{ \mathfrak{S}}_A}} p(\sigma_{1,A},\cdots,\sigma_{n,A})\nonumber\\
&&\textit{\rm and}\;\;  \sigma_{B}\sqsupseteq_{{}_{{ \mathfrak{S}}_B}} p^\ast(\sigma_{1,B},\cdots,\sigma_{n,B}).
\end{eqnarray}
where $p^\ast$ denotes the lattice polynomial obtained from $p$ by dualizing the lattice operations.
\end{lemme}
\begin{proof}
Let us fix $\sigma_{1,A},\cdots,\sigma_{n,A}\in { \mathfrak{S}}_{A}$ and $\sigma_{1,B},\cdots,\sigma_{n,B}\in { \mathfrak{S}}_{B}$ and let us consider 
\begin{eqnarray}
F:=\{\,(\sigma_{A},\sigma_{B})\;\vert\; 
\sigma_{A}\sqsupseteq_{{}_{{ \mathfrak{S}}_A}} p(\sigma_{1,A},\cdots,\sigma_{n,A})\;\;\textit{\rm and}\;\;
\sigma_{B}\sqsupseteq_{{}_{{ \mathfrak{S}}_B}} p^\ast(\sigma_{1,B},\cdots,\sigma_{n,B})\;\;\textit{\rm for some n$-$ary polynomial $p$}\,\}. 
\end{eqnarray}
It is obvious that $F$ contains $(\sigma_{1,A},\sigma_{1,B}),\cdots, (\sigma_{n,A},\sigma_{n,B})$. \\
It is also easy to check that $F$ is a bi-filter.  \\
Endly, we can check that every bi-filter which contains $(\sigma_{1,A},\sigma_{1,B})$, $\cdots$, $(\sigma_{n,A},\sigma_{n,B})$ contains also  $F$. This point can be checked by induction on the complexity of the polynomial $p$ by using the following elementary result, consequence of the bi-filter character of $F$,
\begin{eqnarray}
\forall \sigma_A,\sigma'_A\in { \mathfrak{S}}_A,\sigma_B,\sigma'_B\in { \mathfrak{S}}_B,\;\;\;\; \left( (\sigma_A,\sigma_B),(\sigma'_A,\sigma'_B)\in F \right) &\Rightarrow & \left\{
\begin{array}{l}
(\sigma_A \sqcup_{{}_{{ \mathfrak{S}}_A}} \sigma'_A,\sigma_B \sqcap_{{}_{{ \mathfrak{S}}_B}} \sigma'_B)\in F\nonumber\\
(\sigma_A \sqcap_{{}_{{ \mathfrak{S}}_A}} \sigma'_A,\sigma_B \sqcup_{{}_{{ \mathfrak{S}}_B}} \sigma'_B)\in F
\end{array}\right.
\end{eqnarray}
\end{proof}

\begin{theoreme}
For any ${\mathfrak{l}}_A\in { \mathfrak{E}}_{A}, {\mathfrak{l}}_B\in { \mathfrak{E}}_{B}$ the map
\begin{eqnarray}
\begin{array}{rrcl} 
f^{AB}_{{\mathfrak{l}}_A, {\mathfrak{l}}_B} \; :& { \mathfrak{S}}_{A} \times { \mathfrak{S}}_{B} & \rightarrow & { \mathfrak{B}}\\
  & (\sigma_A,\sigma_B) & \mapsto & {\epsilon}\,{}^{A}_{{\mathfrak{l}}_A} (\sigma_{A}) \bullet {\epsilon}\,{}^{B}_{{\mathfrak{l}}_B} (\sigma_B)
  \end{array}
\end{eqnarray}
is a bi-homomorphism. Then, there exists a unique homomorphism from ${S}_{AB}={ \mathfrak{S}}_{A} \otimes { \mathfrak{S}}_{B}$ to ${ \mathfrak{B}}$, denoted ${\nu}\,{}^{AB}_{{\mathfrak{l}}_A, {\mathfrak{l}}_B}$, and satisfying $f^{AB}_{{\mathfrak{l}}_A, {\mathfrak{l}}_B} = {\nu}\,{}^{AB}_{{\mathfrak{t}}_A, {\mathfrak{t}}_B} \circ \iota$.
Explicitly, we have
\begin{eqnarray}
 {\nu}\,{}^{AB}_{{\mathfrak{l}}_A, {\mathfrak{l}}_B}(\bigsqcap{}^{{}^{{S}_{AB}}}_{i\in I} \sigma_{i,A}\otimes \sigma_{i,B})=\bigwedge{}_{i\in I}\; {\epsilon}\,{}^{A}_{{ \mathfrak{l}}_A}(\sigma_{i,A}) \bullet {\epsilon}\,{}^{B}_{{ \mathfrak{l}}_B}(\sigma_{i,B}).\label{etildemorphism}
\end{eqnarray}
Anticipating the construction of the bipartite effect state, we may denote ${\nu}\,{}^{AB}_{{\mathfrak{l}}_A, {\mathfrak{l}}_B}$ by ${\epsilon}\,{}^{AB}_{{\mathfrak{l}}_A\otimes {\mathfrak{l}}_B}$.
\end{theoreme}
\begin{proof}
The bi-homomorphic property is a direct consequence of (\ref{propconjunctiveprop}) and (\ref{distributivitybullet}).  The existence of ${\nu}\,{}^{AB}_{{\mathfrak{l}}_A, {\mathfrak{l}}_B}$ satisfying $f^{AB}_{{\mathfrak{l}}_A, {\mathfrak{l}}_B} = {\nu}\,{}^{AB}_{{\mathfrak{l}}_A, {\mathfrak{l}}_B} \circ \iota$ is then obtained as a consequence of Theorem \ref{theoremtensorbasic}.
\end{proof}

\begin{theoreme}  
\begin{eqnarray}
&&\forall \sigma_{AB},\sigma'_{AB} \in {{S}_{AB}},\;\;\;\; 
(\,\sigma_{AB} \sqsubseteq_{{}_{{S}_{AB}}} \sigma'_{AB}\,) \;\; \Rightarrow \;\;
(\,\forall {\mathfrak{l}}_{A}\in { \mathfrak{E}}_{A},\forall {\mathfrak{l}}_{B}\in { \mathfrak{E}}_{B},\;\;
{\nu}\,{}^{AB}_{{\mathfrak{l}}_{A}, {\mathfrak{l}}_{B}} (\sigma_{AB})\leq{\nu}\,{}^{AB}_{{\mathfrak{l}}_{A}, {\mathfrak{l}}_{B}} (\sigma'_{AB})\,),\;\;\;\;\;\;\;\;\;\;\;\;\label{tensororder2}\\
&& \forall \{\,\sigma_{i,AB}\;\vert\; i\in I\,\}\subseteq_{fin} {{S}_{AB}},\forall {\mathfrak{l}}_{A}\in { \mathfrak{E}}_{A},\forall {\mathfrak{l}}_{B}\in { \mathfrak{E}}_{B},\;\;\;\;
{\nu}\,{}^{AB}_{{\mathfrak{l}}_{A}, {\mathfrak{l}}_{B}} ( \bigsqcap{}^{{}^{{S}_{AB}}}_{i\in I} \sigma_{i,AB}) =
\bigwedge{}_{i\in I}\,
{\nu}\,{}^{AB}_{{\mathfrak{l}}_{A}, {\mathfrak{l}}_{B}} (\sigma_{i,AB}).\;\;\;\;\;\;\label{tensorinfimum2}
\end{eqnarray}
\end{theoreme}

\subsection{The maximal tensor product} \label{subsectionmaximaltensor}

\begin{defin}
Let us denote by $\widecheck{S}_{AB}$ the set of bimorphisms from ${ \mathfrak{E}}_{{ \mathfrak{S}}_A}\times { \mathfrak{E}}_{{ \mathfrak{S}}_B}$ to ${ \mathfrak{E}}_\bot\cong { \mathfrak{B}}$, i.e. the set of maps $\phi$ satisfying
\begin{eqnarray}
\forall \{{ \mathfrak{l}}_{i,A}\;\vert\; i\in I\}\subseteq { \mathfrak{E}}_{{ \mathfrak{S}}_A},\forall { \mathfrak{l}}_{B} \in { \mathfrak{E}}_{{ \mathfrak{S}}_B}, && \phi(\bigsqcap{}^{{}_{{ \mathfrak{E}}_{{ \mathfrak{S}}_A}}}_{i\in I} { \mathfrak{l}}_{i,A},{ \mathfrak{l}}_{B})= \bigwedge{}_{i\in I} \;\phi( { \mathfrak{l}}_{i,A},{ \mathfrak{l}}_{B})\label{bilinear1}\\
\forall \{{ \mathfrak{l}}_{j,B}\;\vert\; j\in J\}\subseteq { \mathfrak{E}}_{{ \mathfrak{S}}_B},\forall { \mathfrak{l}}_{A} \in { \mathfrak{E}}_{{ \mathfrak{S}}_A}, && \phi({ \mathfrak{l}}_{A},\bigsqcap{}^{{}_{{ \mathfrak{E}}_{{ \mathfrak{S}}_B}}}_{j\in J} { \mathfrak{l}}_{j,B})= \bigwedge{}_{j\in J} \;\phi({ \mathfrak{l}}_{A}, { \mathfrak{l}}_{j,B})\label{bilinear2}
\end{eqnarray}
\end{defin}

\begin{lemme}
$\widecheck{S}_{AB}$ is equipped with the pointwise partial order.\\
$\widecheck{S}_{AB}$ is a down complete Inf semi-lattice with
\begin{eqnarray}
\forall \{\phi_i\;\vert\; i\in I\}\subseteq \widecheck{S}_{AB},&&\forall { \mathfrak{l}}_{A} \in { \mathfrak{E}}_{{ \mathfrak{S}}_A},\forall { \mathfrak{l}}_{B} \in { \mathfrak{E}}_{{ \mathfrak{S}}_B},\;\; (\bigsqcap{}^{{}^{\widecheck{S}_{AB}}}_{i\in I} \phi_i)({ \mathfrak{l}}_{A},{ \mathfrak{l}}_{B}):=\bigwedge{}_{i\in I} \;\phi_i({ \mathfrak{l}}_{A},{ \mathfrak{l}}_{B}).
\end{eqnarray}
It is naturally equipped with the following family of Inf semi-lattice homomorphisms defined for any 
$ {\mathfrak{l}}_{A}\in { \mathfrak{E}}_{A}$ and ${\mathfrak{l}}_{B}\in { \mathfrak{E}}_{B}$ by
\begin{eqnarray}
\begin{array}{rcrcl}
\nu\,{}^{AB}_{{\mathfrak{l}}_A,{\mathfrak{l}}_B} & : & 
\widecheck{S}_{AB}
 & \longrightarrow & { \mathfrak{B}}\\
& & \phi & \mapsto & \nu\,{}^{AB}_{{\mathfrak{l}}_A,{\mathfrak{l}}_B}(\phi) := \phi({\mathfrak{l}}_A,{\mathfrak{l}}_B).
\end{array}
\end{eqnarray}
We note that, for any $\phi$ and $\phi'$ in $\widecheck{S}_{AB}$, we have
\begin{eqnarray}
(\,\forall {\mathfrak{l}}_{A}\in { \mathfrak{E}}_{A},\forall {\mathfrak{l}}_{B}\in { \mathfrak{E}}_{B}, \nu\,{}^{AB}_{{\mathfrak{l}}_A,{\mathfrak{l}}_B}(\phi)=\nu\,{}^{AB}_{{\mathfrak{l}}_A,{\mathfrak{l}}_B}(\phi')\,) &\Leftrightarrow & (\,\phi=\phi'\,).\label{checkB5direct}
\end{eqnarray}
\end{lemme}

\begin{lemme}
$\widecheck{S}_{AB}$ has a bottom element denoted $\bot_{{}_{\widecheck{S}_{AB}}}$ with
\begin{eqnarray}
&&\forall { \mathfrak{l}}_{A} \in { \mathfrak{E}}_{{ \mathfrak{S}}_A},\forall { \mathfrak{l}}_{B} \in { \mathfrak{E}}_{{ \mathfrak{S}}_B},\;\; (\bot_{{}_{\widecheck{S}_{AB}}})({ \mathfrak{l}}_{A},{ \mathfrak{l}}_{B}):=\bot.
\end{eqnarray}
In other words, ${{\widecheck{S}}}_{AB}$ satisfies Axiom {\bf (A2)}. 
\end{lemme}

\begin{defin}
${ \mathcal{P}}({ \mathfrak{E}}_{{ \mathfrak{S}}_A}\times { \mathfrak{E}}_{{ \mathfrak{S}}_B})$ is equipped with the congruence relation $\sim$ defined as follows
\begin{eqnarray}
(\, \{\,({ \mathfrak{l}}_{i,A},{ \mathfrak{l}}_{i,B})\;\vert\; i\in I\,\} \sim \{\,({ \mathfrak{l}}'_{j,A},{ \mathfrak{l}}'_{j,B})\;\vert\; j\in J\,\} \,) & \Leftrightarrow & (\, \forall \phi\in \widecheck{S}_{AB},\;\; \bigwedge{}_{i\in I}\, \phi({ \mathfrak{l}}_{i,A},{ \mathfrak{l}}_{i,B}) =\bigwedge{}_{j\in J}\, \phi({ \mathfrak{l}}'_{j,A},{ \mathfrak{l}}'_{j,B})\,).\;\;\;\;\;\;\;\;\;\;\;\;\;\;\;\;\;
\end{eqnarray}
\end{defin}

\begin{defin}
The space $\widecheck{E}_{AB}$ is defined as the quotient of ${ \mathcal{P}}({ \mathfrak{E}}_{{ \mathfrak{S}}_A}\times { \mathfrak{E}}_{{ \mathfrak{S}}_B})$ by the congruence relation $\sim$.  
\begin{eqnarray}
\forall \lambda_{AB}\in { \mathcal{P}}({ \mathfrak{E}}_{A} \times { \mathfrak{E}}_{B}), && \widecheck{\lambda}_{AB}:=\{\, x_{AB}\in { \mathcal{P}}({ \mathfrak{E}}_{A} \times { \mathfrak{E}}_{B}) \;\vert\; \lambda_{AB}\sim x_{AB}\,\}.
\end{eqnarray}
\end{defin}

\begin{defin}
The evaluation map will be defined as a map from ${{\widecheck{E}}}_{AB}$ to ${ \mathfrak{B}}^{{{\widecheck{S}}}_{AB}}$ by 
\begin{eqnarray}
\forall \lambda_{AB}=\{\,({ \mathfrak{l}}_{i,A},{ \mathfrak{l}}_{i,B})\;\vert\; i\in I\,\}\in { \mathcal{P}}({ \mathfrak{E}}_{A} \times { \mathfrak{E}}_{B}), \forall \phi \in \widecheck{S}_{AB},&& \epsilon\,{}^{{{\widecheck{S}}}_{AB}}_{\widecheck{\lambda}_{AB}}(\phi):= \bigwedge{}_{i\in I}\, \phi({ \mathfrak{l}}_{i,A},{ \mathfrak{l}}_{i,B}).
\end{eqnarray} 
\end{defin}

\begin{lemme}
The evaluation map satisfies trivially
\begin{eqnarray}
\forall \lambda_{AB}\in { \mathcal{P}}({ \mathfrak{E}}_{A} \times { \mathfrak{E}}_{B}), \forall \{\phi_i\;\vert\; i\in I \} \subseteq \widecheck{S}_{AB},&& \epsilon\,{}^{{{\widecheck{S}}}_{AB}}_{\widecheck{\lambda}_{AB}}(\bigsqcap{}^{{}^{\widecheck{S}_{AB}}}_{i\in I}\phi_i)= \bigwedge{}_{i\in I}\;  \epsilon\,{}^{{{\widecheck{S}}}_{AB}}_{\widecheck{\lambda}_{AB}}(\phi_i).
\end{eqnarray} 
In other words, ${{\widecheck{S}}}_{AB}$ satisfies Axiom {\bf (A1)}. 
\end{lemme}

\begin{defin}
${{\widecheck{E}}}_{AB}$ is equipped with a partial order defined according to
\begin{eqnarray}
\forall \widecheck{\lambda}_{AB},\widecheck{\lambda}'_{AB} \in {{{\widecheck{E}}}_{AB}},\;\;\;\; (\,\widecheck{\lambda}_{AB} \sqsubseteq_{{}_{{{\widecheck{S}}}_{AB}}} \widecheck{\lambda}'_{AB}\,) & :\Leftrightarrow &
(\,\forall \phi\in {\widecheck{S}}_{AB},\;\;
{\epsilon}\,{}^{{{\widecheck{S}}}_{AB}}_{\widecheck{\lambda}_{AB}} (\phi)\leq {\epsilon}\,{}^{{{\widecheck{S}}}_{AB}}_{\widecheck{\lambda}\!{}'_{AB}} (\phi)\,).\;\;\;\;\;\;\;\;\;\;\;\;
\end{eqnarray}
\end{defin}

\begin{defin}
We will adopt the following definition
\begin{eqnarray}
\forall \widecheck{\lambda}\in {{\widecheck{E}}}_{AB},\;\;\;\; \langle \widecheck{\lambda} \rangle & := & Max \{\,x\in { \mathcal{P}}({ \mathfrak{E}}_{A} \times { \mathfrak{E}}_{B})\;\vert\; \widecheck{x}  \sqsupseteq_{{}_{{{\widecheck{E}}}_{AB}}} \widecheck{\lambda} \,\}\nonumber\\
& = & \{\,({ \mathfrak{l}}_A,{ \mathfrak{l}}_B)\in { \mathfrak{E}}_{A} \times { \mathfrak{E}}_{B} \;\vert\; \widecheck{({ \mathfrak{l}}_A,{ \mathfrak{l}}_B)}  \sqsupseteq_{{}_{{{\widecheck{E}}}_{AB}}}\widecheck{\lambda} \,\},
\end{eqnarray}
\end{defin}

\begin{lemme}
We have the following Galois relation  
\begin{eqnarray}
\forall \widecheck{\lambda}\in {{\widecheck{E}}}_{AB},\forall x\in { \mathcal{P}}({ \mathfrak{E}}_{A} \times { \mathfrak{E}}_{B}),&&  \langle \widecheck{\lambda} \rangle \supseteq x \;\;\;\Leftrightarrow\;\;\; \widecheck{\lambda} \sqsubseteq_{{}_{{{\widecheck{E}}}_{AB}}} \widecheck{x}.
\end{eqnarray}
\end{lemme}
\begin{proof}
Let us fix $x:=\{\,({ \mathfrak{l}}_{i,A},{ \mathfrak{l}}_{i,B})\;\vert\; i\in I\,\}$.  We derive straightforwardly the following equivalences
\begin{eqnarray}
\langle \widecheck{\lambda} \rangle \supseteq x & \Leftrightarrow & 
\forall i\in I, \widecheck{({ \mathfrak{l}}_{i,A},{ \mathfrak{l}}_{i,B})}  \sqsupseteq_{{}_{{{\widecheck{E}}}_{AB}}}\widecheck{\lambda}\nonumber\\
& \Leftrightarrow & \forall i\in I,  \forall \phi\in \widecheck{S}_{AB}, \;\;\;\;{\epsilon}\,{}^{{{\widecheck{S}}}_{AB}}_{\widecheck{\lambda}} (\phi) \leq
{\nu}\,{}^{AB}_{{\mathfrak{l}}_{i,A},{\mathfrak{l}}_{i,B}} (\phi)\nonumber\\
 & \Leftrightarrow & 
 \forall \phi\in \widecheck{S}_{AB}, \;\;\;\;{\epsilon}\,{}^{{{\widecheck{S}}}_{AB}}_{\widecheck{\lambda}} (\phi) \leq \bigwedge{}_{i\in I}\;
{\nu}\,{}^{AB}_{{\mathfrak{l}}_{i,A},{\mathfrak{l}}_{i,B}} (\phi)=
{\epsilon}\,{}^{{{\widecheck{S}}}_{AB}}_{x} (\phi)\nonumber\\
& \Leftrightarrow & \widecheck{\lambda} \sqsubseteq_{{}_{{{\widecheck{E}}}_{AB}}} \widecheck{x}.
\end{eqnarray}  
\end{proof}

\begin{theoreme}
${{\widecheck{E}}}_{AB}$ is a down-complete Inf semi-lattice with
\begin{eqnarray}
\forall \{\,x_i\;\vert\; i\in I\,\}\subseteq { \mathcal{P}}({ \mathfrak{E}}_{A} \times { \mathfrak{E}}_{B}),&&\bigsqcap{}^{{}^{{{\widecheck{E}}}_{AB}}}_{i\in I} \; \widecheck{x_i}  = \widecheck{\bigcup{}_{i\in I}\; x_i}.\label{tensorinfEABcheck}
\end{eqnarray}
Moreover,  we have
\begin{eqnarray}
\forall \{\,\widecheck{\lambda}_i\;\vert\; i\in I\,\}\subseteq {{\widecheck{E}}}_{AB},\forall \phi\in {{\widecheck{S}}}_{AB}, && {\epsilon}\,{}^{{{\widecheck{S}}}_{AB}}_{\bigsqcap{}^{{}^{{{\widecheck{E}}}_{AB}}}_{i\in I} \widecheck{\lambda}_i}(\phi) =  \bigwedge{}_{\! i\in I}\, {\epsilon}\,{}^{{{\widecheck{S}}}_{AB}}_{\widecheck{\lambda}_i}( \phi )
\end{eqnarray}
In other words, ${{\widecheck{E}}}_{AB}$ satisfies Axiom {\bf (A3)}. 
\end{theoreme}
\begin{proof}
The property (\ref{tensorinfEABcheck}) is a direct consequence of the Galois relation established in previous lemma.
\end{proof}

\begin{remark}
The set of pure states $\widecheck{S}{}_{AB}^{\;{}^{pure}}$ can be easily built as follows. For any map
\begin{eqnarray}
\begin{array}{rcrcl}
\gamma & : & { \mathfrak{E}}_A^{{}^{pure}}\times { \mathfrak{E}}_B^{{}^{pure}} & \longrightarrow & \{\textit{\bf Y},\textit{\bf N}\}
\end{array}
\end{eqnarray}
we define an element $\phi_\gamma$ of $\widecheck{S}{}_{AB}^{\;{}^{pure}}$ by
\begin{eqnarray}
\forall { \mathfrak{l}}_A\in { \mathfrak{E}}_A, \forall { \mathfrak{l}}_B\in { \mathfrak{E}}_B,&& \phi_\gamma ({ \mathfrak{l}}_A,{ \mathfrak{l}}_B):=\bigwedge{}_{{ \mathfrak{u}}_A\in \underline{{ \mathfrak{l}}_A}_{{ \mathfrak{E}}_A}} \bigwedge{}_{{ \mathfrak{u}}_B\in \underline{{ \mathfrak{l}}_B}_{{ \mathfrak{E}}_B}} \gamma ({ \mathfrak{u}}_A,{ \mathfrak{u}}_B).
\end{eqnarray}
It is then rather easy to check axioms {\bf (A4)} and {\bf (A5)}.
\end{remark}

\begin{defin}
The element $\widecheck{{ \mathfrak{l}}}_{AB}\in {{\widecheck{E}}}_{AB}$ associated to the element ${\mathfrak{l}}_{AB}:=\{\, ({ \mathfrak{l}}_{i,A},{ \mathfrak{l}}_{i,B})\;\vert\; i\in I\,\}\in { \mathcal{P}}({ \mathfrak{E}}_{A} \times { \mathfrak{E}}_{B})$ will be denoted $\bigsqcap{}^{{}^{{{\widecheck{E}}}_{AB}}}_{i\in I} { \mathfrak{l}}_{i,A}\widecheck{\otimes}{ \mathfrak{l}}_{i,B}$.
\end{defin}

from previous lemmas, we deduce that the Axioms {\bf (B1)} and {\bf (B2)} are satisfied. \\
The Axiom {\bf (B3)} is trivial by construction.  Axioms {\bf (B3')} and {\bf (B3'')} are direct consequences of properties (\ref{bilinear1}) and (\ref{bilinear2}).\\

Let us now consider the axioms {\bf (B4) (B4')} and {\bf (B4'')}. Inclusion of pure tensors is realized as follows. 
\begin{eqnarray}
\begin{array}{rcrcl}
\iota^{ \mathfrak{S}}_{AB} &:& {{ \mathfrak{S}}_A}\times {{ \mathfrak{S}}_B} & \longrightarrow & \widecheck{S}_{A B}\\
& & (\sigma_A,\sigma_B) & \mapsto & \phi_{(\sigma_A,\sigma_B)}\;\vert\; \forall ({ \mathfrak{l}}_A,{ \mathfrak{l}}_B)\in { \mathfrak{E}}_{{ \mathfrak{S}}_A}\times { \mathfrak{E}}_{{ \mathfrak{S}}_B},\; \phi_{(\sigma_A,\sigma_B)}({ \mathfrak{l}}_A,{ \mathfrak{l}}_B):= \epsilon^{{ \mathfrak{S}}_A}_{{ \mathfrak{l}}_A}(\sigma_A) \bullet \epsilon^{{ \mathfrak{S}}_B}_{{ \mathfrak{l}}_B}(\sigma_B)\;\;\in { \mathfrak{B}}.
\end{array}
\end{eqnarray}
As a consequence, Axiom {\bf (B4)} and {\bf (C)} are satisfied.  Axioms {\bf (B4')} and {\bf (B4'')} are then derived as in proof (\ref{demobifilterproperty}) using the relation {\bf (B5)} that is checked directly in (\ref{checkB5direct}).

\subsection{Candidates for a tensor product}

Let us consider a candidate for the tensor product of the spaces of states ${ \mathfrak{S}}_A$ and ${ \mathfrak{S}}_B$ denoted ${ \mathfrak{S}}_{AB}$. In other words, let us suppose that $({ \mathfrak{S}}_{AB}, { \mathfrak{E}}_{{ \mathfrak{S}}_{AB}},\epsilon^{AB})$ satisfies axioms {\bf (B1)} to {\bf (B5)}. 

\begin{theoreme} We have the following injective Inf semi-lattice homomorphism
\begin{eqnarray}
&&\begin{array}{rcrcl}
\mu & : & {\mathfrak{S}}_{AB} & \hookrightarrow & \widecheck{S}_{AB}\\
& & { \mathfrak{s}}_{AB} & \mapsto & \mu_{{{ \mathfrak{s}}_{AB}}}\;\;\vert\;\; \forall ({ \mathfrak{l}}_A,{ \mathfrak{l}}_B)\in { \mathfrak{E}}_{{ \mathfrak{S}}_A}\times { \mathfrak{E}}_{{ \mathfrak{S}}_B},\;\;\;\; \mu_{{{ \mathfrak{s}}_{AB}}}({ \mathfrak{l}}_A,{ \mathfrak{l}}_B):= \epsilon^{{{\mathfrak{S}}}_{AB}}_{{ \mathfrak{l}}_A\boxtimes { \mathfrak{l}}_B}({ \mathfrak{s}}_{AB}).
\end{array}
\end{eqnarray}
Note that ${ \mathfrak{l}}_A\boxtimes { \mathfrak{l}}_B$ is an element of ${ \mathfrak{S}}_{AB}$ (due to axiom {\bf(B3)}) once $ ({ \mathfrak{l}}_A,{ \mathfrak{l}}_B)$ is fixed in ${ \mathfrak{E}}_{{ \mathfrak{S}}_A}\times { \mathfrak{E}}_{{ \mathfrak{S}}_B}$.
\end{theoreme}
\begin{proof}
We can prove that $\phi_{{ \mathfrak{s}}_{AB}}$ satisfies properties (\ref{bilinear1})(\ref{bilinear2}) 
 using axioms {\bf(B2)} and {\bf (B3')(B3'')}. The map associating $\phi_{{ \mathfrak{s}}_{AB}}$ to ${ \mathfrak{s}}_{AB}$ is injective because of axiom {\bf (B5)}.  This map is an Inf semi-lattice homomorphism because of the axiom {\bf (B1)}.
\end{proof}

\subsection{The basic tensor product}\label{subsectionminimal}

It is now possible to give a definition of the tensor product of ${ \mathfrak{S}}_{A}$ and ${ \mathfrak{S}}_{B}$ that will be defined in reference to the axiomatic relations {\bf (B1) $-$ (B5)} and {\bf (C)}.  This tensor product will be called {\em basic tensor product} and denoted ${{\widetilde{S}}}_{AB}$.  

\begin{defin}\label{definnubullet}
The set ${ \mathcal{P}}({ \mathfrak{S}}_{A} \times { \mathfrak{S}}_{B})$ is equipped with the Inf semi-lattice structure $\cup$ and with the following Inf semi-lattice morphisms defined for any $ {\mathfrak{l}}_{A}\in { \mathfrak{E}}_{A}$ and ${\mathfrak{l}}_{B}\in { \mathfrak{E}}_{B}$,
\begin{eqnarray}
\begin{array}{rcrcl}
\nu\,{}^{AB}_{{\mathfrak{l}}_A,{\mathfrak{l}}_B} & : & { \mathcal{P}}({ \mathfrak{S}}_{A} \times { \mathfrak{S}}_{B}) & \longrightarrow & { \mathfrak{B}}\\
& & \{\,(\sigma_{i,A},\sigma_{i,B})\;\vert\; i\in I\,\} & \mapsto & \nu\,{}^{AB}_{{\mathfrak{l}}_A,{\mathfrak{l}}_B}(\{\,(\sigma_{i,A},\sigma_{i,B})\;\vert\; i\in I\,\}) := \bigwedge{}_{i\in I}\; {\epsilon}\,{}^{{ \mathfrak{S}}_{A}}_{{ \mathfrak{l}}_A}(\sigma_{i,A}) \bullet {\epsilon}\,{}^{{ \mathfrak{S}}_{B}}_{{ \mathfrak{l}}_B}(\sigma_{i,B}).
\end{array}
\end{eqnarray}
\end{defin}

\begin{defin}
${ \mathcal{P}}({ \mathfrak{S}}_{A} \times { \mathfrak{S}}_{B})$ is equipped with a congruence relation defined between any two elements $u_{AB}$ and $u'_{AB}$ of ${ \mathcal{P}}({ \mathfrak{S}}_{A} \times { \mathfrak{S}}_{B})$ by
\begin{eqnarray}
(\,u_{AB} \approx u'_{AB}\,) & :\Leftrightarrow &
(\,\forall {\mathfrak{l}}_A\in { \mathfrak{E}}_{A},\forall {\mathfrak{l}}_B\in { \mathfrak{E}}_{B},\;\;
\nu\,{}^{AB}_{{\mathfrak{l}}_A,{\mathfrak{l}}_B} (u_{AB})=\nu\,{}^{AB}_{{\mathfrak{l}}_A,{\mathfrak{l}}_B} (u'_{AB})\,).\;\;\;\;\;\;\;\;\;\;\;\;
\end{eqnarray}
\end{defin}

\begin{defin}
The space ${{\widetilde{S}}}_{AB}={ \mathfrak{S}}_{A} \widetilde{\otimes} { \mathfrak{S}}_{B}$ is built as the quotient of ${ \mathcal{P}}({ \mathfrak{S}}_{A} \times { \mathfrak{S}}_{B})$ under the congruence relation $\approx$.  
\begin{eqnarray}
\forall \sigma_{AB}\in { \mathcal{P}}({ \mathfrak{S}}_{A} \times { \mathfrak{S}}_{B}), && \widetilde{\sigma_{AB}}:=\{\, u_{AB}\;\vert\; \sigma_{AB}\approx u_{AB}\,\}.
\end{eqnarray}
The map $\nu\,{}^{AB}_{{\mathfrak{l}}_A,{\mathfrak{l}}_B}$ will be abusively defined as a map from ${{\widetilde{S}}}_{AB}$ to ${ \mathfrak{B}}$ by $\nu\,{}^{AB}_{{\mathfrak{l}}_A,{\mathfrak{l}}_B}(\widetilde{\sigma_{AB}}):= \nu\,{}^{AB}_{{\mathfrak{l}}_A,{\mathfrak{l}}_B}({\sigma_{AB}})$ for any $\sigma_{AB}$ in ${ \mathcal{P}}({ \mathfrak{S}}_{A} \times { \mathfrak{S}}_{B})$.
\end{defin}

\begin{defin}\label{defsqsubseteqSAB}
${{\widetilde{S}}}_{AB}$ is equipped with a partial order defined according to
\begin{eqnarray}
\forall \widetilde{\sigma}_{AB},\widetilde{\sigma}'_{AB} \in {{{\widetilde{S}}}_{AB}},\;\;\;\; (\,\widetilde{\sigma}_{AB} \sqsubseteq_{{}_{{{\widetilde{S}}}_{AB}}} \widetilde{\sigma}'_{AB}\,) & :\Leftrightarrow &
(\,\forall {\mathfrak{l}}_A\in { \mathfrak{E}}_{A},\forall {\mathfrak{l}}_B\in { \mathfrak{E}}_{B},\;\;
{\nu}\,{}^{AB}_{{\mathfrak{l}}_A, {\mathfrak{l}}_B} (\widetilde{\sigma}_{AB})\leq {\nu}\,{}^{AB}_{{\mathfrak{l}}_A,{\mathfrak{l}}_B} (\widetilde{\sigma}'_{AB})\,).\;\;\;\;\;\;\;\;\;\;\;\;
\end{eqnarray}
\end{defin}

 
This poset structure can be "explicited" according to following lemma addressing the word problem in ${{\widetilde{S}}}_{AB}$.

\begin{lemme}\label{Lemmadevelopetildeleqetilde}
Let us consider $u_{AB}:=\{\, (\sigma_{i,A},\sigma_{i,B})\;\vert\; i\in I\,\}$ an element of ${ \mathcal{P}}({ \mathfrak{S}}_{A} \times { \mathfrak{S}}_{B})$. We have explicitly, for any $\sigma_{A}\in { \mathfrak{S}}_{A}$ and $\sigma_{B}\in { \mathfrak{S}}_{B}$, the following equivalence
\begin{eqnarray}
\left( \widetilde{u_{AB}} \sqsubseteq_{{}_{{{\widetilde{S}}}_{AB}}} \widetilde{(\sigma_{A}, \sigma_{B})} \right)
&\Leftrightarrow &
\left( (\bigsqcap{}^{{}_{{ \mathfrak{S}}_{A}}}_{k\in I}\; \sigma_{k,A}) \;\sqsubseteq_{{}_{{ \mathfrak{S}}_{A}}}\sigma_{A}
\;\;\textit{\rm and}\;\;
(\bigsqcap{}^{{}_{{ \mathfrak{S}}_{B}}}_{m\in I} \;\sigma_{m,B})\; \sqsubseteq_{{}_{{ \mathfrak{S}}_{B}}} \sigma_{B}
\;\;\textit{\rm and}\;\;\right.\nonumber\\
&&\left.\left(\forall \varnothing \varsubsetneq K  \varsubsetneq I, \;\; (\bigsqcap{}^{{}_{{ \mathfrak{S}}_{A}}}_{k\in K}\; \sigma_{k,A}) \;\sqsubseteq_{{}_{{ \mathfrak{S}}_{A}}}\sigma_{A}\;\;\;\textit{\rm or}\;\;\; (\bigsqcap{}^{{}_{{ \mathfrak{S}}_{B}}}_{m\in I-K} \;\sigma_{m,B})\; \sqsubseteq_{{}_{{ \mathfrak{S}}_{B}}} \sigma_{B}\right) \right).\;\;\;\;\;\;\;\;\;\;\;\;\;\;
\label{developmentetildeordersimplify}
\end{eqnarray}
It is recalled that ${ \mathfrak{S}}_{A}$ and ${ \mathfrak{S}}_{B}$ are down-complete Inf semi-lattice and then the infima in this formula are well-defined.
\end{lemme}
\begin{proof}
We intent to expand the inequality $ \widetilde{u_{AB}} \sqsubseteq_{{}_{{{\widetilde{S}}}_{AB}}} \widetilde{(\sigma_{A}, \sigma_{B})}$. It is equivalent to
\begin{eqnarray}
\forall {\mathfrak{l}}_A\in { \mathfrak{E}}_{A}, \forall {\mathfrak{l}}_B\in { \mathfrak{E}}_{B},&&
\left( \bigwedge{}_{i\in I}\;
{\epsilon}\,{}^{{ \mathfrak{S}}_{A}}_{{\mathfrak{l}}_A} ( \sigma_{i,A})\bullet {\epsilon}\,{}^{{ \mathfrak{S}}_{B}}_{{\mathfrak{l}}_B} ( \sigma_{i,B})\right)
\leq
{\epsilon}\,{}^{{ \mathfrak{S}}_{A}}_{{\mathfrak{l}}_A} (\sigma_{A}) \bullet {\epsilon}\,{}^{{ \mathfrak{S}}_{B}}_{{\mathfrak{l}}_B}(\sigma_{B}).\label{assumptionorder2}
\end{eqnarray}
We intent to choose a pertinent set of effects ${\mathfrak{l}}_A\in { \mathfrak{E}}_{A}$
 and ${\mathfrak{l}}_B\in { \mathfrak{E}}_{B}$ to reformulate this inequality. \\
Let us firstly choose ${\mathfrak{l}}_B={ \mathfrak{Y}}_{{}_{{ \mathfrak{E}}_{B}}}$. Using  (\ref{expressionbullet}), we obtain
\begin{eqnarray}
{\epsilon}\,{}^{{ \mathfrak{S}}_{A}}_{{ \mathfrak{l}}_A}(\bigsqcap{}^{{}^{{ \mathfrak{S}}_{A}}}_{i\in I}\; \sigma_{i,A})\leq {\epsilon}\,{}^{{ \mathfrak{S}}_{A}}_{{ \mathfrak{l}}_A}(\sigma_{A})
,\forall { \mathfrak{l}}_A \in { \mathfrak{E}}_A,
\end{eqnarray} 
whcih leads immediately
\begin{eqnarray}
\bigsqcap{}^{{}^{{ \mathfrak{S}}_{A}}}_{i\in I}\; \sigma_{i,A} \;\sqsubseteq_{{}_{{ \mathfrak{S}}_{A}}} \sigma_{A}.\label{ineqBY}
\end{eqnarray}
Choosing ${\mathfrak{l}}_A={ \mathfrak{Y}}_{{}_{{ \mathfrak{E}}_{A}}}$, we obtain along the same line 
\begin{eqnarray}
\bigsqcap{}^{{}^{{ \mathfrak{S}}_{B}}}_{i\in I} \;\sigma_{i,B}\; \sqsubseteq_{{}_{{ \mathfrak{S}}_{B}}} \sigma_{B}.\label{ineqAY}
\end{eqnarray}
Let us now consider $\varnothing \varsubsetneq K  \varsubsetneq I$ and let us choose ${ \mathfrak{l}}_A$ and ${ \mathfrak{l}}_B$ according to
\begin{eqnarray}
{\epsilon}\,{}^{{ \mathfrak{S}}_{A}}_{{ \mathfrak{l}}_A}(\sigma):=\textit{\bf N},\forall \sigma \sqsupseteq_{{}_{{ \mathfrak{S}}_{A}}} \bigsqcap{}^{{}^{{ \mathfrak{S}}_{A}}}_{k\in K}\; \sigma_{k,A}&\textit{ \rm and}&{\epsilon}\,{}^{{ \mathfrak{S}}_{A}}_{{ \mathfrak{l}}_A}(\sigma):=\bot, \;\;\textit{ \rm elsewhere}, \\
{\epsilon}\,{}^{{ \mathfrak{S}}_{B}}_{{ \mathfrak{l}}_B}(\sigma):=\textit{\bf N},\forall \sigma \sqsupseteq_{{}_{{ \mathfrak{S}}_{B}}} \bigsqcap{}^{{}^{{ \mathfrak{S}}_{B}}}_{m\in I-K}\; \sigma_{m,B}&\textit{ \rm and}& {\epsilon}\,{}^{{ \mathfrak{S}}_{B}}_{{ \mathfrak{l}}_B}(\sigma):=\bot, \;\;\textit{ \rm elsewhere}.
\end{eqnarray}
We deduce, from the assumption (\ref{assumptionorder2}), that, for this $\varnothing \varsubsetneq K  \varsubsetneq I$, we have
\begin{eqnarray}
 &&(\bigsqcap{}^{{}^{{ \mathfrak{S}}_{A}}}_{k\in K}\; \sigma_{k,A} \;\sqsubseteq_{{}_{{ \mathfrak{S}}_{A}}} \sigma_{A}) \;\;\textit{\rm or}\;\; (\bigsqcap{}^{{}^{{ \mathfrak{S}}_{B}}}_{m\in I-K} \;\sigma_{m,B}\; \sqsubseteq_{{}_{{ \mathfrak{S}}_{B}}} \sigma_{B}).\label{ineqN}
\end{eqnarray}
We let the reader check that we have obtained the whole set of independent inequalities reformulating the property (\ref{assumptionorder2}).
\end{proof}

\begin{defin}\label{definlanglesigmqtilderangle}
We will adopt the following definition
\begin{eqnarray}
\forall \widetilde{\sigma}\in {{\widetilde{S}}}_{AB},\;\;\;\; \langle \widetilde{\sigma} \rangle & := & Max \{\,u\in { \mathcal{P}}({ \mathfrak{S}}_{A} \times { \mathfrak{S}}_{B})\;\vert\; \widetilde{u}  \sqsupseteq_{{}_{{{\widetilde{S}}}_{AB}}} \widetilde{\sigma} \,\}\nonumber\\
& = & \{\,(\sigma_A,\sigma_B) \;\vert\; \widetilde{(\sigma_A,\sigma_B)}  \sqsupseteq_{{}_{{{\widetilde{S}}}_{AB}}}\widetilde{\sigma} \,\},
\end{eqnarray}
\end{defin}

\begin{lemme}
We have the following Galois relation
\begin{eqnarray}
\forall \widetilde{\sigma}\in {{\widetilde{S}}}_{AB},\forall u\in { \mathcal{P}}({ \mathfrak{S}}_{A} \times { \mathfrak{S}}_{B}),&&  \langle \widetilde{\sigma} \rangle \supseteq u \;\;\;\Leftrightarrow\;\;\; \widetilde{\sigma} \sqsubseteq_{{}_{{{\widetilde{S}}}_{AB}}} \widetilde{u}.
\end{eqnarray}
\end{lemme}
\begin{proof}
Let us fix $u:=\{\,(\sigma_{i,A},\sigma_{i,B})\;\vert\; i\in I\,\}$.  We derive straightforwardly the following equivalences
\begin{eqnarray}
\langle \widetilde{\sigma} \rangle \supseteq u & \Leftrightarrow & 
\forall i\in I, \widetilde{(\sigma_{i,A},\sigma_{i,B})}  \sqsupseteq_{{}_{{{\widetilde{S}}}_{AB}}}\widetilde{\sigma}\nonumber\\
& \Leftrightarrow & \forall i\in I,  \forall {\mathfrak{l}}_A\in { \mathfrak{E}}_{A},\forall {\mathfrak{l}}_B\in { \mathfrak{E}}_{B}, {\nu}\,{}^{AB}_{{\mathfrak{l}}_A,{\mathfrak{l}}_B} (\widetilde{\sigma}) \leq
{\epsilon}\,{}^{{ \mathfrak{S}}_{A}}_{{\mathfrak{l}}_A} (\sigma_{i,A}) \bullet {\epsilon}\,{}^{{ \mathfrak{S}}_{B}}_{{\mathfrak{l}}_B}(\sigma_{i,B})\nonumber\\
 & \Leftrightarrow & 
\forall {\mathfrak{l}}_A\in { \mathfrak{E}}_{A},\forall {\mathfrak{l}}_B\in { \mathfrak{E}}_{B}, {\nu}\,{}^{AB}_{{\mathfrak{l}}_A, {\mathfrak{l}}_B} (\widetilde{\sigma}) \leq \bigwedge{}_{i\in I}\,
{\epsilon}\,{}^{{ \mathfrak{S}}_{A}}_{{\mathfrak{l}}_A} (\sigma_{i,A}) \bullet {\epsilon}\,{}^{{ \mathfrak{S}}_{B}}_{{\mathfrak{l}}_B}(\sigma_{i,B})\nonumber\\
& \Leftrightarrow & \widetilde{\sigma} \sqsubseteq_{{}_{{{\widetilde{S}}}_{AB}}} \widetilde{u}.
\end{eqnarray}  
\end{proof}

\begin{theoreme} \label{theoremaxiomA1bipartitepre}
${{\widetilde{S}}}_{AB}$ is a down-complete Inf semi-lattice with
\begin{eqnarray}
\forall \{\,u_i\;\vert\; i\in I\,\}\subseteq { \mathcal{P}}({ \mathfrak{S}}_{A} \times { \mathfrak{S}}_{B}),&&\bigsqcap{}^{{}^{{{\widetilde{S}}}_{AB}}}_{i\in I} \; \widetilde{u_i}  = \widetilde{\bigcup{}_{i\in I}\; u_i}.\label{tensorinfSAB}
\end{eqnarray}
Moreover, for any ${\mathfrak{l}}_A\in { \mathfrak{E}}_{A}$ and ${\mathfrak{l}}_B\in { \mathfrak{E}}_{B}$,  we have
\begin{eqnarray}
\forall \{\,\widetilde{u}_i\;\vert\; i\in I\,\}\subseteq {{\widetilde{S}}}_{AB},&& {\nu}\,{}^{AB}_{{\mathfrak{l}}_A,{\mathfrak{l}}_B}(\bigsqcap{}^{{}^{{{\widetilde{S}}}_{AB}}}_{i\in I} \widetilde{u}_i) =  \bigwedge{}_{\! i\in I}\, {\nu}\,{}^{AB}_{{\mathfrak{l}}_A,{\mathfrak{l}}_B}( \widetilde{u}_i)\label{tensorinfcontinuousSAB}
\end{eqnarray}
\end{theoreme}
\begin{proof}
The property (\ref{tensorinfSAB}) is a direct consequence of the Galois relation established in previous lemma.\\
For any ${\mathfrak{l}}_A\in { \mathfrak{E}}_{A}$ and ${\mathfrak{l}}_B\in { \mathfrak{E}}_{B}$, using (\ref{tensorinfSAB}) and the homomorphic property for $\nu\,{}^{AB}_{{\mathfrak{l}}_A,{\mathfrak{l}}_B}$, we have 
\begin{eqnarray}
\forall \{\,u_i\;\vert\; i\in I\,\}\subseteq { \mathcal{P}}({ \mathfrak{S}}_{A} \times { \mathfrak{S}}_{B}),\;\;\;\;{\nu}\,{}^{AB}_{{\mathfrak{l}}_A,{\mathfrak{l}}_B}(\bigsqcap{}^{{}^{{{\widetilde{S}}}_{AB}}}_{i\in I} \;\widetilde{u_{i}}) &= & {\nu}\,{}^{AB}_{{\mathfrak{l}}_A,{\mathfrak{l}}_B}( \widetilde{\bigcup{}_{i\in I} \;u_{i}})\nonumber\\
&=& \nu\,{}^{AB}_{{\mathfrak{l}}_A,{\mathfrak{l}}_B}( {\bigcup{}_{i\in I} \;u_{i}})\nonumber\\
&=&{\bigwedge{}_{\! i\in I}\, \nu\,{}^{AB}_{{\mathfrak{l}}_A,{\mathfrak{l}}_B}(  u_{i}})\nonumber\\
&=&{\bigwedge{}_{\! i\in I}\, {\nu}\,{}^{AB}_{{\mathfrak{l}}_A,{\mathfrak{l}}_B}(  \widetilde{u_{i}}})
\end{eqnarray}
\end{proof}

\begin{defin}
The element $\widetilde{u}\in {{\widetilde{S}}}_{AB}$ associated to the element $u:=\{\, (\sigma_{i,A},\sigma_{i,B})\;\vert\; i\in I\,\}\in { \mathcal{P}}({ \mathfrak{S}}_{A} \times { \mathfrak{S}}_{B})$ will be denoted $\bigsqcap{}^{{}^{{{\widetilde{S}}}_{AB}}}_{i\in I} \sigma_{i,A}\widetilde{\otimes}\sigma_{i,B}$.
\end{defin}

\begin{theoreme} \label{bifilterPtilde}
\begin{eqnarray}
\forall \{\,\sigma_{i,A}\;\vert\; i\in I\,\}\subseteq { \mathfrak{S}}_{A}, \forall \sigma_B\in { \mathfrak{S}}_{B},\;\;\;(\bigsqcap{}^{{}^{{ \mathfrak{S}}_{A}}}_{i\in I}\,\sigma_{i,A})\widetilde{\otimes} \sigma_B & = & \bigsqcap{}^{{}^{{{\widetilde{S}}}_{AB}}}_{i\in I} (\sigma_{i,A}\widetilde{\otimes} \sigma_B),\label{pitensor=tensorpi1true}\\
\forall \{\,\sigma_{i,B}\;\vert\; i\in I\,\}\subseteq { \mathfrak{S}}_{B}, \forall \sigma_A\in { \mathfrak{S}}_{A},\;\;\; \sigma_A \widetilde{\otimes} (\bigsqcap{}^{{}^{{ \mathfrak{S}}_{B}}}_{i\in I}\,\sigma_{i,B})& = & \bigsqcap{}^{{}^{{{\widetilde{S}}}_{AB}}}_{i\in I} (\sigma_A \widetilde{\otimes} \sigma_{i,B}).\label{pitensor=tensorpi2true}
\end{eqnarray}
\end{theoreme}
\begin{proof}
Indeed, using successively properties (\ref{etildemorphism}) (\ref{propconjunctiveprop}) (\ref{distributivitybullet}) and (\ref{etildemorphism}) again,  we deduce that, for any ${\mathfrak{l}}_A\in { \mathfrak{E}}_{A}, {\mathfrak{l}}_B\in { \mathfrak{E}}_{B}$, 
\begin{eqnarray}
\nu\,{}^{AB}_{{\mathfrak{l}}_A,{\mathfrak{l}}_B}(\, (\bigsqcap{}^{{}^{{ \mathfrak{S}}_{A}}}_{i\in I}\sigma_{i,A},\sigma_B)\,)&=&
{\epsilon}\,{}^{{ \mathfrak{S}}_{A}}_{{\mathfrak{l}}_A} (\bigsqcap{}^{{}^{{ \mathfrak{S}}_{A}}}_{i\in I}\sigma_{i,A}) \bullet {\epsilon}\,{}^{{ \mathfrak{S}}_{B}}_{{\mathfrak{l}}_B} (\sigma_B) \nonumber\\
&=&
(\bigwedge{}_{i\in I}\;{\epsilon}\,{}^{{ \mathfrak{S}}_{A}}_{{\mathfrak{l}}_A} (\sigma_{i,A})) \bullet {\epsilon}\,{}^{{ \mathfrak{S}}_{B}}_{{\mathfrak{l}}_B} (\sigma_B) \nonumber\\
&=&
\bigwedge{}_{i\in I}\;
(\,{\epsilon}\,{}^{{ \mathfrak{S}}_{A}}_{{\mathfrak{l}}_A} (\sigma_{i,A})\bullet {\epsilon}\,{}^{{ \mathfrak{S}}_{B}}_{{\mathfrak{l}}_B} (\sigma_B)) \nonumber\\
&=& \nu\,{}^{AB}_{{\mathfrak{l}}_A,{\mathfrak{l}}_B} (\{\,(\sigma_{i,A}, \sigma_B)\;\vert\; i\in I\,\}),\label{demobifilterpropertybasic}
\end{eqnarray}
and then, by definition, we obtain the property 
\begin{eqnarray}
(\bigsqcap{}^{{}^{{ \mathfrak{S}}_{A}}}_{i\in I}\sigma_{i,A},\sigma_B) & \approx & \{\,(\sigma_{i,A}, \sigma_B)\;\vert\; i\in I\,\}
\end{eqnarray}
and then
\begin{eqnarray}
(\bigsqcap{}^{{}^{{ \mathfrak{S}}_{A}}}_{i\in I}\,\sigma_{i,A})\widetilde{\otimes} \sigma_B & = & \bigsqcap{}^{{}^{{\widetilde{S}}_{AB}}}_{i\in I} (\sigma_{i,A}\widetilde{\otimes} \sigma_B).
\end{eqnarray}
We obtain the second property along the same lines of proof.
\end{proof} 

In this subsection, we will assume that $({ \mathfrak{S}}_A,{ \mathfrak{E}}_A,\epsilon^A)$ and $({ \mathfrak{S}}_B,{ \mathfrak{E}}_B,\epsilon^B)$ are valid States/Effects Chu spaces. In other words, they satisfy Axioms {\bf (A1)$-$(A5)}.

\begin{defin}
The evaluation map will be defined as a map 
\begin{eqnarray}
\begin{array}{rcrcl}
\epsilon & : & { \mathcal{P}}({ \mathfrak{E}}_{A} \times { \mathfrak{E}}_{B}) & \longrightarrow & { \mathfrak{B}}^{{{\widetilde{S}}}_{AB}}\\
& & \{({ \mathfrak{l}}_{i,A},{ \mathfrak{l}}_{i,B})\;\vert\; i\in I\} & \mapsto & \epsilon^{{{\widetilde{S}}}_{AB}}_{{}_{\{({ \mathfrak{l}}_{i,A},{ \mathfrak{l}}_{i,B})\;\vert\; i\in I\}}} \;\vert\; \forall \widetilde{\sigma}_{AB}\in {{\widetilde{S}}}_{AB},\; \epsilon^{{{\widetilde{S}}}_{AB}}_{{}_{\{({ \mathfrak{l}}_{i,A},{ \mathfrak{l}}_{i,B})\;\vert\; i\in I\}}}( \widetilde{\sigma}_{AB})=\bigwedge_{i\in I} \nu^{AB}_{{ \mathfrak{l}}_{i,A},{ \mathfrak{l}}_{i,B}}(\widetilde{\sigma}_{AB}).
\end{array}
\end{eqnarray}
\end{defin}

\begin{defin}
${ \mathcal{P}}({ \mathfrak{E}}_{A} \times { \mathfrak{E}}_{B})$ is equipped with a congruence relation defined between any two elements $x_{AB}$ and $x'_{AB}$ of ${ \mathcal{P}}({ \mathfrak{E}}_{A} \times { \mathfrak{E}}_{B})$ by
\begin{eqnarray}
(\,x_{AB} \simeq x'_{AB}\,) & :\Leftrightarrow &
(\,\forall \widetilde{\sigma}_{AB}\in \widetilde{S}_{AB},\;\;
\epsilon\,{}^{{{\widetilde{S}}}_{AB}}_{x_{AB}}(\widetilde{\sigma}_{AB})=\epsilon\,{}^{{{\widetilde{S}}}_{AB}}_{x'_{AB}}(\widetilde{\sigma}_{AB})\,).\;\;\;\;\;\;\;\;\;\;\;\;
\end{eqnarray}
\end{defin}

\begin{defin}
The space ${{\widetilde{E}}}_{AB}$ is built as the quotient of ${ \mathcal{P}}({ \mathfrak{E}}_{A} \times { \mathfrak{E}}_{B})$ under the congruence relation $\simeq$.  
\begin{eqnarray}
\forall \lambda_{AB}\in { \mathcal{P}}({ \mathfrak{E}}_{A} \times { \mathfrak{E}}_{B}), && \widetilde{\lambda_{AB}}:=\{\, x_{AB}\in { \mathcal{P}}({ \mathfrak{E}}_{A} \times { \mathfrak{E}}_{B})\;\vert\; \lambda_{AB}\simeq x_{AB}\,\}.
\end{eqnarray}
The evaluation map will be defined as a map from ${{\widetilde{E}}}_{AB}$ to ${ \mathfrak{B}}^{{{\widetilde{S}}}_{AB}}$ by $\epsilon\,{}^{{{\widetilde{S}}}_{AB}}_{\widetilde{\lambda_{AB}}}:=\epsilon\,{}^{{{\widetilde{S}}}_{AB}}_{{\lambda_{AB}}}$ for any $\lambda_{AB}\in { \mathcal{P}}({ \mathfrak{E}}_{A} \times { \mathfrak{E}}_{B})$.
\end{defin}

\begin{defin}\label{defsqsubseteqSAB}
${{\widetilde{E}}}_{AB}$ is equipped with a partial order defined according to
\begin{eqnarray}
\forall \widetilde{\lambda}_{AB},\widetilde{\lambda}'_{AB} \in {{{\widetilde{E}}}_{AB}},\;\;\;\; (\,\widetilde{\lambda}_{AB} \sqsubseteq_{{}_{{{\widetilde{E}}}_{AB}}} \widetilde{\lambda}'_{AB}\,) & :\Leftrightarrow &
(\,\forall \widetilde{\sigma}_{AB}\in {\widetilde{S}}_{AB},\;\;
{\epsilon}\,{}^{{{\widetilde{S}}}_{AB}}_{\widetilde{\lambda}_{AB}} (\widetilde{\sigma}_{AB})\leq {\epsilon}\,{}^{{{\widetilde{S}}}_{AB}}_{\widetilde{\lambda}\!{}'_{AB}} (\widetilde{\sigma}_{AB})\,).\;\;\;\;\;\;\;\;\;\;\;\;
\end{eqnarray}
\end{defin}

\begin{defin}\label{definlanglelambdatilderangle}
We will adopt the following definition
\begin{eqnarray}
\forall \widetilde{\lambda}\in {{\widetilde{E}}}_{AB},\;\;\;\; \langle \widetilde{\lambda} \rangle & := & Max \{\,x\in { \mathcal{P}}({ \mathfrak{E}}_{A} \times { \mathfrak{E}}_{B})\;\vert\; \widetilde{x}  \sqsupseteq_{{}_{{{\widetilde{E}}}_{AB}}} \widetilde{\lambda} \,\}\nonumber\\
& = & \{\,({ \mathfrak{l}}_A,{ \mathfrak{l}}_B) \;\vert\; \widetilde{({ \mathfrak{l}}_A,{ \mathfrak{l}}_B)}  \sqsupseteq_{{}_{{{\widetilde{E}}}_{AB}}}\widetilde{\lambda} \,\},
\end{eqnarray}
\end{defin}

\begin{lemme}
We have the following Galois relation
\begin{eqnarray}
\forall \widetilde{\lambda}\in {{\widetilde{E}}}_{AB},\forall x\in { \mathcal{P}}({ \mathfrak{E}}_{A} \times { \mathfrak{E}}_{B}),&&  \langle \widetilde{\lambda} \rangle \supseteq x \;\;\;\Leftrightarrow\;\;\; \widetilde{\lambda} \sqsubseteq_{{}_{{{\widetilde{E}}}_{AB}}} \widetilde{x}.
\end{eqnarray}
\end{lemme}
\begin{proof}
Let us fix $x:=\{\,({ \mathfrak{l}}_{i,A},{ \mathfrak{l}}_{i,B})\;\vert\; i\in I\,\}$.  We derive straightforwardly the following equivalences
\begin{eqnarray}
\langle \widetilde{\lambda} \rangle \supseteq x & \Leftrightarrow & 
\forall i\in I, \widetilde{({ \mathfrak{l}}_{i,A},{ \mathfrak{l}}_{i,B})}  \sqsupseteq_{{}_{{{\widetilde{E}}}_{AB}}}\widetilde{\lambda}\nonumber\\
& \Leftrightarrow & \forall i\in I,  \forall \widetilde{\sigma}_{AB}\in \widetilde{S}_{AB}, \;\;\;\;{\epsilon}\,{}^{{{\widetilde{S}}}_{AB}}_{\widetilde{\lambda}} (\widetilde{\sigma}_{AB}) \leq
{\nu}\,{}^{AB}_{{\mathfrak{l}}_{i,A},{\mathfrak{l}}_{i,B}} (\widetilde{\sigma}_{AB})\nonumber\\
 & \Leftrightarrow & 
 \forall \widetilde{\sigma}_{AB}\in \widetilde{S}_{AB}, \;\;\;\;{\epsilon}\,{}^{{{\widetilde{S}}}_{AB}}_{\widetilde{\lambda}} (\widetilde{\sigma}_{AB}) \leq \bigwedge{}_{i\in I}\;
{\nu}\,{}^{AB}_{{\mathfrak{l}}_{i,A},{\mathfrak{l}}_{i,B}} (\widetilde{\sigma}_{AB})=
{\epsilon}\,{}^{{{\widetilde{S}}}_{AB}}_{x} (\widetilde{\sigma}_{AB})\nonumber\\
& \Leftrightarrow & \widetilde{\lambda} \sqsubseteq_{{}_{{{\widetilde{E}}}_{AB}}} \widetilde{x}.
\end{eqnarray}  
\end{proof}

\begin{theoreme} \label{theoremaxiomA2bipartitepre}
${{\widetilde{E}}}_{AB}$ is a down-complete Inf semi-lattice with
\begin{eqnarray}
\forall \{\,x_i\;\vert\; i\in I\,\}\subseteq { \mathcal{P}}({ \mathfrak{E}}_{A} \times { \mathfrak{E}}_{B}),&&\bigsqcap{}^{{}^{{{\widetilde{E}}}_{AB}}}_{i\in I} \; \widetilde{x_i}  = \widetilde{\bigcup{}_{i\in I}\; x_i}.\label{tensorinfEAB}
\end{eqnarray}
Moreover,  we have
\begin{eqnarray}
\forall \{\,\widetilde{\lambda}_i\;\vert\; i\in I\,\}\subseteq {{\widetilde{E}}}_{AB},\forall \widetilde{\sigma}_{AB}\in{{\widetilde{S}}}_{AB}, && {\epsilon}\,{}^{{{\widetilde{S}}}_{AB}}_{\bigsqcap{}^{{}^{{{\widetilde{E}}}_{AB}}}_{i\in I} \widetilde{\lambda}_i}(\widetilde{\sigma}_{AB}) =  \bigwedge{}_{\! i\in I}\, {\epsilon}\,{}^{{{\widetilde{S}}}_{AB}}_{\widetilde{\lambda}_i}( \widetilde{\sigma}_{AB} )\label{tensorinfcontinuousEAB}
\end{eqnarray}
In other words, ${{\widetilde{E}}}_{AB}$ satisfies Axiom {\bf (A3)}. 
\end{theoreme}
\begin{proof}
The property (\ref{tensorinfEAB}) is a direct consequence of the Galois relation established in previous lemma.\\
For any $\widetilde{\sigma}_{AB}\in \widetilde{S}_{AB}$ we have 
\begin{eqnarray}
\forall \{\,x_i\;\vert\; i\in I\,\}\subseteq { \mathcal{P}}({ \mathfrak{E}}_{A} \times { \mathfrak{E}}_{B}),\;\;\;\;{\epsilon}\,{}^{{{\widetilde{S}}}_{AB}}_{\bigsqcap{}^{{}^{{{\widetilde{S}}}_{AB}}}_{i\in I} \;\widetilde{x_{i}}}(\widetilde{\sigma}_{AB}) &= & {\epsilon}\,{}^{{{\widetilde{S}}}_{AB}}_{ \widetilde{\bigcup{}_{i\in I} \;x_{i}}}(\widetilde{\sigma}_{AB})\nonumber\\
&=& \epsilon\,{}^{{{\widetilde{S}}}_{AB}}_{{\bigcup{}_{i\in I} \;x_{i}}}(\widetilde{\sigma}_{AB})\nonumber\\
&=&{\bigwedge{}_{\! i\in I}\, \epsilon\,{}^{{{\widetilde{S}}}_{AB}}_{x_{i}}}(\widetilde{\sigma}_{AB})\nonumber\\
&=&{\bigwedge{}_{\! i\in I}\, {\epsilon}\,{}^{{{\widetilde{S}}}_{AB}}_{\widetilde{x_{i}}}}(\widetilde{\sigma}_{AB})
\end{eqnarray}
\end{proof}

\begin{defin}
The element $\widetilde{\,{ \mathfrak{l}}_{AB}}\in {{\widetilde{E}}}_{AB}$ associated to the element ${\mathfrak{l}}_{AB}:=\{\, ({ \mathfrak{l}}_{i,A},{ \mathfrak{l}}_{i,B})\;\vert\; i\in I\,\}\in { \mathcal{P}}({ \mathfrak{E}}_{A} \times { \mathfrak{E}}_{B})$ will be denoted $\bigsqcap{}^{{}^{{{\widetilde{E}}}_{AB}}}_{i\in I} { \mathfrak{l}}_{i,A}\widetilde{\otimes}{ \mathfrak{l}}_{i,B}$.
\end{defin}

\begin{theoreme}
${{\widetilde{S}}}_{AB}$ satisfies Axiom {\bf (A1)}. Explicitly, ${{\widetilde{S}}}_{AB}$ is a down-complete Inf semi-lattice. Moreover, we have
\begin{eqnarray}
\forall \{\,\widetilde{\sigma}_{i,AB}\;\vert\; i\in I\,\}\subseteq {{\widetilde{S}}}_{AB},\forall \widetilde{\lambda}_{AB}\in {{\widetilde{E}}}_{AB},&& {\epsilon}\,{}^{{{\widetilde{S}}}_{AB}}_{\widetilde{\lambda}_{AB}}(\bigsqcap{}^{{}^{{{\widetilde{S}}}_{AB}}}_{i\in I} \widetilde{\sigma}_{i,AB}) =  \bigwedge{}_{\! i\in I}\, {\epsilon}\,{}^{{{\widetilde{S}}}_{AB}}_{\widetilde{\lambda}_{AB}}( \widetilde{\sigma}_{i,AB})\label{tensorinfcontinuousSABpost}
\end{eqnarray}
\end{theoreme}
\begin{proof}
Direct consequence of Theorem \ref{theoremaxiomA1bipartitepre} with property (\ref{tensorinfcontinuousEAB}).   
\end{proof}

\begin{theoreme}
If ${ \mathfrak{S}}_{A}$ and ${ \mathfrak{S}}_{B}$ satisfy the axiom {\bf (A2)}, then
${{\widetilde{S}}}_{AB}$ satisfies the axiom {\bf (A2)} as well : the bottom element of ${{\widetilde{S}}}_{AB}$ is explicitly given by $\bot_{{}_{{\mathfrak{S}}_A}}\widetilde{\otimes}\bot_{{}_{{\mathfrak{S}}_B}}$.
\end{theoreme}
\begin{proof}
Trivial using the expansion (\ref{developmentetildeordersimplify}).
\end{proof}

\begin{theoreme}\label{theorempuretilde}
\begin{eqnarray}
{{\widetilde{S}}}^{\;{}^{pure}}_{AB} & = & \{\, \sigma_A \widetilde{\otimes} \sigma_B\;\vert\; \sigma_A \in { \mathfrak{S}}^{\;{}^{pure}}_{A}, \sigma_B \in { \mathfrak{S}}^{\;{}^{pure}}_{B}\,\}
\end{eqnarray}
Moreover, ${{\widetilde{S}}}_{AB}={ \mathfrak{S}}_{A}\widetilde{\otimes} { \mathfrak{S}}_{B}$ satisfies the axiom {\bf (A5)}, i.e.  ${{\widetilde{S}}}^{\;{}^{pure}}_{AB}=Max({{\widetilde{S}}}_{AB})$.
\end{theoreme}
\begin{proof}
First of all, it is a trivial fact that the completely meet-irreducible elements of  ${{\widetilde{S}}}_{AB}$ are necessarily pure tensors of  ${{\widetilde{S}}}_{AB}$, i.e.  elements of the form $\sigma_A \widetilde{\otimes} \sigma_B$. \\
Let us then consider $\sigma_A \widetilde{\otimes} \sigma_B$ a completely meet-irreducible element of ${{\widetilde{S}}}_{AB}$ and let us assume that $\sigma_A  = \bigsqcap{}^{{}^{{ \mathfrak{S}}_{A}}}_{i\in I} \sigma_{i,A}$ for $ \sigma_{i,A}\in { \mathfrak{S}}_{A}$ for any $i\in I$. We have then $(\sigma_A \widetilde{\otimes} \sigma_B) = ((\bigsqcap{}^{{}^{{ \mathfrak{S}}_{A}}}_{i\in I} \sigma_{i,A})\widetilde{\otimes}  \sigma_B)=\bigsqcap{}^{{}^{{{\widetilde{S}}}_{AB}}}_{i\in I} ( \sigma_{i,A}\widetilde{\otimes}  \sigma_B)$.  On another part, $\sigma_A \widetilde{\otimes} \sigma_B$ being completely meet-irreducible in ${{\widetilde{S}}}_{AB}$, there exists $k\in I$ such that $\sigma_A \widetilde{\otimes} \sigma_B=\sigma_{k,A} \widetilde{\otimes} \sigma_B$, i.e, $\sigma_A=\sigma_{k,A}$. As a conclusion, $\sigma_A$ is completely meet-irreducible. In the same way, $\sigma_B$ is completely meet-irreducible.  As a first result, pure states of ${{\widetilde{S}}}_{AB}$ are necessarily of the form $\sigma_A \widetilde{\otimes} \sigma_B$ with $\sigma_A \in { \mathfrak{S}}^{\;{}^{pure}}_{A}, \sigma_B \in { \mathfrak{S}}^{\;{}^{pure}}_{B}$.\\
Conversely, let us consider $\sigma_A$ a pure state of ${ \mathfrak{S}}_{A}$ and $\sigma_B$ a pure state of ${ \mathfrak{S}}_{B}$, and let us suppose that $ (\bigsqcap{}^{{}^{{{\widetilde{S}}}_{AB}}}_{i\in I} \sigma_{i,A}\widetilde{\otimes}  \sigma_{i,B})  =  (\sigma_A \widetilde{\otimes} \sigma_B)$ with $\sigma_{i,A}\in { \mathfrak{S}}_{A}$ and $\sigma_{i,B}\in { \mathfrak{S}}_{B}$ for any $i\in I$. We now exploit the two conditions $(\bigsqcap{}^{{}_{{ \mathfrak{S}}_{A}}}_{k\in I}\; \sigma_{k,A}) = \sigma_{A}$ and $(\bigsqcap{}^{{}_{{ \mathfrak{S}}_{B}}}_{m\in I} \;\sigma_{m,B}) = \sigma_{B}$ derived from the expansion (\ref{developmentetildeordersimplify}).  From $\sigma_A\in Max({ \mathfrak{S}}_{A})$ and $\sigma_B\in Max({ \mathfrak{S}}_{B})$, we deduce that $\sigma_{i,A} = \sigma_A$ and $\sigma_{j,B} = \sigma_B$ for any $i,j\in I$. As a second result, we have then obtained that the state $(\sigma_A \widetilde{\otimes} \sigma_B)$, with $\sigma_A$ a pure state of ${ \mathfrak{S}}_{A}$ and $\sigma_B$ a pure state of ${ \mathfrak{S}}_{B}$, is completely meet-irreducible.\\
From the expansion (\ref{developmentetildeordersimplify}), we deduce also immediately that $(\sigma_A \widetilde{\otimes} \sigma_B)\in Max({{\widetilde{S}}}_{AB})$ as long as $\sigma_A\in Max({ \mathfrak{S}}_{A})$ and $\sigma_B\in Max({ \mathfrak{S}}_{B})$.
\end{proof}

\begin{theoreme} \label{theoremA4Ptilde}
${{\widetilde{S}}}_{AB}={ \mathfrak{S}}_{A}\widetilde{\otimes} { \mathfrak{S}}_{B}$ satisfies the axiom {\bf (A4)}. Explicitly, 
\begin{eqnarray}
&&\forall \sigma \in {{\widetilde{S}}}_{AB}, \;\; \sigma= \bigsqcap{}^{{}^{{{\widetilde{S}}}_{AB}}}  \underline{\sigma}_{{}_{ {{\widetilde{S}}}_{AB}}},\;\;\textit{\rm where}\;\;
\underline{\sigma}_{{}_{ {{\widetilde{S}}}_{AB}}}=
({{\widetilde{S}}}{}_{AB}^{\,{}^{pure}} \cap (\uparrow^{{}^{{{\widetilde{S}}}_{AB}}}\!\!\!\! \sigma) ).
\end{eqnarray}
\end{theoreme}
\begin{proof}
Let us fix $\sigma \in {{\widetilde{S}}}_{AB}$.\\ 
We note that $\sigma \sqsubseteq_{{}_{{{\widetilde{S}}}_{AB}}} \sigma'$ for any $\sigma'\in ({{\widetilde{S}}}{}_{AB}^{\,{}^{pure}} \cap (\uparrow^{{}^{{{\widetilde{S}}}_{AB}}}\!\!\!\! \sigma)) $ and then $\sigma \sqsubseteq_{{}_{{{\widetilde{S}}}_{AB}}} \bigsqcap{}^{{}^{{{\widetilde{S}}}_{AB}}}  \underline{\sigma}_{{}_{ {{\widetilde{S}}}_{AB}}}$. \\
Secondly,  denoting  
$\sigma:=(\bigsqcap{}^{{}^{{{\widetilde{S}}}_{AB}}}_{i\in I} \sigma_{i,A}\widetilde{\otimes}  \sigma_{i,B})$, we note immediately that, for any 
$\sigma_{A} \in { \mathfrak{S}}_{A}^{pure}$ and $\sigma_{B} \in { \mathfrak{S}}_{B}^{pure}$, if 
$\sigma_{A}\sqsupseteq_{{}_{{ \mathfrak{S}}_{A}}} \sigma_{i,A}$ and $ \sigma_{B}\sqsupseteq_{{}_{{ \mathfrak{S}}_{B}}} \sigma_{i,B}$, then $(\sigma_{A}\widetilde{\otimes}  \sigma_{B}) \sqsupseteq_{{}_{{ \mathfrak{S}}_{AB}}} \sigma$, i.e. $(\sigma_{A}\widetilde{\otimes}  \sigma_{B}) \in \underline{\sigma}_{{}_{ {{\widetilde{S}}}_{AB}}}$. As a consequence, we have
\begin{eqnarray}
(\bigsqcap{}^{{}^{{{\widetilde{S}}}_{AB}}}_{i\in I}\bigsqcap{}^{{}^{{{\widetilde{S}}}_{AB}}}_{\sigma_{A} \in { \mathfrak{S}}_{A}^{pure}\;\vert\; \sigma_{A}\sqsupseteq_{{}_{{ \mathfrak{S}}_{A}}} \sigma_{i,A}}\bigsqcap{}^{{}^{{{\widetilde{S}}}_{AB}}}_{\sigma_{B} \in { \mathfrak{S}}_{B}^{pure}\;\vert\; \sigma_{B}\sqsupseteq_{{}_{{ \mathfrak{S}}_{B}}} \sigma_{i,B}} \sigma_{A}\widetilde{\otimes}  \sigma_{B}) \sqsupseteq_{{}_{{{\widetilde{S}}}_{AB}}} \bigsqcap{}^{{}^{{{\widetilde{S}}}_{AB}}}  \underline{\sigma}_{{}_{ {{\widetilde{S}}}_{AB}}}.
\end{eqnarray}
Endly, using Theorem \ref{bifilterPtilde},we have
\begin{eqnarray}
\sigma=\bigsqcap{}^{{}^{{{\widetilde{S}}}_{AB}}}_{i\in I} \sigma_{i,A}\widetilde{\otimes}  \sigma_{i,B} &= & 
\bigsqcap{}^{{}^{{{\widetilde{S}}}_{AB}}}_{i\in I} (\bigsqcap{}^{{}^{{{\widetilde{S}}}_{AB}}}_{\sigma_{A} \in { \mathfrak{S}}_{A}^{pure}\;\vert\; \sigma_{A}\sqsupseteq_{{}_{{ \mathfrak{S}}_{A}}} \sigma_{i,A}}\sigma_{A})\widetilde{\otimes}  (\bigsqcap{}^{{}^{{{\widetilde{S}}}_{AB}}}_{\sigma_{B} \in { \mathfrak{S}}_{B}^{pure}\;\vert\; \sigma_{B}\sqsupseteq_{{}_{{ \mathfrak{S}}_{B}}} \sigma_{i,B}}\sigma_{B})\nonumber\\
&= & 
\bigsqcap{}^{{}^{{{\widetilde{S}}}_{AB}}}_{i\in I}\bigsqcap{}^{{}^{{{\widetilde{S}}}_{AB}}}_{\sigma_{A} \in { \mathfrak{S}}_{A}^{pure}\;\vert\; \sigma_{A}\sqsupseteq_{{}_{{ \mathfrak{S}}_{A}}} \sigma_{i,A}}\bigsqcap{}^{{}^{{{\widetilde{S}}}_{AB}}}_{\sigma_{B} \in { \mathfrak{S}}_{B}^{pure}\;\vert\; \sigma_{B}\sqsupseteq_{{}_{{ \mathfrak{S}}_{B}}} \sigma_{i,B}} \sigma_{A}\widetilde{\otimes}  \sigma_{B}.
\end{eqnarray}
As a final conclusion, we obtain
\begin{eqnarray}
\sigma = (\bigsqcap{}^{{}^{{{\widetilde{S}}}_{AB}}}_{i\in I}\bigsqcap{}^{{}^{{{\widetilde{S}}}_{AB}}}_{\sigma_{A} \in { \mathfrak{S}}_{A}^{pure}\;\vert\; \sigma_{A}\sqsupseteq_{{}_{{ \mathfrak{S}}_{A}}} \sigma_{i,A}}\bigsqcap{}^{{}^{{{\widetilde{S}}}_{AB}}}_{\sigma_{B} \in { \mathfrak{S}}_{B}^{pure}\;\vert\; \sigma_{B}\sqsupseteq_{{}_{{ \mathfrak{S}}_{B}}} \sigma_{i,B}} \sigma_{A}\widetilde{\otimes}  \sigma_{B}) = \bigsqcap{}^{{}^{{{\widetilde{S}}}_{AB}}}  \underline{\sigma}_{{}_{ {{\widetilde{S}}}_{AB}}}.
\end{eqnarray}
\end{proof}

As a conclusion of previous theorems, we have also obtained that ${{\widetilde{S}}}_{AB}$ is a valid space of states and ${{\widetilde{E}}}_{AB}$ is a valid space of effects satisfying axioms {\bf (A1) $-$ (A5)} and {\bf extensionality$+$separation}.  As a consequence, they satisfy axioms {\bf (B1)} and {\bf (B2)}. \\ Axioms {\bf (B3)} and {\bf (B4)} are also trivial by construction.\\
By construction of the basic tensor product, it will also satisfy the axiom {\bf (B5)}, i.e.
\begin{eqnarray}
\forall \widetilde{\sigma}_{AB},  \widetilde{\sigma}'_{AB}\in {{\widetilde{S}}}_{AB},\;\;\;\;\;
(\,\forall {\mathfrak{l}}_A\in { \mathfrak{E}}_{A},\forall {\mathfrak{l}}_B\in { \mathfrak{E}}_{B},\;\;
{\epsilon}\,{}^{{{\widetilde{S}}}_{AB}}_{{\mathfrak{l}}_A\widetilde{\otimes}{\mathfrak{l}}_B} (\widetilde{\sigma}_{AB})={\epsilon}\,{}^{{{\widetilde{S}}}_{AB}}_{{\mathfrak{l}}_A\widetilde{\otimes} {\mathfrak{l}}_B} (\widetilde{\sigma}'_{AB})\,) & \Leftrightarrow &
(\, \widetilde{\sigma}_{AB} =  \widetilde{\sigma}'_{AB}\,) .\;\;\;\;\;\;\;\;\;\;\;\;\label{epsilon=epsilonimplies=}
\end{eqnarray}

Endly, Definition \ref{definnubullet} has been chosen in such a way that we obtain trivially the axiom {\bf (C)}, i.e. 
\begin{eqnarray}
&& \forall \sigma_A\in { \mathfrak{S}}_{A},\forall \sigma_B\in { \mathfrak{S}}_{B}, \forall {\mathfrak{l}}_A\in { \mathfrak{E}}_{A},\forall {\mathfrak{l}}_B\in { \mathfrak{E}}_{B}, \;\;\;\;{\epsilon}\,{}^{{{\widetilde{S}}}_{AB}}_{{\mathfrak{l}}_A\widetilde{\otimes} {\mathfrak{l}}_B} (\sigma_A\widetilde{\otimes} \sigma_B)  =  {\epsilon}\,{}^{{\mathfrak{S}}_A}_{{\mathfrak{l}}_A} (\sigma_A) \bullet {\epsilon}\,{}^{{ \mathfrak{S}}_B}_{{\mathfrak{l}}_B} (\sigma_B). 
\end{eqnarray}

\subsection{Multipartite experiments defined by the basic tensor product}

Let ${ \mathfrak{S}}_A, { \mathfrak{S}}_B, { \mathfrak{S}}_C$ be three spaces of states. We intent to define the tripartite state space ${ \mathfrak{S}}_{ABC}$ ? Clearly one option is to first form the bipartite state space ${ \mathfrak{S}}_A \widetilde{\otimes} { \mathfrak{S}}_B$ and then tensor the result with ${ \mathfrak{S}}_C$, so that we get $({ \mathfrak{S}}_A \widetilde{\otimes}  { \mathfrak{S}}_B) \widetilde{\otimes} { \mathfrak{S}}_C$. Another way to build these tripartite experiments is to first form ${ \mathfrak{S}}_B \widetilde{\otimes} { \mathfrak{S}}_C$  and then tensor with ${ \mathfrak{S}}_A$ to obtain ${ \mathfrak{S}}_A \widetilde{\otimes}  ({ \mathfrak{S}}_B \widetilde{\otimes} { \mathfrak{S}}_C)$. It is natural to require that both of these constructions yield the same result.

\begin{theoreme}
The basic tensor product of state spaces is associative, i.e., we must have
\begin{eqnarray}
({ \mathfrak{S}}_A \widetilde{\otimes}  { \mathfrak{S}}_B) \widetilde{\otimes} { \mathfrak{S}}_C &=& { \mathfrak{S}}_A \widetilde{\otimes}  ({ \mathfrak{S}}_B \widetilde{\otimes} { \mathfrak{S}}_C).
\end{eqnarray}
\end{theoreme}
\begin{proof}
$({ \mathfrak{S}}_A \widetilde{\otimes}  { \mathfrak{S}}_B) \widetilde{\otimes} { \mathfrak{S}}_C$ is defined as the quotient of ${\mathcal{P}}({ \mathfrak{S}}_A \times  { \mathfrak{S}}_B  \times  { \mathfrak{S}}_C)$ by the congruence relation defined for any $u_{ABC},u'_{ABC}\in {\mathcal{P}}({ \mathfrak{S}}_A \times  { \mathfrak{S}}_B  \times  { \mathfrak{S}}_C)$
\begin{eqnarray}
(u_{ABC} \approx_{{}_{(AB)C}} u'_{ABC}) & :\Leftrightarrow & (\,\forall {\mathfrak{l}}_{AB}\in \widetilde{E}_{AB},{\mathfrak{l}}_{C}\in {\mathfrak{E}}_{C},\;\;\;\; \nu_{{\mathfrak{l}}_{AB},{\mathfrak{l}}_{C}}(u_{ABC})=\nu_{{\mathfrak{l}}_{AB},{\mathfrak{l}}_{C}}(u'_{ABC})\,)\\
& \Leftrightarrow & (\,\forall {\mathfrak{l}}_{A}\in {\mathfrak{E}}_{A},{\mathfrak{l}}_{B}\in {\mathfrak{E}}_{B},{\mathfrak{l}}_{C}\in {\mathfrak{E}}_{C},\;\;\;\; \nu_{{\mathfrak{l}}_{A},{\mathfrak{l}}_{B},{\mathfrak{l}}_{C}}(u_{ABC})=\nu_{{\mathfrak{l}}_{A},{\mathfrak{l}}_{B},{\mathfrak{l}}_{C}}(u'_{ABC})\,)
\end{eqnarray}
where
\begin{eqnarray}
\nu_{{\mathfrak{l}}_{A},{\mathfrak{l}}_{B},{\mathfrak{l}}_{C}}(\{(\sigma_{i,A},\sigma_{i,B},\sigma_{i,C})\;\vert\; i\in I\}) & := & \bigwedge_{i\in I} \epsilon^{{ \mathfrak{S}}_A}_{{\mathfrak{l}}_{A}}(\sigma_{i,A}) \bullet \epsilon^{{ \mathfrak{S}}_B}_{{\mathfrak{l}}_{B}}(\sigma_{i,B}) \bullet \epsilon^{{ \mathfrak{S}}_C}_{{\mathfrak{l}}_{C}}(\sigma_{i,C})
\end{eqnarray}
In the same way we have that ${ \mathfrak{S}}_A \widetilde{\otimes}  ({ \mathfrak{S}}_B \widetilde{\otimes} { \mathfrak{S}}_C)$ is defined as the quotient of ${\mathcal{P}}({ \mathfrak{S}}_A \times  { \mathfrak{S}}_B  \times  { \mathfrak{S}}_C)$ by the congruence relation defined for any $u_{ABC},u'_{ABC}\in {\mathcal{P}}({ \mathfrak{S}}_A \times  { \mathfrak{S}}_B  \times  { \mathfrak{S}}_C)$
\begin{eqnarray}
(u_{ABC} \approx_{{}_{A(BC)}} u'_{ABC}) & :\Leftrightarrow & (\,\forall {\mathfrak{l}}_{BC}\in \widetilde{E}_{BC},{\mathfrak{l}}_{A}\in {\mathfrak{E}}_{A},\;\;\;\; \nu_{{\mathfrak{l}}_{A},{\mathfrak{l}}_{BC}}(u_{ABC})=\nu_{{\mathfrak{l}}_{A},{\mathfrak{l}}_{BC}}(u'_{ABC})\,)\\
& \Leftrightarrow & (\,\forall {\mathfrak{l}}_{A}\in {\mathfrak{E}}_{A},{\mathfrak{l}}_{B}\in {\mathfrak{E}}_{B},{\mathfrak{l}}_{C}\in {\mathfrak{E}}_{C},\;\;\;\; \nu_{{\mathfrak{l}}_{A},{\mathfrak{l}}_{B},{\mathfrak{l}}_{C}}(u_{ABC})=\nu_{{\mathfrak{l}}_{A},{\mathfrak{l}}_{B},{\mathfrak{l}}_{C}}(u'_{ABC})\,).
\end{eqnarray}
The announced equality is then proved.
\end{proof}

We can then define a multiple tensor product of spaces of states.  
\begin{defin}
The set ${\mathcal{P}}(\prod_{j\in J}{ \mathfrak{S}}^{(j)})$ is equipped with the Inf semi-lattice structure $\cup$ and with the following Inf semi-lattice morphisms defined for any $({ \mathfrak{l}}^{(j)})_{i\in J}$ with ${ \mathfrak{l}}^{(j)}\in { \mathfrak{E}}^{(j)}$ :
\begin{eqnarray}
\begin{array}{rcccc}
\nu_{({ \mathfrak{l}}^{(j)})_{j\in J}} & : & {\mathcal{P}}(\prod_{j\in J}{ \mathfrak{S}}^{(j)}) & \longrightarrow & { \mathfrak{B}}\\
& & \{\, (\sigma_i^{(j)})_{j\in J}\;\vert\; i\in I\,\} & \mapsto & \bigwedge_{i\in I} \bigodot{}_{{}_{j\in J}} \epsilon_{{ \mathfrak{l}}^{(j)}}(\sigma_i^{(j)})
\end{array}
\end{eqnarray}
where we have used the symbol $\bigodot$ to denote the multiple $\bullet$ product.
\end{defin}

\begin{defin}
${\mathcal{P}}(\prod_{j\in J}{ \mathfrak{S}}^{(j)})$ is equipped with a congruence relation defined between any two elements $u,u'\in {\mathcal{P}}(\prod_{j\in J}{ \mathfrak{S}}^{(j)})$ by
\begin{eqnarray}
u \approx u' & :\Leftrightarrow & \forall ({ \mathfrak{l}}^{(j)})_{j\in J}\in \prod{}_{j\in J}{ \mathfrak{E}}^{(j)},\; \nu_{({ \mathfrak{l}}^{(j)})_{j\in J}}(u)=\nu_{({ \mathfrak{l}}^{(j)})_{j\in J}}(u').
\end{eqnarray}
\end{defin}

\begin{defin}

The multiple tensor product $\widetilde{\bigotimes}_{j\in J}{ \mathfrak{S}}^{(j)}$ is defined as the quotient of the set ${\mathcal{P}}(\prod_{j\in J}{ \mathfrak{S}}^{(j)})$ under the congruence relation $\approx$.
\begin{eqnarray}
\forall \sigma \in {\mathcal{P}}(\prod_{j\in J}{ \mathfrak{S}}^{(j)}),&& \widetilde{\; \sigma\;} := \{\, u \; \vert\; \sigma \approx u \,\}.
\end{eqnarray}
The element $\widetilde{\; u\;}\in \widetilde{S}:=\widetilde{\bigotimes}_{j\in J}{ \mathfrak{S}}^{(j)}$ associated to the element $u:=\{\, (\sigma_i^{(j)})_{j\in J}\;\vert\; i\in I\,\} \in {\mathcal{P}}(\prod_{j\in J}{ \mathfrak{S}}^{(j)})$ will be denoted $\bigsqcap{}_{i\in I}^{\widetilde{S}} \widetilde{\bigotimes}_{j\in J} \sigma_i^{(j)}$.
\end{defin}

\begin{theoreme}
Let us introduce the following notation
\begin{eqnarray}
{ \mathcal{K}}^{(J)}_{I}&:=&\{\, (K^{(j)})_{j\in J}\;\vert\; ( K^{(j)}\subseteq I,\;\forall j\in J)\;\textit{\rm and}\; ( K^{(j)}\cap K^{(j')}=\varnothing,\;\forall j,j'\in J ) \;\textit{\rm and}\; (\bigcup{}_{j\in J} K^{(j)} =I )\}.\;\;\;\;\;\;\;\;\;\;\;\;
\end{eqnarray}
The poset structure on $\widetilde{S}:=\widetilde{\bigotimes}_{j\in J}{ \mathfrak{S}}^{(j)}$ is defined according to
\begin{eqnarray}
&&\hspace{-2cm}  (\bigsqcap{}_{{}_{i\in I}}^{{}^{\widetilde{S}}} \widetilde{\bigotimes}_{\!{}_{j\in J}} \sigma_i^{(j)}) \sqsubseteq_{{}_{\widetilde{S}}} \widetilde{\bigotimes}_{\!{}_{j\in J}} \sigma^{(j)}\nonumber \\
&&\Leftrightarrow  (\, \forall (K^{(j)})_{j\in J}\in { \mathcal{K}}^{(J)}_{I},\; \exists j\in J\;\vert\; (\, K^{(j)}\not= \varnothing \;\textit{\rm and}\; \sigma^{(j)}\sqsupseteq_{{}_{{ \mathfrak{S}}^{(j)}}} \bigsqcap{}^{{}^{{ \mathfrak{S}}^{(j)}}}_{{}_{k\in K^{(j)}}} \sigma_k^{(j)}\,) \,)\;\;\;\;\;\;\;\;\;\;\;\;\;\;\;
\end{eqnarray}
\end{theoreme}
\begin{proof}
The proof follows the same line as in Lemma \ref{Lemmadevelopetildeleqetilde}. 
\end{proof}


\subsection{Canonical vs basic tensor product}\label{subsectioncomparison}

\begin{lemme}\label{theoremexpressionexplicitlanglesigmatilderangle}
For any $\widetilde{\sigma}$ in ${\widetilde{S}}_{AB}$, $\langle\widetilde{\sigma}\rangle$ is a bi-filter of ${ \mathfrak{S}}_A\times { \mathfrak{S}}_B$ and we have explicitly
\begin{eqnarray}
\hspace{-1cm}\langle\; \bigsqcap{}^{{}^{{{\widetilde{S}}}_{AB}}}_{i\in I} (\sigma_{i,A} \widetilde{\otimes} \sigma_{i,B})\;\rangle
&= &\{\, (\sigma_A,\sigma_B)\;\vert\; 
(\bigsqcap{}^{{}_{{ \mathfrak{S}}_{A}}}_{k\in I}\; \sigma_{k,A}) \;\sqsubseteq_{{}_{{ \mathfrak{S}}_{A}}}\sigma_{A}
\;\;\textit{\rm and}\;\;
(\bigsqcap{}^{{}_{{ \mathfrak{S}}_{B}}}_{m\in I} \;\sigma_{m,B})\; \sqsubseteq_{{}_{{ \mathfrak{S}}_{B}}} \sigma_{B}
\;\;\textit{\rm and}\nonumber\\
&&\left(\forall \varnothing \varsubsetneq K  \varsubsetneq I, \;\; (\bigsqcap{}^{{}_{{ \mathfrak{S}}_{A}}}_{k\in K}\; \sigma_{k,A}) \;\sqsubseteq_{{}_{{ \mathfrak{S}}_{A}}}\sigma_{A}\;\;\;\textit{\rm or}\;\;\; (\bigsqcap{}^{{}_{{ \mathfrak{S}}_{B}}}_{m\in I-K} \;\sigma_{m,B})\; \sqsubseteq_{{}_{{ \mathfrak{S}}_{B}}} \sigma_{B}\right) \,\}.\;\;\;\;\;\;\;\;\;\;\;\;\;\;\label{expressionexplicitlanglesigmatilderangle}\\
& = &
\left\{ (\sigma_{A},\sigma_{B})\;\vert\; 
\exists { \mathcal{K}},{ \mathcal{K}}'\subseteq 2^I\;\;\textit{\rm with}\;\; { \mathcal{K}}\cup { \mathcal{K}}'=2^I, { \mathcal{K}}\cap { \mathcal{K}}'=\varnothing, \{\varnothing\}\in { \mathcal{K}}', I\in { \mathcal{K}}, \right. \;\;\nonumber\\
&& \left. (\bigsqcup{}^{{}_{{ \mathfrak{S}}_{A}}}_{K\in { \mathcal{K}}}\bigsqcap{}^{{}_{{ \mathfrak{S}}_{A}}}_{k\in K}\; \sigma_{k,A}) \;\sqsubseteq_{{}_{{ \mathfrak{S}}_{A}}}\sigma_{A}\;\;\;\textit{\rm and}\;\;\; (\bigsqcup{}^{{}_{{ \mathfrak{S}}_{A}}}_{K'\in { \mathcal{K}}'}\bigsqcap{}^{{}_{{ \mathfrak{S}}_{B}}}_{m\in I-K'} \;\sigma_{m,B})\; \sqsubseteq_{{}_{{ \mathfrak{S}}_{B}}} \sigma_{B}\right\}.\label{developmentetildeordersimplifypost}
\end{eqnarray}
We will also use the following notation $\widetilde{ \mathfrak{F}}\{(\sigma_{i,A},\sigma_{i,B})\;\vert\; i\in I\}:=\langle\; \bigsqcap{}^{{}^{{{\widetilde{S}}}_{AB}}}_{i\in I} (\sigma_{i,A} \widetilde{\otimes} \sigma_{i,B})\;\rangle$.
\end{lemme}
\begin{proof}
From Definition \ref{definlanglesigmqtilderangle} and Lemma \ref{Lemmadevelopetildeleqetilde} we deduce immediately the expression (\ref{expressionexplicitlanglesigmatilderangle}).\\ 
Let us now check the bi-filter properties.\\
The property (\ref{defbifilter1}) is trivially obtained from the expression (\ref{expressionexplicitlanglesigmatilderangle}).\\
Let us now consider that $(\sigma'_{1,A},\sigma'_{B}),(\sigma'_{2,A},\sigma'_{B})\in \langle\; \bigsqcap{}^{{}^{{{\widetilde{S}}}_{AB}}}_{i\in I} (\sigma_{i,A} \widetilde{\otimes} \sigma_{i,B})\;\rangle$. In other words, we have for any ${\mathfrak{l}}_A\in { \mathfrak{E}}_{A}$ and ${\mathfrak{l}}_B\in { \mathfrak{E}}_{B}$ : 
${\nu}\,{}^{AB}_{{\mathfrak{l}}_{A},{\mathfrak{l}}_B}((\sigma'_{1,A},\sigma'_{B}))\geq
{\nu}\,{}^{AB}_{{\mathfrak{l}}_{A},{\mathfrak{l}}_B} (\{\,(\sigma_{i,A}, \sigma_{i,B})\;\vert\; i\in I\,\})$ and ${\nu}\,{}^{AB}_{{\mathfrak{l}}_{A},{\mathfrak{l}}_B}((\sigma'_{2,A},\sigma'_{B}))\geq
{\nu}\,{}^{AB}_{{\mathfrak{l}}_{A},{\mathfrak{l}}_B} (\{\,(\sigma_{i,A}, \sigma_{i,B})\;\vert\; i\in I\,\})$. Moreover, we have proved in (\ref{demobifilterproperty}) that ${\nu}\,{}^{AB}_{{\mathfrak{l}}_{A},{\mathfrak{l}}_B}((\sigma'_{1,A},\sigma'_{B})) \wedge {\nu}\,{}^{AB}_{{\mathfrak{l}}_{A},{\mathfrak{l}}_B}((\sigma'_{2,A},\sigma'_{B})) = {\nu}\,{}^{AB}_{{\mathfrak{l}}_{A},{\mathfrak{l}}_B}(\, (\sigma'_{1,A}\sqcap_{{}_{{ \mathfrak{S}}_A}}\sigma'_{2,A},\sigma'_{B})\,)$. As a consequence, we obtain ${\nu}\,{}^{AB}_{{\mathfrak{l}}_{A},{\mathfrak{l}}_B}(\, (\sigma'_{1,A}\sqcap_{{}_{{ \mathfrak{S}}_A}}\sigma'_{2,A},\sigma'_{B})\,) \geq {\nu}\,{}^{AB}_{{\mathfrak{l}}_{A},{\mathfrak{l}}_B} (\{\,(\sigma_{i,A}, \sigma_{i,B})\;\vert\; i\in I\,\})$ for any ${\mathfrak{l}}_A\in { \mathfrak{E}}_{A}$ and ${\mathfrak{l}}_B\in { \mathfrak{E}}_{B}$. As a result, we obtain that $(\sigma'_{1,A}\sqcap_{{}_{{ \mathfrak{S}}_A}}\sigma'_{2,A},\sigma'_{B})\in \langle\; \bigsqcap{}^{{}^{{{\widetilde{S}}}_{AB}}}_{i\in I} (\sigma_{i,A} \widetilde{\otimes} \sigma_{i,B})\;\rangle$. We have then proved property (\ref{defbifilter2}).\\
The property (\ref{defbifilter3}) is proved along the same lines.\\
The expression (\ref{developmentetildeordersimplifypost}) is a trivial reformulation of (\ref{expressionexplicitlanglesigmatilderangle}).
\end{proof}

\begin{defin}
We denote ${\widetilde{S}}{}^{fin}_{AB}$ the sub-poset of ${\widetilde{S}}_{AB}$ defined as follows :
\begin{eqnarray}
{\widetilde{S}}{}^{fin}_{AB} &:=& \{ \widetilde{u}\;\vert\; u \subseteq_{fin} { \mathfrak{S}}_A\times { \mathfrak{S}}_B\,\}.
\end{eqnarray}
It is also a sub- Inf semi-lattice of ${\widetilde{S}}_{AB}$.
\end{defin}

\begin{theoreme} \label{theoremsqsubseteqPABgeqSAB} 
We have the following obvious property relating the partial orders of ${{\widetilde{S}}}{}^{fin}_{AB}$ and ${S}_{AB}$. For any $\{(\sigma_{i,A},\sigma_{i,B})\;\vert\; i\in I\}\subseteq_{fin} { \mathfrak{S}}_{A}\times { \mathfrak{S}}_{B}$,
\begin{eqnarray}
(\bigsqcap{}^{{}^{{S}_{AB}}}_{i\in I} \sigma_{i,A}\otimes \sigma_{i,B}) \sqsubseteq_{{}_{{S}_{AB}}}  \sigma'_{A}\otimes \sigma'_{B}  & \Rightarrow & (\bigsqcap{}^{{}^{{\widetilde{S}}_{AB}}}_{i\in I} \sigma_{i,A}\widetilde{\otimes} \sigma_{i,B}) \sqsubseteq_{{}_{{\widetilde{S}}_{AB}}}   \sigma'_{A}\widetilde{\otimes} \sigma'_{B}.\;\;\;\;\;\;\;\;\;\;\;
\end{eqnarray}
\end{theoreme}
\begin{proof}
We intent to prove ${ \mathfrak{F}}\{(\sigma_{i,A},\sigma_{i,B})\;\vert\; i\in I\} \subseteq \widetilde{ \mathfrak{F}}\{(\sigma_{i,A},\sigma_{i,B})\;\vert\; i\in I\}$ for any $\{(\sigma_{i,A},\sigma_{i,B})\;\vert\; i\in I\}\subseteq_{fin} { \mathfrak{S}}_{A}\times { \mathfrak{S}}_{B}$ (we recall that we have adopted the notation $\widetilde{ \mathfrak{F}}\{(\sigma_{i,A},\sigma_{i,B})\;\vert\; i\in I\}:=\langle\; \bigsqcap{}^{{}^{{{\widetilde{S}}}_{AB}}}_{i\in I} (\sigma_{i,A} \widetilde{\otimes} \sigma_{i,B})\;\rangle$).\\
First of all, it is recalled from Lemma \ref{theoremexpressionexplicitlanglesigmatilderangle} that $\widetilde{ \mathfrak{F}}\{(\sigma_{i,A},\sigma_{i,B})\;\vert\; i\in I\}$ is a bi-filter.\\ 
Secondly, it is easy to check that $(\sigma_{k,A},\sigma_{k,B})\in \widetilde{ \mathfrak{F}}\{(\sigma_{i,A},\sigma_{i,B})\;\vert\; i\in I\}$ for any $k\in I$ using the expression (\ref{expressionexplicitlanglesigmatilderangle}). Indeed, for any $K\subseteq I$, if $k\in K$ we have $(\bigsqcap{}^{{}_{{ \mathfrak{S}}_{A}}}_{l\in K}\; \sigma_{l,A}) \;\sqsubseteq_{{}_{{ \mathfrak{S}}_{A}}}\sigma_{k,A} $ and if $k\notin K$ we have $(\bigsqcap{}^{{}_{{ \mathfrak{S}}_{B}}}_{m\in I-K} \;\sigma_{m,B})\; \sqsubseteq_{{}_{{ \mathfrak{S}}_{B}}} \sigma_{k,B}$. \\
As a conclusion,  and by definition of ${ \mathfrak{F}}\{(\sigma_{i,A},\sigma_{i,B})\;\vert\; i\in I\}$ as the intersection of all bi-filters containing $(\sigma_{i,A},\sigma_{i,B})$ for any $i\in I$, we have then 
$\widetilde{ \mathfrak{F}}\{(\sigma_{i,A},\sigma_{i,B})\;\vert\; i\in I\} \supseteq  { \mathfrak{F}}\{(\sigma_{i,A},\sigma_{i,B})\;\vert\; i\in I\}$.\\
We now use Lemma \ref{sigmainFinequality} and Definition \ref{definlanglesigmqtilderangle} to obtain the announced result.
\end{proof}

\begin{theoreme} \label{theoremsqsubseteqPAB=SAB}
If ${ \mathfrak{S}}_{A}$ \underline{or} ${ \mathfrak{S}}_{B}$ are distributive, then ${{\widetilde{S}}}{}^{fin}_{AB}$ and ${S}_{AB}$ are in fact isomorphic posets.  \\
As shown in Remark \ref{remarkdistributivity}, the distributivity of ${ \mathfrak{S}}_{A}$ or ${ \mathfrak{S}}_{B}$ is a key condition for this isomorphism to be valid.
\end{theoreme}
\begin{proof}
We now suppose that ${ \mathfrak{S}}_{A}$ or ${ \mathfrak{S}}_{B}$ is distributive and we intent to prove that ${ \mathfrak{F}}\{(\sigma_{i,A},\sigma_{i,B})\;\vert\; i\in I\}=\widetilde{ \mathfrak{F}}\{(\sigma_{i,A},\sigma_{i,B})\;\vert\; i\in I\}$ for any $\{(\sigma_{i,A},\sigma_{i,B})\;\vert\; i\in I\}\subseteq_{fin} { \mathfrak{S}}_{A}\times { \mathfrak{S}}_{B}$.\\
Let us prove the following fact : every bi-filter $F$ which contains $(\sigma_{k,A},\sigma_{k,B})$ for any $k\in I$ contains also $\widetilde{ \mathfrak{F}}\{(\sigma_{i,A},\sigma_{i,B})\;\vert\; i\in I\}$.  In fact, we can show that, for any bi-filter $F$ we have
\begin{eqnarray}
(\forall k\in I,\;(\sigma_{k,A},\sigma_{k,B})\in F) & \Rightarrow & (\bigsqcup{}^{{}_{{ \mathfrak{S}}_{A}}}_{K\in { \mathcal{K}}}\bigsqcap{}^{{}_{{ \mathfrak{S}}_{A}}}_{k\in K}\; \sigma_{k,A}, \bigsqcup{}^{{}_{{ \mathfrak{S}}_{A}}}_{K'\in { \mathcal{K}}'}\bigsqcap{}^{{}_{{ \mathfrak{S}}_{B}}}_{m\in I-K'} \;\sigma_{m,B})\in F, \nonumber\\
&& \forall { \mathcal{K}},{ \mathcal{K}}'\subseteq 2^I,{ \mathcal{K}}\cup { \mathcal{K}}'=2^I, { \mathcal{K}}\cap { \mathcal{K}}'=\varnothing, \{\varnothing\}\in { \mathcal{K}}', I\in { \mathcal{K}}.\;\;\;\;\;\;\;\;\;\;\;\;\;\label{intermediatedistrib}
\end{eqnarray}

The first step towards (\ref{intermediatedistrib}) is obtained by checking that $\forall { \mathcal{K}},{ \mathcal{K}}'\subseteq 2^I,{ \mathcal{K}}\cup { \mathcal{K}}'=2^I, { \mathcal{K}}\cap { \mathcal{K}}'=\varnothing, \{\varnothing\}\in { \mathcal{K}}', I\in { \mathcal{K}}$,
\begin{eqnarray}
 && (\bigsqcup{}^{{}^{{ \mathfrak{S}}}}_{K'\in { \mathcal{K}}'}\bigsqcap{}^{{}^{{ \mathfrak{S}}}}_{m\in I-K'} \;\sigma_{m}) \sqsupseteq_{{}_{ \mathfrak{S}}}
(\bigsqcap{}^{{}^{{ \mathfrak{S}}}}_{K\in { \mathcal{K}}}\bigsqcup{}^{{}^{{ \mathfrak{S}}}}_{k\in K}\; \sigma_{k})\label{firststep}
\end{eqnarray}
for any distributive ${ \mathfrak{S}}$ and any collection of elements of ${ \mathfrak{S}}$  denoted $\sigma_k$ for $k\in I$ for which these two sides of inequality exist.  To check this fact, we have to note that,  using \cite[Lemma 8 p. 50]{Balbes1977}, we have first of all
\begin{eqnarray}
(\bigsqcap{}^{{}^{{ \mathfrak{S}}}}_{K\in { \mathcal{K}}}\bigsqcup{}^{{}^{{ \mathfrak{S}}}}_{k\in K}\; \sigma_{k})= \bigsqcup{}^{{}^{{ \mathfrak{S}}}}\!\!\! \left\{ \bigsqcap{}^{{}^{{ \mathfrak{S}}}}_{K\in { \mathcal{K}}} \pi_K(A) \;\vert\; A\in \prod_{K\in { \mathcal{K}}} K \right\},
\end{eqnarray}
where $\pi_K$ denotes the projection of the component indexed by K in the cardinal product $\prod_{K\in { \mathcal{K}}} K$. Moreover,  for any $A\in \prod_{K\in { \mathcal{K}}} K$,  there exists $L\in { \mathcal{K}}'$ such that $\bigcup \{ \pi_K(A) \;\vert\; K\in { \mathcal{K}}\}\;\supseteq\; (I\smallsetminus L)$ and then $(\bigsqcap{}^{{}^{{ \mathfrak{S}}}}_{K\in { \mathcal{K}}} \pi_K(A)) \sqsubseteq_{{}_{ \mathfrak{S}}} (\bigsqcap{}^{{}^{{ \mathfrak{S}}}}_{m\in I-L} \;\sigma_{m}) \sqsubseteq_{{}_{ \mathfrak{S}}} (\bigsqcup{}^{{}^{{ \mathfrak{S}}}}_{K'\in { \mathcal{K}}'}\bigsqcap{}^{{}^{{ \mathfrak{S}}}}_{m\in I-K'} \;\sigma_{m})$. As a result, we obtain the property (\ref{firststep}).\\

The second step towards (\ref{intermediatedistrib}) consists in showing that 
\begin{eqnarray}
(\forall k\in I,\;(\sigma_{k,A},\sigma_{k,B})\in F) & \Rightarrow & (\bigsqcup{}^{{}_{{ \mathfrak{S}}_{A}}}_{K\in { \mathcal{K}}}\bigsqcap{}^{{}_{{ \mathfrak{S}}_{A}}}_{k\in K}\; \sigma_{k,A},  \bigsqcap{}^{{}_{{ \mathfrak{S}}_{B}}}_{K\in { \mathcal{K}}}\bigsqcup{}^{{}_{{ \mathfrak{S}}_{B}}}_{k\in K}\; \sigma_{k,B})\;\in F
\end{eqnarray}
for any ${ \mathcal{K}}\subseteq 2^I$. This intermediary result is obtained by induction on the complexity of the polynomial $(\bigsqcup{}^{{}_{{ \mathfrak{S}}_{A}}}_{K\in { \mathcal{K}}}\bigsqcap{}^{{}_{{ \mathfrak{S}}_{A}}}_{k\in K}\; \sigma_{k,A})$ by using the following elementary result
\begin{eqnarray}
\forall \sigma_A,\sigma'_A\in { \mathfrak{S}}_A,\sigma_B,\sigma'_B\in { \mathfrak{S}}_B,\;\;\;\; \left( (\sigma_A,\sigma_B),(\sigma'_A,\sigma'_B)\in F \right) &\Rightarrow & \left\{
\begin{array}{l}
(\sigma_A \sqcup_{{}_{{ \mathfrak{S}}_A}} \sigma'_A,\sigma_B \sqcap_{{}_{{ \mathfrak{S}}_B}} \sigma'_B)\in F\nonumber\\
(\sigma_A \sqcap_{{}_{{ \mathfrak{S}}_A}} \sigma'_A,\sigma_B \sqcup_{{}_{{ \mathfrak{S}}_B}} \sigma'_B)\in F
\end{array}\right.
\end{eqnarray}
trivially deduced using the bi-filter character of $F$, i.e. properties (\ref{defbifilter1})(\ref{defbifilter2})(\ref{defbifilter3}). \\

As a final conclusion,  using the explicit definition of ${ \mathfrak{F}}\{(\sigma_{i,A},\sigma_{i,B})\;\vert\; i\in I\}$ as the intersection of all bi-ideals containing $(\sigma_{k,A},\sigma_{k,B})$ for any $k\in I$,  we obtain $\widetilde{ \mathfrak{F}}\{(\sigma_{i,A},\sigma_{i,B})\;\vert\; i\in I\}={ \mathfrak{F}}\{(\sigma_{i,A},\sigma_{i,B})\;\vert\; i\in I\}$.\\

${{\widetilde{S}}}{}^{fin}_{AB}$ and ${S}_{AB}$ are then isomorphic posets.
\end{proof}

\begin{remark}\label{remarkdistributivity}
We note that the distributivity property is a key condition to obtain previous isomorphism between  ${{\widetilde{S}}}{}^{fin}_{AB}$ and ${S}_{AB}$. Indeed, let us consider that ${ \mathfrak{S}}_A$ and ${ \mathfrak{S}}_B$ are both defined as the lattice associated to the following Hasse diagram:
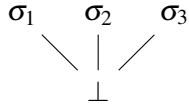
\begin{center}\begin{tikzpicture}
  \node (bot) at (0,0) {$\bot$};
  \node (1) at (-1,1) {$\sigma_1$};
  \node (2) at (0,1) {$\sigma_2$};
  \node (3) at (1,1) {$\sigma_3$};
  \draw (bot) -- (1)  (bot) -- (2) (bot) -- (3) ;
\end{tikzpicture}\end{center}
According to (\ref{developmentetildeordersimplify}), we have $(\bot_{{}_{{ \mathfrak{S}}_A}}, \bot_{{}_{{ \mathfrak{S}}_B}})\in \widetilde{ \mathfrak{F}}\{(\sigma_{1},\sigma_{1}),(\sigma_{2},\sigma_{2}),(\sigma_{3},\sigma_{3})\}.$ However, we have obviously $(\bot_{{}_{{ \mathfrak{S}}_A}}, \bot_{{}_{{ \mathfrak{S}}_B}})\notin { \mathfrak{F}}\{(\sigma_{1},\sigma_{1}),(\sigma_{2},\sigma_{2}),(\sigma_{3},\sigma_{3})\}.$
\end{remark}

\subsection{Properties of the basic tensor product} \label{subsectionremarks}

\begin{theoreme}\label{formulacupStilde}
Let $\widetilde{\sigma}_{AB}$ and $\widetilde{\sigma}'_{AB}$ be two elements of ${{\widetilde{S}}}_{AB}$ having a common upper-bound. Then the supremum of $\{\widetilde{\sigma}_{AB},\widetilde{\sigma}'_{AB}\}$ exists in ${{\widetilde{S}}}_{AB}$ and its expression is given by
\begin{eqnarray}
\widetilde{\sigma}_{AB} \sqcup_{{}_{{{\widetilde{S}}}_{AB}}} \widetilde{\sigma}'_{AB} = \bigsqcap{}^{{}^{{{\widetilde{S}}}_{AB}}}_{\widetilde{\sigma} \in (\underline{\widetilde{\sigma}_{AB}}_{{}_{ {{\widetilde{S}}}_{AB}}}\!\!\!\!\!\!\!\!\!\cap\; \underline{\widetilde{\sigma}'_{AB}}_{{}_{ {{\widetilde{S}}}_{AB}}}\!\!\!\!\!\!\!\!\!)} \; \widetilde{\sigma}\label{formulacupPtilde2}
\end{eqnarray}
\end{theoreme}

\begin{theoreme}\label{SASBditribandcup}
If ${ \mathfrak{S}}_A$ \underline{and} ${ \mathfrak{S}}_B$ are distributive (cf. Definition \ref{defdistrib}), then ${{\widetilde{S}}}_{AB}$ is also distributive.  \\
Note, using Theorem \ref{theoremsqsubseteqPAB=SAB},  that, in this situation, we have also ${{\widetilde{S}}}{}^{fin}_{AB}={S}_{AB}$.\\
In that case, the explicit expression for the supremum of two elements in ${{\widetilde{S}}}{}^{fin}_{AB}$ is given by
\begin{eqnarray}
(\bigsqcap{}^{{}^{{{\widetilde{S}}}_{AB}}}_{i\in I} \sigma_{i,A}\widetilde{\otimes} \sigma_{i,B}) \sqcup{}_{{}_{{{\widetilde{S}}}_{AB}}} (\bigsqcap{}^{{}^{{{\widetilde{S}}}_{AB}}}_{j\in J} \sigma'_{j,A}\widetilde{\otimes} \sigma'_{j,B}) = \bigsqcap{}^{{}^{{{\widetilde{S}}}_{AB}}}_{i\in I,\;j\in J}\; (\sigma_{i,A}\sqcup_{{}_{{ \mathfrak{S}}_A}}\sigma'_{j,A}) \widetilde{\otimes} (\sigma_{i,B} \sqcup_{{}_{{ \mathfrak{S}}_B}} \sigma'_{j,B}). \label{formulacupSAB}
\end{eqnarray}
\end{theoreme}
\begin{proof}
First of all, using Theorem \ref{theoremsqsubseteqPAB=SAB}, we note that, as soon as ${ \mathfrak{S}}_A$ or ${ \mathfrak{S}}_B$ is distributive, we have ${{\widetilde{S}}}_{AB}={S}_{AB}$ as Inf semi-lattices.  We are then reduced to prove the distributivity of ${S}_{AB}$. \\
In reference to the definition of distributivity of an Inf semi-lattice given in Definition \ref{defdistrib}, we have then to prove that if $\bigsqcap{}^{{}^{{S}_{AB}}}_{1\leq i\leq n} \sigma_{i,A}{\otimes} \sigma_{i,B} \sqsubseteq_{{}_{{S}_{AB}}}\sigma_{A}{\otimes} \sigma_{B}$, then there exists $\sigma'_{i,A}{\otimes} \sigma'_{i,B} \sqsupseteq_{{}_{{S}_{AB}}}\sigma_{i,A}{\otimes} \sigma_{i,B}$ for any $1\leq i\leq n$ such that $\bigsqcap{}^{{}^{{S}_{AB}}}_{1\leq i\leq n} \sigma'_{i,A}{\otimes} \sigma'_{i,B} = \sigma_{A}{\otimes} \sigma_{B}$.  From Lemma \ref{orderimpliespolynomial}, we conclude that it is sufficient to prove that, for any $n-$ary polynomial $p$, if $\sigma_{A} \sqsupseteq_{{}_{{ \mathfrak{S}}_{A}}} p(\sigma_{1,A},\cdots,\sigma_{n,A})$ and $\sigma_{B} \sqsupseteq_{{}_{{ \mathfrak{S}}_{B}}} p^\ast(\sigma_{1,B},\cdots,\sigma_{n,B})$, then there exist $\sigma'_{i,A} \sqsupseteq_{{}_{{ \mathfrak{S}}_{A}}}\sigma_{i,A}$ and $\sigma'_{i,B}\sqsupseteq_{{}_{{ \mathfrak{S}}_{B}}}  \sigma_{i,B}$ for $1\leq i\leq n$ such that $\sigma_{A} \sqsupseteq_{{}_{{ \mathfrak{S}}_{A}}} p(\sigma'_{1,A},\cdots,\sigma'_{n,A})$ and $\sigma_{B} \sqsupseteq_{{}_{{ \mathfrak{S}}_{B}}} p^\ast(\sigma'_{1,B},\cdots,\sigma'_{n,B})$, and $\sigma'_{i,A} \sqsupseteq_{{}_{{ \mathfrak{S}}_{A}}}\sigma_{A}$ and $\sigma'_{i,B}\sqsupseteq_{{}_{{ \mathfrak{S}}_{B}}}  \sigma_{B}$ for $1\leq i\leq n$. \\ 
The proof of this fact is sketched in \cite[Theorem 3]{Fraser1978}, and we give here a developed version of it.  \\

Let us prove the following statement for any $n-$ary polynomial $p$ : 
\begin{eqnarray}
\hspace{-1cm} \sigma_{A} \sqsupseteq_{{}_{{ \mathfrak{S}}_{A}}} p(\sigma_{1,A},\cdots,\sigma_{n,A}) & \Rightarrow & \exists \sigma'_{i,A} \sqsupseteq_{{}_{{ \mathfrak{S}}_{A}}}\sigma_{i,A}, \forall 1\leq i\leq n \;\vert\; \left( \sigma_{A} \sqsupseteq_{{}_{{ \mathfrak{S}}_{A}}} p(\sigma'_{1,A},\cdots,\sigma'_{n,A}) \;\textit{\rm and}\; \sigma'_{i,A} \sqsupseteq_{{}_{{ \mathfrak{S}}_{A}}}\sigma_{A}, \forall 1\leq i\leq n\right).\;\;\;\;\;\;\;\;\;\;\;\;\;\;
\end{eqnarray}

This statement is obviously true for $p(\sigma_{1,A},\cdots,\sigma_{n,A}):=\sigma_{k,A}$, it suffices to chose $\sigma_{k,A}=\sigma_A$. \\ 

Let us assume that the induction statement is true for two $n-$ary polynomials $p$ and $q$, and let us prove the statement is also true for $(p\sqcap q)$.\\
We will assume 
$\sigma_{A} \sqsupseteq_{{}_{{ \mathfrak{S}}_{A}}} p(\sigma_{1,A},\cdots,\sigma_{n,A}) \sqcap_{{}_{{ \mathfrak{S}}_A}} q(\sigma_{1,A},\cdots,\sigma_{n,A})$. Then, there exist $\gamma_A,\delta_A\in { \mathfrak{S}}_A$ such that $\sigma_{A} \sqsupseteq_{{}_{{ \mathfrak{S}}_{A}}} (\gamma_A\sqcap_{{}_{{ \mathfrak{S}}_A}} \delta_A)$ and $\gamma_{A} \sqsupseteq_{{}_{{ \mathfrak{S}}_{A}}} p(\sigma_{1,A},\cdots,\sigma_{n,A})$ and $\delta_{A} \sqsupseteq_{{}_{{ \mathfrak{S}}_{A}}} q(\sigma_{1,A},\cdots,\sigma_{n,A})$. \\
From distributivity of ${ \mathfrak{S}}_{A}$, we deduce that there exist $\gamma'_A$ and $\delta'_A$ such that $\sigma_{A} = (\gamma'_A\sqcap_{{}_{{ \mathfrak{S}}_A}} \delta'_A)$ and $\gamma'_A \sqsupseteq_{{}_{{ \mathfrak{S}}_{A}}}  \gamma_A$ and $\delta'_A\sqsupseteq_{{}_{{ \mathfrak{S}}_{A}}}  \delta_A$. As a result, we have $\gamma'_A \sqsupseteq_{{}_{{ \mathfrak{S}}_{A}}} p(\sigma_{1,A},\cdots,\sigma_{n,A})$ and $\delta'_A \sqsupseteq_{{}_{{ \mathfrak{S}}_{A}}} q(\sigma_{1,A},\cdots,\sigma_{n,A})$.\\
By assumption, there exist $\sigma'_{i,A}  \sqsupseteq_{{}_{{ \mathfrak{S}}_{A}}} \sigma_{i,A}$ and $\sigma''_{i,A}  \sqsupseteq_{{}_{{ \mathfrak{S}}_{A}}} \sigma_{i,A}$ for $1\leq i\leq n$ with $\gamma'_A \sqsupseteq_{{}_{{ \mathfrak{S}}_{A}}} p(\sigma'_{1,A},\cdots,\sigma'_{n,A})$ and $\delta'_A \sqsupseteq_{{}_{{ \mathfrak{S}}_{A}}} q(\sigma''_{1,A},\cdots,\sigma''_{n,A})$,  and with $\sigma'_{i,A}  \sqsupseteq_{{}_{{ \mathfrak{S}}_{A}}} \gamma'_{A}$ and $\sigma''_{i,A}  \sqsupseteq_{{}_{{ \mathfrak{S}}_{A}}} \delta'_{A}$ for $1\leq i\leq n$. \\
Let us denote $\overline{\sigma}_{i,A}:=\sigma'_{i,A}\sqcap_{{}_{{ \mathfrak{S}}_{A}}} \sigma''_{i,A}$. \\
We first note that $\overline{\sigma}_{i,A}\sqsupseteq_{{}_{{ \mathfrak{S}}_{A}}} \sigma_{i,A}$ for $1\leq i\leq n$.\\
From $\overline{\sigma}_{i,A} \sqsubseteq_{{}_{{ \mathfrak{S}}_{A}}} {\sigma}'_{i,A}$ and $\overline{\sigma}_{i,A} \sqsubseteq_{{}_{{ \mathfrak{S}}_{A}}} {\sigma}''_{i,A}$ for any $1\leq i\leq n$, and $\gamma'_{A} \sqsupseteq_{{}_{{ \mathfrak{S}}_{A}}} p(\sigma'_{1,A},\cdots,\sigma'_{n,A})$ and $\delta'_{A} \sqsupseteq_{{}_{{ \mathfrak{S}}_{A}}} q(\sigma''_{1,A},\cdots,\sigma''_{n,A})$, we deduce 
$\gamma'_{A} \sqsupseteq_{{}_{{ \mathfrak{S}}_{A}}} p(\overline{\sigma}_{1,A},\cdots,\overline{\sigma}_{n,A})$ and $\delta'_{A} \sqsupseteq_{{}_{{ \mathfrak{S}}_{A}}} q(\overline{\sigma}_{1,A},\cdots,\overline{\sigma}_{n,A})$. As a consequence,  $\sigma_A=(\gamma'_{A}\sqcap_{{}_{{ \mathfrak{S}}_{A}}} \delta'_{A}) \sqsupseteq_{{}_{{ \mathfrak{S}}_{A}}} p(\overline{\sigma}_{1,A},\cdots,\overline{\sigma}_{n,A})\sqcap_{{}_{{ \mathfrak{S}}_{A}}} q(\overline{\sigma}_{1,A},\cdots,\overline{\sigma}_{n,A})$.\\
From $\sigma'_{i,A}  \sqsupseteq_{{}_{{ \mathfrak{S}}_{A}}} \gamma'_{A}$ and $\sigma''_{i,A}  \sqsupseteq_{{}_{{ \mathfrak{S}}_{A}}} \delta'_{A}$ for $1\leq i\leq n$, we deduce also $\overline{\sigma}_{i,A}  \sqsupseteq_{{}_{{ \mathfrak{S}}_{A}}} \gamma'_{A}\sqcap_{{}_{{ \mathfrak{S}}_{A}}} \delta'_{A}=\sigma_A$ for $1\leq i\leq n$.\\
As a summary, there exist $\overline{\sigma}_{i,A}\sqsupseteq_{{}_{{ \mathfrak{S}}_{A}}} \sigma_{i,A}$ for $1\leq i\leq n$,  such that $\sigma_A \sqsupseteq_{{}_{{ \mathfrak{S}}_{A}}} p(\overline{\sigma}_{1,A},\cdots,\overline{\sigma}_{n,A})\sqcap_{{}_{{ \mathfrak{S}}_{A}}} q(\overline{\sigma}_{1,A},\cdots,\overline{\sigma}_{n,A})$, and $\overline{\sigma}_{i,A}  \sqsupseteq_{{}_{{ \mathfrak{S}}_{A}}} \sigma_A$ for $1\leq i\leq n$. In other words, the $n-$ary polynomial $(p\sqcap q)$ satisfies also the induction assumption.\\

Let us assume that the induction statement is true for two $n-$ary polynomials $p$ and $q$, and let us now prove the statement is also true for $(p\sqcup q)$.\\
We will assume 
$\sigma_{A} \sqsupseteq_{{}_{{ \mathfrak{S}}_{A}}} p(\sigma_{1,A},\cdots,\sigma_{n,A}) \sqcup_{{}_{{ \mathfrak{S}}_A}} q(\sigma_{1,A},\cdots,\sigma_{n,A})$. Then,  we have $\sigma_{A} \sqsupseteq_{{}_{{ \mathfrak{S}}_{A}}} p(\sigma_{1,A},\cdots,\sigma_{n,A})$ and $\sigma_{A} \sqsupseteq_{{}_{{ \mathfrak{S}}_{A}}} q(\sigma_{1,A},\cdots,\sigma_{n,A})$. \\
By assumption, there exist $\sigma'_{i,A}  \sqsupseteq_{{}_{{ \mathfrak{S}}_{A}}} \sigma_{i,A}$ and $\sigma''_{i,A}  \sqsupseteq_{{}_{{ \mathfrak{S}}_{A}}} \sigma_{i,A}$ for $1\leq i\leq n$ with $\sigma_A \sqsupseteq_{{}_{{ \mathfrak{S}}_{A}}} p(\sigma'_{1,A},\cdots,\sigma'_{n,A})$ and $\sigma_A \sqsupseteq_{{}_{{ \mathfrak{S}}_{A}}} q(\sigma''_{1,A},\cdots,\sigma''_{n,A})$,  and with $\sigma'_{i,A}  \sqsupseteq_{{}_{{ \mathfrak{S}}_{A}}} \sigma_{A}$ and $\sigma''_{i,A}  \sqsupseteq_{{}_{{ \mathfrak{S}}_{A}}} \sigma_{A}$ for $1\leq i\leq n$. \\
Let us denote $\overline{\sigma}_{i,A}:=\sigma'_{i,A}\sqcap_{{}_{{ \mathfrak{S}}_{A}}} \sigma''_{i,A}$. \\
We first note that $\overline{\sigma}_{i,A}\sqsupseteq_{{}_{{ \mathfrak{S}}_{A}}} \sigma_{i,A}$ for $1\leq i\leq n$.\\
From $\overline{\sigma}_{i,A} \sqsubseteq_{{}_{{ \mathfrak{S}}_{A}}} {\sigma}'_{i,A}$ and $\overline{\sigma}_{i,A} \sqsubseteq_{{}_{{ \mathfrak{S}}_{A}}} {\sigma}''_{i,A}$ for any $1\leq i\leq n$, and $\sigma_{A} \sqsupseteq_{{}_{{ \mathfrak{S}}_{A}}} p(\sigma'_{1,A},\cdots,\sigma'_{n,A})$ and $\sigma_{A} \sqsupseteq_{{}_{{ \mathfrak{S}}_{A}}} q(\sigma''_{1,A},\cdots,\sigma''_{n,A})$, we deduce 
$\sigma_{A} \sqsupseteq_{{}_{{ \mathfrak{S}}_{A}}} p(\overline{\sigma}_{1,A},\cdots,\overline{\sigma}_{n,A})$ and $\sigma_{A} \sqsupseteq_{{}_{{ \mathfrak{S}}_{A}}} q(\overline{\sigma}_{1,A},\cdots,\overline{\sigma}_{n,A})$. As a consequence,  $\sigma_A\sqsupseteq_{{}_{{ \mathfrak{S}}_{A}}} p(\overline{\sigma}_{1,A},\cdots,\overline{\sigma}_{n,A})\sqcup_{{}_{{ \mathfrak{S}}_{A}}} q(\overline{\sigma}_{1,A},\cdots,\overline{\sigma}_{n,A})$.\\
From $\sigma'_{i,A}  \sqsupseteq_{{}_{{ \mathfrak{S}}_{A}}} \sigma_{A}$ and $\sigma''_{i,A}  \sqsupseteq_{{}_{{ \mathfrak{S}}_{A}}} \sigma_{A}$ for $1\leq i\leq n$, we deduce also $\overline{\sigma}_{i,A}  \sqsupseteq_{{}_{{ \mathfrak{S}}_{A}}} \sigma_A$ for $1\leq i\leq n$.\\
As a summary, there exist $\overline{\sigma}_{i,A}\sqsupseteq_{{}_{{ \mathfrak{S}}_{A}}} \sigma_{i,A}$ for $1\leq i\leq n$,  such that $\sigma_A \sqsupseteq_{{}_{{ \mathfrak{S}}_{A}}} p(\overline{\sigma}_{1,A},\cdots,\overline{\sigma}_{n,A})\sqcup_{{}_{{ \mathfrak{S}}_{A}}} q(\overline{\sigma}_{1,A},\cdots,\overline{\sigma}_{n,A})$, and $\overline{\sigma}_{i,A}  \sqsupseteq_{{}_{{ \mathfrak{S}}_{A}}} \sigma_A$ for $1\leq i\leq n$. In other words, the $n-$ary polynomial $(p\sqcup q)$ satisfies also the induction assumption.\\

By induction on the complexity of the $n-$ary polynomial $p$ we have then proved the statement. As a final consequence, ${S}_{AB}$ and then also ${{\widetilde{S}}}_{AB}$ is a distributive Inf semi-lattice.\\

As a consequence of this distributivity property, we obtain the following simplification
\begin{eqnarray}
(\bigsqcap{}^{{}^{{{\widetilde{S}}}_{AB}}}_{i\in I} \sigma_{i,A}\widetilde{\otimes} \sigma_{i,B})
\sqcup_{{}_{{{\widetilde{S}}}_{AB}}}
(\bigsqcap{}^{{}^{{{\widetilde{S}}}_{AB}}}_{j\in J} \sigma'_{j,A}\widetilde{\otimes} \sigma'_{j,B}) & = & \bigsqcap{}^{{}^{{{\widetilde{S}}}_{AB}}}_{i\in I}\bigsqcap{}^{{}^{{{\widetilde{S}}}_{AB}}}_{j\in J} \left( (\sigma_{i,A}\widetilde{\otimes} \sigma_{i,B})\sqcup_{{}_{{{\widetilde{S}}}_{AB}}} (\sigma'_{j,A}\widetilde{\otimes} \sigma'_{j,B}) \right).\;\;\;\;\;\;\;\;\;\;\;
\end{eqnarray}
Using the expansion (\ref{developmentetildeordersimplify}), we know that
\begin{eqnarray}
(\sigma_{i,A}\widetilde{\otimes} \sigma_{i,B})\sqcup_{{}_{{{\widetilde{S}}}_{AB}}} (\sigma'_{j,A}\widetilde{\otimes} \sigma'_{j,B}) & = & (\sigma_{i,A} \sqcup_{{}_{{ \mathfrak{S}}_{A}}} \sigma'_{j,A})\widetilde{\otimes} (\sigma_{i,B}\sqcup_{{}_{{ \mathfrak{S}}_{B}}} \sigma'_{j,B}) 
\end{eqnarray}
This concludes the proof of the formula (\ref{formulacupSAB}).
\end{proof}

\begin{theoreme}
If ${ \mathfrak{S}}_{A}$ and ${ \mathfrak{S}}_{B}$ are atomic, then ${{\widetilde{S}}}_{AB}$ is also atomic, i.e. 
\begin{eqnarray}
\exists { \mathcal{A}}_{{{\widetilde{S}}}_{AB}}\subseteq {{\widetilde{S}}}_{AB} & \vert &\forall \alpha_{AB}\in { \mathcal{A}}_{{{\widetilde{S}}}_{AB}},\;\; (\bot_{{}_{{ \mathfrak{S}}_{A}}}\widetilde{\otimes} \bot_{{}_{{ \mathfrak{S}}_{B}}}) \sqcoversubset_{{}_{{{\widetilde{S}}}_{AB}}} \alpha_{AB},\\
&&\forall \sigma_{AB} \in {{\widetilde{S}}}_{AB}, \;\; \exists \alpha_{AB}\in { \mathcal{A}}_{{{\widetilde{S}}}_{AB}}\smallsetminus \{\bot_{{}_{{ \mathfrak{S}}_{A}}}\widetilde{\otimes} \bot_{{}_{{ \mathfrak{S}}_{B}}} \}\;\vert\; \alpha_{AB}\sqsubseteq_{{}_{{{\widetilde{S}}}_{AB}}} \sigma_{AB}.
\end{eqnarray}
Here, we denote $\sigma \sqcoversubset \sigma'$ iff $(\sigma \sqsubset \sigma'\;\textit{\rm and} (\sigma \sqsubseteq \sigma''\sqsubseteq \sigma' \;\Leftrightarrow\; (\sigma = \sigma'' \;\textit{\rm or}\; \sigma''=\sigma')))$.\\
The set of atoms of ${{\widetilde{S}}}_{AB}$ is indeed defined by
\begin{eqnarray}
{ \mathcal{A}}_{{{\widetilde{S}}}_{AB}} & := & \{\, (\alpha_A \widetilde{\otimes}\bot_{{}_{{ \mathfrak{S}}_{B}}})\sqcap_{{}_{{{\widetilde{S}}}_{AB}}} (\bot_{{}_{{ \mathfrak{S}}_{A}}}\widetilde{\otimes} \alpha_B)\;\vert\; \alpha_A\in { \mathcal{A}}_{{ \mathfrak{S}}_{A}},\; \alpha_B\in { \mathcal{A}}_{{ \mathfrak{S}}_{B}}\,\}.
\end{eqnarray}
\end{theoreme}
\begin{proof}
Using the expansion (\ref{developmentetildeordersimplify}), we deduce immediately 
\begin{eqnarray}
\forall \alpha_A\in { \mathcal{A}}_{{ \mathfrak{S}}_{A}},\forall \alpha_B\in { \mathcal{A}}_{{ \mathfrak{S}}_{B}}, && (\alpha_A \widetilde{\otimes}\bot_{{}_{{ \mathfrak{S}}_{B}}})\sqcap_{{}_{{{\widetilde{S}}}_{AB}}} (\bot_{{}_{{ \mathfrak{S}}_{A}}}\widetilde{\otimes} \alpha_B) \;\not\sqsubseteq_{{}_{{{\widetilde{S}}}_{AB}}} \; \bot_{{}_{{ \mathfrak{S}}_{A}}} \widetilde{\otimes}\bot_{{}_{{ \mathfrak{S}}_{B}}}.
\end{eqnarray}
In other words,  $\bot_{{}_{{ \mathfrak{S}}_{A}}} \widetilde{\otimes}\bot_{{}_{{ \mathfrak{S}}_{B}}} \sqsubset_{{}_{{{\widetilde{S}}}_{AB}}}  
(\alpha_A \widetilde{\otimes}\bot_{{}_{{ \mathfrak{S}}_{B}}})\sqcap_{{}_{{{\widetilde{S}}}_{AB}}} (\bot_{{}_{{ \mathfrak{S}}_{A}}}\widetilde{\otimes} \alpha_B)$.\\

Secondly, let us show that, for any $\sigma_{AB}:=(\bigsqcap{}^{{}^{{{\widetilde{S}}}_{AB}}}_{i\in I} \sigma_{i,A}\widetilde{\otimes} \sigma_{i,B})$ distinct from $\bot_{{}_{{ \mathfrak{S}}_{A}}} \widetilde{\otimes}\bot_{{}_{{ \mathfrak{S}}_{B}}}$, there exist $\alpha_A\in { \mathcal{A}}_{{ \mathfrak{S}}_{A}}$ and $\alpha_B\in { \mathcal{A}}_{{ \mathfrak{S}}_{B}}$ such that $((\alpha_A \widetilde{\otimes}\bot_{{}_{{ \mathfrak{S}}_{B}}})\sqcap_{{}_{{{\widetilde{S}}}_{AB}}} (\bot_{{}_{{ \mathfrak{S}}_{A}}}\widetilde{\otimes} \alpha_B)) \sqsubseteq_{{}_{{{\widetilde{S}}}_{AB}}}  \sigma_{AB}$. Using once again the expansion (\ref{developmentetildeordersimplify}), we know that $\sigma_{AB} \sqsupset_{{}_{{{\widetilde{S}}}_{AB}}}  
\bot_{{}_{{ \mathfrak{S}}_{A}}} \widetilde{\otimes}\bot_{{}_{{ \mathfrak{S}}_{B}}}$ (or, in other words, $\sigma_{AB} \not\sqsubseteq_{{}_{{{\widetilde{S}}}_{AB}}}  
\bot_{{}_{{ \mathfrak{S}}_{A}}} \widetilde{\otimes}\bot_{{}_{{ \mathfrak{S}}_{B}}}$) implies that there exists $\varnothing \subseteq K \subseteq I$ such that $(\bigsqcap{}^{{}_{{ \mathfrak{S}}_{A}}}_{k\in K}\; \sigma_{k,A}) \;\sqsupset_{{}_{{ \mathfrak{S}}_{A}}} \bot_{{}_{{ \mathfrak{S}}_{A}}}$ and $(\bigsqcap{}^{{}_{{ \mathfrak{S}}_{B}}}_{m\in I-K} \;\sigma_{m,B})\; \sqsupset_{{}_{{ \mathfrak{S}}_{B}}} \bot_{{}_{{ \mathfrak{S}}_{B}}} $. Let us fix such a $K$ and let us choose $\alpha_A\in { \mathcal{A}}_{{ \mathfrak{S}}_{A}}$ and $\alpha_B\in { \mathcal{A}}_{{ \mathfrak{S}}_{B}}$ such that $(\bigsqcap{}^{{}_{{ \mathfrak{S}}_{A}}}_{k\in K}\; \sigma_{k,A}) \;\sqsupseteq_{{}_{{ \mathfrak{S}}_{A}}} \alpha_A$ and $(\bigsqcap{}^{{}_{{ \mathfrak{S}}_{B}}}_{m\in I-K} \;\sigma_{m,B})\; \sqsupseteq_{{}_{{ \mathfrak{S}}_{B}}} \alpha_B$.  We obtain $(\bigsqcap{}^{{}^{{{\widetilde{S}}}_{AB}}}_{i\in K} \sigma_{i,A}\widetilde{\otimes} \sigma_{i,B}) \;\sqsupseteq_{{}_{{{\widetilde{S}}}_{AB}}} (\alpha_A \widetilde{\otimes} \bot_{{}_{{ \mathfrak{S}}_{B}}})$ and $(\bigsqcap{}^{{}^{{{\widetilde{S}}}_{AB}}}_{i\in I-K} \sigma_{i,A}\widetilde{\otimes} \sigma_{i,B}) \;\sqsupseteq_{{}_{{{\widetilde{S}}}_{AB}}} (\bot_{{}_{{ \mathfrak{S}}_{A}}} \widetilde{\otimes} \alpha_B)$. As a first conclusion, we obtain $((\alpha_A \widetilde{\otimes}\bot_{{}_{{ \mathfrak{S}}_{B}}})\sqcap_{{}_{{{\widetilde{S}}}_{AB}}} (\bot_{{}_{{ \mathfrak{S}}_{A}}}\widetilde{\otimes} \alpha_B)) \sqsubseteq_{{}_{{{\widetilde{S}}}_{AB}}}  \sigma_{AB}$. \\

Thirdly, let us consider $\sigma_{AB}:=(\bigsqcap{}^{{}^{{{\widetilde{S}}}_{AB}}}_{i\in I} \sigma_{i,A}\widetilde{\otimes} \sigma_{i,B})$ such that $\sigma_{AB} \sqsubseteq_{{}_{{{\widetilde{S}}}_{AB}}}  
(\alpha_A \widetilde{\otimes}\bot_{{}_{{ \mathfrak{S}}_{B}}})\sqcap_{{}_{{{\widetilde{S}}}_{AB}}} (\bot_{{}_{{ \mathfrak{S}}_{A}}}\widetilde{\otimes} \alpha_B)$. As a first case, we may have obviously $\sigma_{AB}=\bot_{{}_{{ \mathfrak{S}}_{A}}} \widetilde{\otimes}\bot_{{}_{{ \mathfrak{S}}_{B}}}$. If however $\sigma_{AB}\not= \bot_{{}_{{ \mathfrak{S}}_{A}}} \widetilde{\otimes}\bot_{{}_{{ \mathfrak{S}}_{B}}}$, the previous result implies that there exist $\alpha'_A\in { \mathcal{A}}_{{ \mathfrak{S}}_{A}}$ and $\alpha'_B\in { \mathcal{A}}_{{ \mathfrak{S}}_{B}}$ such that $((\alpha'_A \widetilde{\otimes}\bot_{{}_{{ \mathfrak{S}}_{B}}})\sqcap_{{}_{{{\widetilde{S}}}_{AB}}} (\bot_{{}_{{ \mathfrak{S}}_{A}}}\widetilde{\otimes} \alpha'_B)) \sqsubseteq_{{}_{{{\widetilde{S}}}_{AB}}}  \sigma_{AB}$.  Using once again the expansion (\ref{developmentetildeordersimplify}), we deduce immediately that $\alpha_A=\alpha'_A$ and $\alpha_B=\alpha'_B$. As a result, we obtain
\begin{eqnarray}
\sigma_{AB} \sqsubseteq_{{}_{{{\widetilde{S}}}_{AB}}}  
(\alpha_A \widetilde{\otimes}\bot_{{}_{{ \mathfrak{S}}_{B}}})\sqcap_{{}_{{{\widetilde{S}}}_{AB}}} (\bot_{{}_{{ \mathfrak{S}}_{A}}}\widetilde{\otimes} \alpha_B) &\Rightarrow & \left( \sigma_{AB}=\bot_{{}_{{ \mathfrak{S}}_{A}}} \widetilde{\otimes}\bot_{{}_{{ \mathfrak{S}}_{B}}} \;\;\textit{\rm or}\;\; \sigma_{AB}=(\alpha_A \widetilde{\otimes}\bot_{{}_{{ \mathfrak{S}}_{B}}})\sqcap_{{}_{{{\widetilde{S}}}_{AB}}} (\bot_{{}_{{ \mathfrak{S}}_{A}}}\widetilde{\otimes} \alpha_B) \right).\;\;\;\;\;\;\;\;\;\;\;\;\;\;
\end{eqnarray}
As a second conclusion, we then obtain $\bot_{{}_{{{\widetilde{S}}}_{AB}}}\sqcoversubset_{{}_{{{\widetilde{S}}}_{AB}}}  
(\alpha_A \widetilde{\otimes}\bot_{{}_{{ \mathfrak{S}}_{B}}})\sqcap_{{}_{{{\widetilde{S}}}_{AB}}} (\bot_{{}_{{ \mathfrak{S}}_{A}}}\widetilde{\otimes} \alpha_B)$.
\end{proof} 

\begin{remark}
Let us consider the following orthocomplemented space of states
\begin{eqnarray}
&&\;\;\;\;\; S_4 :=\begin{tikzpicture}  \node (bot) at (0,0) {$\bot$};  \node (1) at (-2,1) {$\alpha_1$};  \node (2) at (-1,1) {$\alpha_2$};  \node (3) at (1,1) {$\alpha^\star_1$};  \node (4) at (2,1) {$\alpha^\star_2$};  \draw (bot) -- (1)  (bot) -- (2)    (bot) -- (3) (bot) -- (4) ; \end{tikzpicture} 
\end{eqnarray}
It is easy to check that ${\widetilde{S}}':=S_4 \widetilde{\otimes} S_4$ is NOT orthocomplemented. Indeed, from the result above, the single candidate for $(\alpha_1\widetilde{\otimes}\alpha_1)^\star$ is obviously $\alpha_1^\star \widetilde{\otimes} \bot \sqcap_{{}_{{\widetilde{S}}'}} \bot \widetilde{\otimes} \alpha_1^\star$. However, we check immediately, using the expansion (\ref{developmentetildeordersimplify}),  that the two elements $\alpha_1^\star \widetilde{\otimes} \bot \sqcap_{{}_{{\widetilde{S}}'}} \bot \widetilde{\otimes} \alpha_1^\star$ and $\alpha_1\widetilde{\otimes}\alpha_1 \sqcap_{{}_{{\widetilde{S}}'}} \alpha_2\widetilde{\otimes}\alpha_2$ have no common upper-bound : this point contradicts the condition (\ref{starcomplement}) for the definition of $\star$ on ${\widetilde{S}}'$.\\
\end{remark}

\begin{remark}
According to \cite[Axiom 9 and Lemma 40]{Buffenoir2021}, we can introduce the following notion : the space of states ${ \mathfrak{S}}$ is said to be {\em irreducible} iff
\begin{eqnarray} \forall \sigma_1,\sigma_2\in { \mathfrak{S}}^{{}^{pure}},&&\{\sigma_1,\sigma_2\} \varsubsetneq\underline{ \sigma_1\sqcap_{{}_{ \mathfrak{S}}}\sigma_2}_{{}_{ \mathfrak{S}}}.\end{eqnarray} 
Then, it is important to remark that, even if ${ \mathfrak{S}}_A$ and ${ \mathfrak{S}}_B$ are both irreducible, the tensor product ${\widetilde{S}}_{AB}={ \mathfrak{S}}_A \widetilde{\otimes} { \mathfrak{S}}_B$ appears to be NEVER irreducible. 
Indeed,  from the expansion (\ref{developmentetildeordersimplify}), we deduce
\begin{eqnarray}
\forall \sigma_1,\sigma'_1\in { \mathfrak{S}}_A^{{}^{pure}},\forall \sigma_2,\sigma'_2\in { \mathfrak{S}}_B^{{}^{pure}}\;\vert \sigma_1\not=\sigma'_1,\sigma_2\not=\sigma'_2, && \underline{\sigma_1 \widetilde{\otimes} \sigma_2 \sqcap_{{}_{{\widetilde{S}}_{AB}}} \sigma'_1 \widetilde{\otimes} \sigma'_2}_{{}_{{\widetilde{S}}_{AB}}}=\{\sigma_1 \widetilde{\otimes} \sigma_2, \sigma'_1 \widetilde{\otimes} \sigma'_2\}.\;\;\;\;\;\;\;\;\;\;\;\;\;\;
\end{eqnarray}
\end{remark}

\subsection{The star tensor product}\label{subsectionstartensor}

In the present subsection, we will assume the space of states ${ \mathfrak{S}}$ (this remark concerns ${ \mathfrak{S}}_A$ and ${ \mathfrak{S}}_B$ in the following) to be atomistic and equipped with a star map denoted $\star$ (i.e.  a map from ${ \mathfrak{S}},\smallsetminus \{\bot_{{}_{ \mathfrak{S}}}\}$ to itself satisfying (\ref{involutive})(\ref{orderreversing})(\ref{inconsistent})(\ref{staratom})). We will try to build a tensor product of ${ \mathfrak{S}}_A$ and ${ \mathfrak{S}}_B$ denoted ${ \mathfrak{S}}_A\overline{\otimes} { \mathfrak{S}}_B$ by exploiting this extra-structure. 

\begin{defin}
We define $\widehat{S}_{AB}$ to be the set of bimorphisms from ${ \mathfrak{S}}_A\times { \mathfrak{S}}_B$ to ${ \mathfrak{B}}$. Equipped with the pointwise poset structure, this is an Inf semi-lattice.
\end{defin}

\begin{defin}
Using the maps $\langle\cdot,\cdot\rangle_A$ and $\langle\cdot,\cdot\rangle_B$, respectively associated to the star maps on ${ \mathfrak{S}}_A$ and ${ \mathfrak{S}}_B$ by (\ref{explicitomega1}), we can define the following bimorphisms denoted $\alpha_{A}\overline{\otimes}\beta_{B}$ for any $(\alpha_{A},\beta_{B})\in { \mathfrak{S}}^{{}^{pure}}_A\times { \mathfrak{S}}^{{}^{pure}}_B$ :
\begin{eqnarray}
&& \begin{array}{rcrcl}
\alpha_{A}\overline{\otimes}\beta_{B} &:& { \mathfrak{S}}_A\times { \mathfrak{S}}_B & \longrightarrow & { \mathfrak{B}}\\
& & (\sigma_A,\sigma_B) & \mapsto & \langle \alpha_{A},\sigma_A\rangle_A\bullet  \langle \beta_{B},\sigma_B\rangle_B.
\end{array}
\end{eqnarray} 
The tensor product ${ \mathfrak{S}}_A\overline{\otimes} { \mathfrak{S}}_B$ is defined as a sub Inf semi-lattice of $\widehat{S}_{AB}$ : this is the Inf semi-lattice generated by the bimorphisms $\alpha_{A}\overline{\otimes}\beta_{B}$ associated to any $(\alpha_{A},\beta_{B})\in { \mathfrak{S}}^{{}^{pure}}_A\times { \mathfrak{S}}^{{}^{pure}}_B$. \\
We have obviously, for any $\{\,(\alpha_{i,A},\beta_{i,B})\;\vert\; i\in I\,\} \subseteq { \mathfrak{S}}^{{}^{pure}}_A\times { \mathfrak{S}}^{{}^{pure}}_B$
\begin{eqnarray}
\forall (\sigma_A,\sigma_B)\in { \mathfrak{S}}_A\times { \mathfrak{S}}_B,\;\;\;\;
(\bigsqcap{}^{{}^{\overline{S}_{AB}}}_{i\in I} \alpha_{i,A}\overline{\otimes}\beta_{i,B})(\sigma_A,\sigma_B) & := & \bigwedge{}_{i\in I}\;\langle \alpha_{i,A},\sigma_A\rangle_A\bullet  \langle \beta_{i,B},\sigma_B\rangle_B.\;\;\;\;\;\;\;\;\label{explicitsigmapre}
\end{eqnarray}
$\overline{S}_{AB}:={ \mathfrak{S}}_A\overline{\otimes} { \mathfrak{S}}_B$ is chosen as our space of states.\\
For any $(\sigma_A,\sigma_B)\in { \mathfrak{S}}_A\times { \mathfrak{S}}_B$, we will denote
\begin{eqnarray}
\sigma_A\overline{\otimes} \sigma_B & := & \bigsqcap{}^{{}^{\overline{S}_{AB}}}_{\alpha\in \underline{\sigma_A},\beta\in \underline{\sigma_B}}\alpha \overline{\otimes} \beta.\label{explicitsigma}
\end{eqnarray}
\end{defin}

\begin{defin}\label{defindoublelangle}
We define a bimorphism from $\overline{S}_{AB}\times \overline{S}_{AB}$ to ${ \mathfrak{B}}$, denoted $\langle\langle \cdot,\cdot\rangle\rangle$, by
\begin{eqnarray}
\begin{array}{rcccc}
\langle\langle \cdot,\cdot\rangle\rangle & : & \overline{S}_{AB}\times \overline{S}_{AB} & \longrightarrow & { \mathfrak{B}}\\
 & & (\bigsqcap{}^{{}^{\overline{S}_{AB}}}_{i\in I} \alpha_{i,A}\overline{\otimes}\beta_{i,B},
 \bigsqcap{}^{{}^{\overline{S}_{AB}}}_{j\in J} \alpha'_{j,A}\overline{\otimes}\beta'_{j,B}) & \mapsto & \bigwedge{}_{i\in I}\bigwedge{}_{j\in J} \;\langle\alpha_{i,A},\alpha'_{j,A}\rangle_A \bullet  \langle \beta_{i,B},\beta'_{j,B}\rangle_B.\label{expressiondoublelangle}
\end{array}
\end{eqnarray}
As a consequence of the symmetry of $\langle \cdot, \cdot\rangle_A$ and $\langle \cdot, \cdot\rangle_B$ mentioned in (\ref{symmetrylangle}), we note the following property
\begin{eqnarray}
\forall (\sigma_{AB},\sigma_{AB}')\in \overline{S}_{AB} \times \overline{S}_{AB}, && \langle\langle \sigma_{AB},\sigma_{AB}'\rangle\rangle = \langle\langle \sigma_{AB}',\sigma_{AB}\rangle\rangle \label{symmetrydoublelangle}
\end{eqnarray}
\end{defin}

\begin{defin}
The space of effects $\overline{E}_{AB}$ is chosen to be given by ${ \mathfrak{S}}_A\overline{\otimes} { \mathfrak{S}}_B$ as well.  The evaluation map $\epsilon^{\overline{S}_{AB}}$ is defined by
\begin{eqnarray}
\forall \sigma_{AB}\in \overline{S}_{AB},\forall { \mathfrak{l}}_{AB}\in \overline{E}_{AB},&& 
\epsilon^{\overline{S}_{AB}}_{{ \mathfrak{l}}_{AB}}(\sigma_{AB}):=\langle\langle { \mathfrak{l}}_{AB}, \sigma_{AB}\rangle\rangle.
\end{eqnarray}
\end{defin}

\begin{lemme}
The axiom of bi-extensionality is verified.
\end{lemme}
\begin{proof}
By symmetry, it is the same to check extensionality and/or separation.  Now, for any ${ \mathfrak{l}}_{AB}$ and ${ \mathfrak{l}}_{AB}'$ in $\overline{S}_{AB}$, we have
\begin{eqnarray}
(\forall \sigma_{AB}\in \overline{S}_{AB},\;\; \epsilon^{\overline{S}_{AB}}_{{ \mathfrak{l}}_{AB}}(\sigma_{AB})=\epsilon^{\overline{S}_{AB}}_{{ \mathfrak{l}}'_{AB}}(\sigma_{AB})) & \Rightarrow & 
(\,\forall (\sigma_{A},\sigma_B)\in { \mathfrak{S}}_A\times { \mathfrak{S}}_B,\;\; \epsilon^{\overline{S}_{AB}}_{{ \mathfrak{l}}_{AB}}(\sigma_A\overline{\otimes} \sigma_B)=\epsilon^{\overline{S}_{AB}}_{{ \mathfrak{l}}'_{AB}}(\sigma_A\overline{\otimes} \sigma_B)\,)\nonumber\\
& \Rightarrow & 
(\,\forall (\sigma_{A},\sigma_B)\in { \mathfrak{S}}_A\times { \mathfrak{S}}_B,\;\; { \mathfrak{l}}_{AB}(\sigma_{A},\sigma_B)={ \mathfrak{l}}'_{AB}(\sigma_{A},\sigma_B)\,)\nonumber\\
& \Rightarrow & (\, { \mathfrak{l}}_{AB}={ \mathfrak{l}}_{AB}'\,).
\end{eqnarray}
\end{proof}

\begin{lemme}
Axioms {\bf (A1)} and {\bf (A3)} are satisfied by construction. {\bf (B1)} and {\bf (B2)} are then satisfied as well.
\end{lemme}

\begin{lemme}
Axiom {\bf (A2)} is satisfied.
\end{lemme}
\begin{proof}
The map associating to any $(\sigma_A,\sigma_B)\in { \mathfrak{S}}_A\times { \mathfrak{S}}_B$ the element $\bot$ is an element of $\overline{S}_{AB}$. It is indeed given by $\bot_A\overline{\otimes}\bot_B$.
\end{proof}

\begin{lemme}
Axioms {\bf (A4)} and {\bf (A5)} are satisfied by construction. We have
\begin{eqnarray}
\overline{S}_{AB}\!\!\!\!\!\!{}^{{}^{pure}} &= & \{\, \alpha \overline{\otimes}\beta \;\vert\; \alpha\in { \mathfrak{S}}_A^{{}^{pure}},\beta\in { \mathfrak{S}}_B^{{}^{pure}}\,\}.
\end{eqnarray}
\end{lemme}

\begin{lemme}
Axioms {\bf (B3)(B3')(B3'')} and {\bf (B4)(B4')(B4'')} are satisfied.
\end{lemme}
\begin{proof}
Axioms {\bf (B3)} and {\bf (B4)} are checked according to the mapping (\ref{explicitsigma}). 
Axioms {\bf (B3')(B3'')} and {\bf (B4')(B4'')} are checked directly from the expressions  (\ref{explicitsigma}) and (\ref{explicitsigmapre}) using the properties of $\langle\cdot,\cdot\rangle$. 
\end{proof}

\begin{lemme}
Axiom {\bf (B5)} is satisfied.
\end{lemme}
\begin{proof}
For any $\sigma_{AB}$ and $\sigma_{AB}'$ in $\overline{S}_{AB}$, we have
\begin{eqnarray}
(\forall { \mathfrak{l}}_{AB}\in \overline{E}_{AB},\;\; \epsilon^{\overline{S}_{AB}}_{{ \mathfrak{l}}_{A}\overline{\otimes}{ \mathfrak{l}}_{B}}(\sigma_{AB})=\epsilon^{\overline{S}_{AB}}_{{ \mathfrak{l}}_{A}\overline{\otimes}{ \mathfrak{l}}_{B}}(\sigma'_{AB})) & \Leftrightarrow & 
(\,\forall ({ \mathfrak{l}}_{A},{ \mathfrak{l}}_{B})\in { \mathfrak{S}}_A\times { \mathfrak{S}}_B,\;\; \sigma_{AB}({ \mathfrak{l}}_{A},{ \mathfrak{l}}_{B})=\sigma_{AB}({ \mathfrak{l}}_{A},{ \mathfrak{l}}_{B})
\,)\nonumber\\
& \Leftrightarrow & 
 (\, \sigma_{AB}=\sigma_{AB}'\,).
\end{eqnarray}
\end{proof}

\subsection{Properties of the star tensor product}\label{subsectionpropertiesstartensor}

Our first task is to check the following result on the atomicity of $\overline{S}_{AB}$ :
\begin{theoreme}
$\overline{S}_{AB}$ is atomic, the set of atoms of $\overline{S}_{AB}$ denoted $At(\overline{S}_{AB})$ satisfies
\begin{eqnarray}
At(\overline{S}_{AB}) &=& \{\, (\sigma_A \overline{\otimes} \bot_{{}_{ \mathfrak{S}_B}} \sqcap_{{}_{\overline{S}_{AB}}} \bot_{{}_{ \mathfrak{S}_A}}\overline{\otimes}\sigma_B)\;\vert\; \sigma_A\in At({ \mathfrak{S}_A}),\sigma_B\in At({ \mathfrak{S}_B})\,\}.
\end{eqnarray}
\end{theoreme}
\begin{proof}
First of all, we note that $(\sigma_A \overline{\otimes} \bot_{{}_{ \mathfrak{S}_B}} \sqcap_{{}_{\overline{S}_{AB}}} \bot_{{}_{ \mathfrak{S}_A}}\overline{\otimes}\sigma_B) \sqsupset_{{}_{\overline{S}_{AB}}} \bot_{{}_{\overline{S}_{AB}}}$ for any $\sigma_A\in At({ \mathfrak{S}_A}),\sigma_B\in At({ \mathfrak{S}_B})$.  Indeed, it suffices to choose $\alpha_A= \sigma^\star_A \in { \mathfrak{S}}_A^{{}^{pure}}$ and $\alpha_B= \sigma^\star_B \in { \mathfrak{S}}_B^{{}^{pure}}$ to observe that 
\begin{eqnarray}
(\sigma_A \overline{\otimes} \bot_{{}_{ \mathfrak{S}_B}} \sqcap_{{}_{\overline{S}_{AB}}} \bot_{{}_{ \mathfrak{S}_A}}\overline{\otimes}\sigma_B)(\alpha_A,\alpha_B)=\textit{\bf N}\bullet \bot \wedge \bot\bullet\textit{\bf N}=\textit{\bf N} &>& \bot = \bot_{{}_{\overline{S}_{AB}}}(\alpha_A,\alpha_B).
\end{eqnarray}
Here we have used the expression (\ref{explicitsigmapre}) with (\ref{explicitomega1}).\\
Secondly, we note easily that 
\begin{eqnarray}
\forall \sigma_{AB}\in {\overline{S}_{AB}},\exists \sigma_A\in At({ \mathfrak{S}_A}),\sigma_B\in At({ \mathfrak{S}_B}) & \vert &\sigma_{AB} \sqsupseteq_{{}_{\overline{S}_{AB}}}
 (\sigma_A \overline{\otimes} \bot_{{}_{ \mathfrak{S}_B}} \sqcap_{{}_{\overline{S}_{AB}}} \bot_{{}_{ \mathfrak{S}_A}}\overline{\otimes}\sigma_B).
\end{eqnarray} 
It suffices to choose $\sigma_A$ and $\sigma_B$ such that $\sigma_{AB}(\sigma_A^\star,\sigma_B^\star)=\textit{\bf N}$.\\
Endly, let us consider $(\sigma_A \overline{\otimes} \bot_{{}_{ \mathfrak{S}_B}} \sqcap_{{}_{\overline{S}_{AB}}} \bot_{{}_{ \mathfrak{S}_A}}\overline{\otimes}\sigma_B)
\sqsubseteq_{{}_{\overline{S}_{AB}}}
(\sigma'_A \overline{\otimes} \bot_{{}_{ \mathfrak{S}_B}} \sqcap_{{}_{\overline{S}_{AB}}} \bot_{{}_{ \mathfrak{S}_A}}\overline{\otimes}\sigma'_B)$ and let us apply this relation on the couple $(\sigma_A^\star,\sigma_B^\star)$, we deduce that $\sigma'_A=\sigma_A$ and $\sigma'_B=\sigma_B$. \\
As a final conclusion, we deduce that $\overline{S}_{AB}$ is atomic and that the elements $(\sigma_A \overline{\otimes} \bot_{{}_{ \mathfrak{S}_B}} \sqcap_{{}_{\overline{S}_{AB}}} \bot_{{}_{ \mathfrak{S}_A}}\overline{\otimes}\sigma_B)$ (where $\sigma_A\in At({ \mathfrak{S}_A})$ and $\sigma_B\in At({ \mathfrak{S}_B})$) are the atoms.
\end{proof}

Our second task is to define a star map on $\overline{S}_{AB}$.

\begin{defin}
We can define a binary relation denoted $\underline{\perp}$ on $\overline{S}_{AB}$ as follows :
\begin{eqnarray}
\forall \sigma_{AB},\sigma_{AB}'\in \overline{S}_{AB},\;\;\;\; \sigma_{AB}\; \underline{\perp}\; \sigma_{AB}' & :\Leftrightarrow & \langle\langle \sigma_{AB},\sigma_{AB}'\rangle\rangle =\textit{\bf N}.
\end{eqnarray}
\end{defin}

\begin{lemme}
The binary relation $\underline{\perp}$ is an orthogonality relation. In other words, it is symmetric and irreflexive.
\end{lemme}
\begin{proof}
The symmetry of $\underline{\perp}$ is due to the symmetry of $\langle\langle \cdot,\cdot\rangle\rangle$ mentioned in (\ref{symmetrydoublelangle}).\\
The irreflexivity is checked directly on the expression (\ref{expressiondoublelangle}) (we check in fact easily $\sigma_{AB}\; \not\!\!\underline{\perp}\; \sigma_{AB}$).
\end{proof}

\begin{defin}
We define as usual a star operation associated to the orthogonality relation :
\begin{eqnarray}
\forall \sigma_{AB}\in \overline{S}_{AB}, \;\;\;\; \sigma_{AB}^\star &:= & \bigsqcap{}^{{}^{\overline{S}_{AB}}} \{\, \sigma_{AB}'\;\vert\; \sigma_{AB}'\; \underline{\perp}\; \sigma_{AB}\,\}.
\end{eqnarray}
It is involutive and order-reversing.
\end{defin}

\begin{lemme}
We have explicitly
\begin{eqnarray}
\forall \alpha\in { \mathfrak{S}}^{{}^{pure}}_A, \forall \beta\in { \mathfrak{S}}^{{}^{pure}}_B,\;\;\;\; (\alpha \overline{\otimes}\beta)^\star &=& \alpha^\star \overline{\otimes} \bot_{{}_{ \mathfrak{S}_B}} \sqcap_{{}_{\overline{S}_{AB}}} \bot_{{}_{ \mathfrak{S}_A}} \overline{\otimes}\beta^\star.
\end{eqnarray}
\end{lemme}
\begin{proof}
We note the following facts.\\
First, 
\begin{eqnarray}
 (\alpha^\star \overline{\otimes} \bot_{{}_{ \mathfrak{S}_B}} \sqcap_{{}_{\overline{S}_{AB}}} \bot_{{}_{ \mathfrak{S}_A}} \overline{\otimes}\beta^\star) &\in & (\alpha \overline{\otimes}\beta)^{\underline{\perp}}.
\end{eqnarray}
Indeed,  $\langle\langle (\alpha^\star \overline{\otimes} \bot_{{}_{ \mathfrak{S}_B}} \sqcap_{{}_{\overline{S}_{AB}}} \bot_{{}_{ \mathfrak{S}_A}} \overline{\otimes}\beta^\star)\;,\; \alpha \overline{\otimes}\beta\rangle\rangle = \langle\alpha,\alpha^\star\rangle_A\bullet \bot \wedge \bot \bullet \langle\beta,\beta^\star\rangle_B=\textit{\bf N}\bullet \bot \wedge \bot \bullet \textit{\bf N}=\textit{\bf N}$.\\
Second, we have obviously
\begin{eqnarray}
 \bot_{{}_{\overline{S}_{AB}}} &\notin & (\alpha \overline{\otimes}\beta)^{\underline{\perp}}.
\end{eqnarray}
Third, $(\alpha^\star \overline{\otimes} \bot_{{}_{ \mathfrak{S}_B}} \sqcap_{{}_{\overline{S}_{AB}}} \bot_{{}_{ \mathfrak{S}_A}} \overline{\otimes}\beta^\star)$ is an atom.\\
Fourth, $(\alpha \overline{\otimes}\beta)^{\underline{\perp}}$ is a filter.\\
This concludes the proof.
\end{proof}

\begin{lemme}
If the space of states ${ \mathfrak{S}}_B$ is orthocomplemented and the space of states ${ \mathfrak{S}}_A$ is orthogonal,  then the spaces of states ${ \mathfrak{S}}_B\overline{\otimes} { \mathfrak{S}}_A$ and ${ \mathfrak{S}}_A \overline{\otimes} { \mathfrak{S}}_B$ are orthocomplemented. \\
The star map defined on ${\overline{S}}_{AB}:={ \mathfrak{S}}_A\overline{\otimes} { \mathfrak{S}}_B$ will be denoted $\star$ as well. This map is built according to 
\begin{eqnarray}
 ({\alpha} \widetilde{\otimes}{\beta})^\star:= {\alpha}^\star \widetilde{\otimes} \bot_{{}_{ \mathfrak{S}_B}} \sqcap_{{}_{{\overline{S}}}} \bot_{{}_{ \mathfrak{S}_A}} \overline{\otimes} {\beta}^\star, && \forall {\alpha}\overline{\otimes} {\beta}\in {\overline{S}}{}^{\;{}^{pure}}_{AB},\\ 
(\bigsqcap{}^{{}^{{\overline{S}}}} U )^\star := \bigsqcap{}^{{}^{{\overline{S}}}}_{{}_{\alpha \in \bigcap_{\sigma \in U} \underline{\sigma^\star}_{{}_{{\overline{S}}}}}} \alpha,
&& \forall U \subseteq  {\overline{S}}{}^{\;{}^{pure}}_{AB}.
\end{eqnarray}
We have the same formulas for ${ \mathfrak{S}}_B \overline{\otimes} { \mathfrak{S}}_A$.
\end{lemme}
\begin{proof}
The main point to check is the property (\ref{starcomplement}) for the star map on ${\overline{S}}_{AB}$. \\
To begin, we have to check that $({\alpha} \overline{\otimes}{\beta}) \not\sqsupseteq_{{}_{{\overline{S}_{AB}}}}({\alpha}^\star \overline{\otimes} \bot_{{}_{ \mathfrak{S}_B}}\!\! \sqcap_{{}_{{\overline{S}_{AB}}}}\!\!\! \bot_{{}_{ \mathfrak{S}_A}} \overline{\otimes} {\beta}^\star)$ 
and that 
$u:={\alpha}\overline{\otimes} {\beta} \sqcap_{{}_{{\overline{S}_{AB}}}} {\alpha}\overline{\otimes} {\beta}'$,  $v:={\alpha}\overline{\otimes} {\beta} \sqcap_{{}_{{\overline{S}_{AB}}}} {\alpha}'\overline{\otimes} {\beta}''$ and $w:=  {\alpha}^\star \overline{\otimes} \bot_{{}_{ \mathfrak{S}_B}}\!\! \sqcap_{{}_{{\overline{S}_{AB}}}} \!\!\! \bot_{{}_{ \mathfrak{S}_A}} \overline{\otimes} {\beta}^\star$ satisfy $\widehat{uw}{}^{{}^{{\overline{S}_{AB}}}}$ and $\widehat{vw}{}^{{}^{{\overline{S}_{AB}}}}$ for any ${\alpha},z'_A\in { \mathfrak{S}}_A$ and ${\beta},{\beta}',{\beta}''\in { \mathfrak{S}}_B$ with ${\beta}\not= {\beta}'$ and ${\alpha}\not= \alpha'$. \\ 
We check the first property by recalling that the map $\sigma_{AB}\mapsto \langle\langle \sigma_{AB}\;,\; {\alpha} \overline{\otimes}{\beta}\rangle\rangle$ is order-preserving and by observing simply that $\langle\langle {\alpha} \overline{\otimes}{\beta}\;,\; {\alpha} \overline{\otimes}{\beta}\rangle\rangle =\textit{\bf Y} \ngeqslant \textit{\bf N}= \langle\langle ({\alpha} \overline{\otimes}{\beta}) \;,\; ({\alpha}^\star \overline{\otimes} \bot_{{}_{ \mathfrak{S}_B}}\!\! \sqcap_{{}_{{\overline{S}_{AB}}}}\!\!\! \bot_{{}_{ \mathfrak{S}_A}} \overline{\otimes} {\beta}^\star)\rangle\rangle$. \\
Secondly, $\widehat{uw}{}^{{}^{{\overline{S}_{AB}}}}$ is obtained quite easily as follows. We have (i) ${\alpha}\overline{\otimes} {\beta} \!\sqcap_{{}_{{\overline{S}_{AB}}}} \!\!\!{\alpha}\overline{\otimes} {\beta}'=
{\alpha}\overline{\otimes} ({\beta} \sqcap_{{}_{{\mathfrak{S}}}} {\beta}')$ and (ii) $({\beta} \sqcap_{{}_{{\mathfrak{S}}}} {\beta}')$ and ${\beta}^\star$ have a common upper-bound denoted $z$ (because of the property (\ref{starcomplement}) applied to ${ \mathfrak{S}}_B$). From (i) and (ii) we deduce that $u$ and $w$ have ${\alpha}\overline{\otimes}z$ as common upper-bound. \\ $\widehat{vw}{}^{{}^{{\overline{S}_{AB}}}}$ is obtained because ${\alpha}' \overline{\otimes}{\beta}''$ is a common upper-bound of $v$ and $w$ (this point is guarantied by the orthogonality of ${ \mathfrak{S}}_A$ and the fact that ${\alpha}'\not={\alpha}$).\\
The rest of the proof follows the same analysis.
\end{proof}

\begin{lemme}
If ${ \mathfrak{S}}_A$ and ${ \mathfrak{S}}_B$ are orthocomplemented, but neither ${ \mathfrak{S}}_A$ nor ${ \mathfrak{S}}_B$ are orthogonal, then $\overline{S}_{AB}$ is NOT orthocomplemented.
\end{lemme}
\begin{proof}
In order to check the orthocomplemented character of $\overline{S}_{AB}$, we should check the validity of the relation
\begin{eqnarray}
(\alpha^\star \overline{\otimes} \bot_{{}_{ \mathfrak{S}_B}} \!\!\sqcap_{{}_{\overline{S}_{AB}}}\!\!\! \bot_{{}_{ \mathfrak{S}_A}} \overline{\otimes}\beta^\star) &\widecheck{\bowtie} & (\alpha \overline{\otimes}\beta)\label{checkbowtie}
\end{eqnarray}
For any $\alpha'\in { \mathfrak{S}}^{{}^{pure}}_A\smallsetminus \{\alpha\}$ and $\beta' \in { \mathfrak{S}}^{{}^{pure}}_B\smallsetminus\{\beta\}$, we observe that $(\alpha^\star \overline{\otimes} \bot_{{}_{ \mathfrak{S}_B}} \sqcap_{{}_{\overline{S}_{AB}}} \bot_{{}_{ \mathfrak{S}_A}} \overline{\otimes}\beta^\star)$ and $(\alpha \overline{\otimes}\beta \sqcap_{{}_{\overline{S}_{AB}}} \alpha' \overline{\otimes}\beta')$ have no common upper-bound in $\overline{S}_{AB}$, which invalidates the relation (\ref{checkbowtie}). To check this last point, we just note that 
\begin{eqnarray}
\underline{(\alpha \overline{\otimes}\beta \sqcap_{{}_{\overline{S}_{AB}}}\!\! \alpha' \overline{\otimes}\beta')}_{{}_{\overline{S}_{AB}}} &= & \{\, \alpha \overline{\otimes}\beta\;,\; \alpha' \overline{\otimes}\beta'\,\}.\label{irreducibleSbar}
\end{eqnarray}
The property (\ref{irreducibleSbar}) is easily proved using the expression of $(\alpha \overline{\otimes}\beta \sqcap_{{}_{\overline{S}_{AB}}}\!\! \alpha' \overline{\otimes}\beta')(\sigma_A,\sigma_B)$ inherited from (\ref{explicitsigmapre}). Indeed, it suffices to choose $(\sigma_A,\sigma_B):=(\alpha^\star,\beta'{}^\star)$ and $(\sigma_A,\sigma_B):=(\alpha'{}^\star,\beta^\star)$ in the inequality $(\alpha \overline{\otimes}\beta \sqcap_{{}_{\overline{S}_{AB}}} \!\!\! \alpha' \overline{\otimes}\beta')(\sigma_A,\sigma_B) \leq (\alpha'' \overline{\otimes}\beta'')(\sigma_A,\sigma_B)$ to conclude that $(\alpha'',\beta'')=(\alpha,\beta)$ or $(\alpha'',\beta'')=(\alpha',\beta')$.\\
Endly, we note that $\alpha' \overline{\otimes}\beta'$ satisfies necessarily $(\alpha' \overline{\otimes}\beta') \sqsupseteq_{{}_{\overline{S}_{AB}}} (\alpha^\star \overline{\otimes} \bot_{{}_{ \mathfrak{S}_B}}\!\! \sqcap_{{}_{\overline{S}_{AB}}} \!\!\!\bot_{{}_{ \mathfrak{S}_A}} \overline{\otimes}\beta^\star)$ for any $\alpha'\in { \mathfrak{S}}^{{}^{pure}}_A\smallsetminus \{\alpha\}$ and $\beta' \in { \mathfrak{S}}^{{}^{pure}}_B\smallsetminus\{\beta\}$ 
if and only if ${ \mathfrak{S}}_A$ is orthogonal or ${ \mathfrak{S}}_B$ is orthogonal.
\end{proof}

\begin{remark}
Previous results strongly suggest to study the existence of a suitable completion of $\overline{S}_{AB}$ in $\widehat{S}_{AB}$ (we recall that $\widehat{S}_{AB}$ is the Inf semi-lattice constituted by bimorphisms from ${ \mathfrak{S}}_A\times { \mathfrak{S}}_B$ to ${ \mathfrak{B}}$) : a completion in which the orthocomplementation property (\ref{checkbowtie}) would be restored.  To sketch what we have in mind, let us consider any $\alpha,\alpha'\in { \mathfrak{S}}_A^{{}^{pure}}$ and $\beta,\beta'\in { \mathfrak{S}}_B^{{}^{pure}}$ with $\alpha'\not\in \{\alpha\}\cup\underline{\alpha^\star}_{{}_{ \mathfrak{S}_A}}$ and $\beta'\not\in\{\beta\}\cup \underline{\beta^\star}_{{}_{ \mathfrak{S}_B}}$ (during this remark we will suppose that we are able to choose $\alpha$ and $\alpha'$ such that $\alpha \sqcap_{{}_{{\mathfrak{S}}_A}}\alpha'\not=\bot_{{}_{{\mathfrak{S}}_A}}$; the trivial case where ${\mathfrak{S}}_A$ is a flat domain can be treated separately).  
We have noted that $(\alpha \overline{\otimes}\beta \sqcap_{{}_{\overline{S}_{AB}}}\!\!\!\! \alpha' \overline{\otimes}\beta')$ and $(\alpha \overline{\otimes}\beta)^\star=(\alpha^\star \overline{\otimes} \bot_{{}_{ \mathfrak{S}_B}} \sqcap_{{}_{\overline{S}_{AB}}}\!\!\!\! \bot_{{}_{ \mathfrak{S}_A}} \overline{\otimes}\beta^\star)$ have no common upper-bound in $\overline{S}_{AB}$.  This is in fact the precise point where the orthocomplementation property (\ref{checkbowtie}) appears to be "broken". We could however hope to define the supremum $\Delta:=(\alpha \overline{\otimes}\beta \sqcap_{{}_{\overline{S}_{AB}}}\!\!\!\! \alpha' \overline{\otimes}\beta')\sqcup_{{}_{\widehat{S}_{AB}}}\!\!\!\! (\alpha \overline{\otimes}\beta)^\star$ as an element of $\widehat{S}_{AB}$.  Indeed, we can observe that the two properties $(\alpha^\star \overline{\otimes} \bot_{{}_{ \mathfrak{S}_B}} \sqcap_{{}_{\overline{S}_{AB}}}\!\!\!\! \bot_{{}_{ \mathfrak{S}_A}} \overline{\otimes}\beta^\star)(\sigma_A,\sigma_B) \in \{ \textit{\bf N},\bot\}$ and  $(\alpha \overline{\otimes}\beta \sqcap_{{}_{\overline{S}_{AB}}}\!\!\!\! \alpha' \overline{\otimes}\beta')(\sigma_A,\sigma_B) \in \{ \textit{\bf N},\bot\}$ for any $(\sigma_A,\sigma_B)\in { \mathfrak{S}}_A\times { \mathfrak{S}}_B$ allow us to define $\Delta (\sigma_A,\sigma_B)$ as an element of ${ \mathfrak{B}}$ for any $(\sigma_A,\sigma_B)\in { \mathfrak{S}}_A\times { \mathfrak{S}}_B$. Sadly, we can easily invalidate the bimorphic property. Indeed, we remark that $(\alpha^\star \overline{\otimes} \bot_{{}_{ \mathfrak{S}_B}} \sqcap_{{}_{\overline{S}_{AB}}}\!\!\!\! \bot_{{}_{ \mathfrak{S}_A}} \overline{\otimes}\beta^\star)(\alpha,\beta) = \textit{\bf N}$ and $(\alpha \overline{\otimes}\beta \sqcap_{{}_{\overline{S}_{AB}}}\!\!\!\! \alpha' \overline{\otimes}\beta')(\alpha^\star \sqcup_{{}_{{ \mathfrak{S}}_A}}\!\!\alpha'{}^\star,\beta)=\textit{\bf N}$, but  (1) $(\alpha^\star \overline{\otimes} \bot_{{}_{ \mathfrak{S}_B}} \sqcap_{{}_{\overline{S}_{AB}}}\!\!\!\! \bot_{{}_{ \mathfrak{S}_A}} \overline{\otimes}\beta^\star)(\alpha \sqcap_{{}_{{ \mathfrak{S}}_A}}\!\!\!\! (\alpha^\star \sqcup_{{}_{{ \mathfrak{S}}_A}}\!\!\alpha'{}^\star),\beta) = \bot$ because $(\alpha \sqcap_{{}_{{ \mathfrak{S}}_A}}\!\!\!\! (\alpha^\star \sqcup_{{}_{{ \mathfrak{S}}_A}}\!\!\alpha'{}^\star)) \sqsubset_{{}_{{ \mathfrak{S}}_A}} \alpha$, and (2) $(\alpha \overline{\otimes}\beta \sqcap_{{}_{\overline{S}_{AB}}}\!\!\!\! \alpha' \overline{\otimes}\beta')(\alpha \sqcap_{{}_{{ \mathfrak{S}}_A}}\!\!\!\! (\alpha^\star \sqcup_{{}_{{ \mathfrak{S}}_A}}\!\!\alpha'{}^\star),\beta)=\bot$ because $(\alpha \sqcap_{{}_{{ \mathfrak{S}}_A}}\!\!\!\! (\alpha^\star \sqcup_{{}_{{ \mathfrak{S}}_A}}\!\!\alpha'{}^\star)) \not\sqsupseteq_{{}_{{ \mathfrak{S}}_A}} \alpha'{}^\star$ and $\beta'\not\sqsupseteq_{{}_{{ \mathfrak{S}}_B}} \beta{}^\star$. As a result, we have 
\begin{eqnarray}
\Delta (\alpha \sqcap_{{}_{{ \mathfrak{S}}_A}}\!\!\!\! (\alpha^\star \sqcup_{{}_{{ \mathfrak{S}}_A}}\!\!\alpha'{}^\star),\beta)=\bot 
\;\;\;\; &\not= &\;\;\;\; \Delta (\alpha,\beta) \wedge 
\Delta (\alpha^\star \sqcup_{{}_{{ \mathfrak{S}}_A}}\!\!\alpha'{}^\star,\beta)=\textit{\bf N}\wedge \textit{\bf N}=\textit{\bf N}
\end{eqnarray}
which invalidates the bimorphic property.\\
As a conclusion, the attempt to build an interesting completion of $\overline{S}_{AB}$ in $\widehat{S}_{AB}$, allowing to restore the orthocompletion property for the star tensor product of generic orthocomplemented spaces of states, seems to be a dead end. Another construction must then be adopted to find the tensor product adapted to orthocomplementation, and then to quantum systems.
\end{remark}

\section{Conclusion}

Inspired by the {\em operational quantum logic program}, we suspect that probabilities can be viewed as a derived concept, even in a reconstruction program of Quantum Mechanics. The already cited remark of S. Abramsky \cite[Theorem 4.4]{Abramsky2012} can be viewed as another justification of this perspective on quantum mechanics.  These two perspectives have stimulated our desire to build an operational description based on a possibilistic semantic (in a sense, the 'probabilities' are replaced by statements associated to a semantic domain made of three values 'indeterminate', 'definitely YES', 'definitely NO'). The present paper intents to develop such an operational formalism. It will be called Generalized possibilistic Theory (GpT) as it is partly inspired by the formalism of Generalized Probabilistic Theory (GPT).  We note that we are also indebted to the work of Abramsky \cite{Abramsky2012} for our choice to give to Chu duality a central role in our construction, in replacement of traditional duality between states and effects. \\
The section 2 is devoted to a brief summary of the axiomatic relative to the space of states (subsection \ref{subsectionstates}), the space of effects (subsection \ref{subsectioneffects}), the set of pure states (subsection \ref{subsectionpurestates}), and the notion of "channels" or symmetries for our theory (subsection \ref{subsectionsymmetries}).  The section 2 collects and completes some elements already developed in our previous work \cite{Buffenoir2021}.  The convexity requirements imposed traditionally in GPT on the space of states and space of effects are naturally replaced  by Inf semi-lattice structures on these spaces in GpT, the set of pure states being naturally associated to completely meet-irreductible elements of the space of states.  Our central point is the Chu duality imposed between the space of states and the space of effects, with an evaluation space given by the three elements domain associated to possibilistic statements of the observer.  This Chu duality is sufficient to deduce the whole set of properties of the channels which are viewed as Chu morphisms. \\ 
The section 3 represents a first attempt to the construction of bipartite experiments on compound systems. This point is central because it has been the main obstacle on the pathway towards a complete reconstruction of quantum mechanics along the operational quantum logic program. The central problem in our perspective is the construction of a tensor product for our space of states and space of effects.  It is well known that this tensor product notion is ambiguous in GPT program \cite[Section 5]{Plavala} although this choice of tensor product is the key point for the choice of theory we want to describe within GPT. The traditional construction of tensor product of Inf semi-lattices should have been of some help for our work \cite{Fraser1976}, it is succinctly recalled in subsection \ref{subsectionbasic} and called {\em canonical tensor product}.  The tensor product, naturally build from the Chu construction \cite{Pratt1999}, may also have played a role here.  Surprisingly, the natural axiomatic for bipartite experiments, proposed in subsection \ref{subsectionaxiomatic},  leads to alternative constructions for the tensor product of Inf semi-lattices (the construction of the symmetries associated to the bipartite space of states is completed in subsection \ref{subsectionsymmetriesbipartite}).  To begin, the different tensor products are placed with respect to the largest one, which is called {\em maximal tensor product} and described in subsection \ref{subsectionmaximaltensor}.  The more simple construction is presented in subsection \ref{subsectionminimal} and called {\em basic tensor product}. The comparison between basic tensor product and canonical tensor product is made in subsection \ref{subsectioncomparison} and some remarks concerning the specific properties of the basic tensor product are made in subsection \ref{subsectionremarks}.  The subsection \ref{subsectionremarks} comes however to a sad conclusion : the basic tensor product of orthocomplemented spaces of states is eventually not orthocomplemented.  This point is crucial in order to use GpT for a reconstruction program of quantum theory, as it appears clear in our previous work \cite{Buffenoir2021}.  The basic tensor product appears then to correspond to a theory which is not quantum.  Another construction is detailed in subsection \ref{subsectionstartensor}. The new tensor product called {\em star tensor product} is defined in intimate connection with the star structures defined on the spaces of states which are tensored.  As shown at the end of subsection \ref{subsectionpropertiesstartensor},  this proposal is also not adequate to obtain an orthocomplementation property on the tensor product of orthocomplemented spaces of states. We recognize here, even if we have adopted a rather different perspective, the recurrent problem in defining a tensor product for quantum logic; this essential problem has already been noticed a long time ago by D. Foulis and C. Randall in \cite{randall1979tensor}. We intent to come back to this subject in a forthcoming paper.

\end{document}